\documentclass[12pt]{article}
\pdfoutput=1
\usepackage{jheppub}
\usepackage{subcaption}
\usepackage{pifont}

\usepackage{mathrsfs}
\usepackage{tikz}
\usepackage{subfiles}

\def\d{\operatorname{d}}

\newcommand{\cd}{{\cal D}}

\newcommand{\cn}{{\cal N}}

\newcommand{\cl}{{\cal L}}

\newcommand{\cj}{{\cal J}}

\newcommand{\ch}{{\cal H}}

\newcommand{\co}{{\cal O}}

\newcommand{\cz}{{\cal Z}}
\newcommand{\nn}{\nonumber}

\def\bal#1\eal{\begin{align}#1\end{align}}
\def\alp[#1]{\begin{align}#1\end{align}}

\def\secnum[#1]{\texorpdfstring{$#1$}{TEXT}}

\def\secnuml#1\secnumr{\texorpdfstring{$#1$}{TEXT}}

\def\eqa{\begin{eqnarray}}
\def\eqae{\end{eqnarray}}
\def\eq{\begin{equation}}
\def\eqe{\end{equation}}
\def\be{\begin{equation}}
\def\ee{\end{equation}}
\def\bea{\begin{eqnarray}}
\def\eea{\end{eqnarray}}
\def\ba{\begin{array}}
\def\ea{\end{array}}
\def\bd{\begin{displaymath}}
\def\ed{\end{displaymath}}

\def\Tr{{\rm Tr}}

\def\>{\rangle}
\def\<{\langle}
\def\a{\alpha}

\def\f{\phi}

\def\l{\lambda}
\def\m{\mu}
\def\n{\nu}

\def\p{\pi}
\def\q{\theta}

\def\s{\sigma}
\def\t{\tau}
\def\u{\upsilon}

\def\F{\Phi}

\def\L{\Lambda}

\def\S{\Sigma}

\def\pa{\partial}

\newcommand{\br}{{\mathbb{R}}}

\def\({\left(}
\def\){\right)}

\def\d{\text{d}}
\def\im{\text{i}}

\def\wh{\widehat}

\title{Half-Wormholes and Ensemble Averages}

\author{ Cheng Peng, Jia Tian and Yingyu Yang}
\affiliation{Kavli Institute for Theoretical Sciences (KITS),\\
University of Chinese Academy of Sciences (UCAS), Beijing 100190, China}
\emailAdd{pengcheng@ucas.ac.cn, wukongjiaozi@ucas.ac.cn, yangyingyu18@mails.ucas.ac.cn  }

\abstract{We study ``half-wormhole-like" saddle point contributions to spectral correlators in a variety of ensemble average models, including various statistical models, generalized 0d SYK models, 1d Brownian SYK models and an extension of it. 
	In statistical ensemble models, where more general distributions of the random variables could be studied in great details, we find the  accuracy of the previously proposed approximation for the half-wormholes could be improved when the distribution of the random variables deviate significantly from Gaussian distributions. We propose a modified approximation scheme of the half-wormhole contributions that also work well in these more general theories. In various generalized 0d SYK models we identify new half-wormhole-like saddle point contributions.
	In the 0d SYK model and 1d Brownian SYK model, apart from the wormhole and half-wormhole saddles, we find new non-trivial saddles in the spectral correlators that would potentially give contributions of the same order as the trivial self-averaging saddles. 
	However after a careful Lefschetz-thimble analysis we show that these non-trivial saddles should not be included. 
	We also clarify the difference between ``linked half-wormholes" and ``unlinked half-wormholes" in some models. 
}

\begin{document}

\maketitle

\section{Introduction}
The AdS/CFT correspondence \cite{Maldacena:1997re,Witten:1998qj,Gubser:1998bc} provides a non-perturbative definition of quantum gravity. An important lesson from the recently progress in understanding the black hole information paradox is that a summation of different configurations in the  semi-classical gravitational path integral is crucial to probe some quantum mechanical properties of the system, such as the Page curve~\cite{Penington:2019npb,Almheiri:2019psf,Almheiri:2019hni,Penington:2019kki}, the late-time behavior of the spectral form factor~\cite{Saad:2019lba,Saad:2018bqo}, and correlation functions~\cite{Saad:2019pqd,Yan:2022nod}, see also a recent review in~\cite{Bousso:2022ntt}. However, the inclusion of spacetime wormholes leads to an apparent factorization puzzle~\cite{Maldacena:2004rf}; a holographic computation of the correlation functions of field theory partition functions living on different boundaries gives non-factorized results, i.e.~$\<Z_{L}Z_{R}\>\neq \<Z_L\> \times \<Z_R\>$, which is in tension with the general expectation on the field theory side. 
This revitalizes the hypothetical connection between wormholes and ensemble averages~\cite{Coleman:1988cy,Giddings:1988wv,Giddings:1988cx,Polchinski:1994zs}, and motivates an appealing conjectural duality between a bulk gravitational theory and (the average of) an ensemble of theories on the boundary~\cite{Saad:2019lba,Stanford:2019vob, Iliesiu:2019lfc, Kapec:2019ecr,Maxfield:2020ale, Witten:2020wvy, Mefford:2020vde, Altland:2020ccq,Eberhardt:2021jvj,Stanford:2021bhl, Arefeva:2019buu, Betzios:2020nry, Anninos:2020ccj, Berkooz:2020uly, Mertens:2020hbs, Turiaci:2020fjj, Anninos:2020geh, Gao:2021uro, Godet:2021cdl, Johnson:2021owr, Blommaert:2021etf, Okuyama:2019xbv,Forste:2021roo, Maloney:2020nni, Afkhami-Jeddi:2020ezh,Cotler:2020ugk,Benjamin:2021wzr,Perez:2020klz,Cotler:2020hgz,Ashwinkumar:2021kav,Afkhami-Jeddi:2021qkf,Collier:2021rsn, Benjamin:2021ygh,Dong:2021wot,Dymarsky:2020pzc,Meruliya:2021utr,Bousso:2020kmy,Janssen:2021stl,Cotler:2021cqa,Marolf:2020xie,Balasubramanian:2020jhl,Gardiner:2020vjp,Belin:2020hea,Belin:2020jxr,Altland:2021rqn,Belin:2021ibv,Peng:2021vhs,Banerjee:2022pmw,Heckman:2021vzx, Johnson:2022wsr,Collier:2022emf,Chandra:2022bqq,Schlenker:2022dyo}, whose prototype is
the by-now well known duality between the two-dimensional Jackiw-Teitelboim (JT) gravity \cite{Jackiw:1984je,Teitelboim:1983ux} and the Schwarzian sector of the Sachdev-Ye-Kitaev (SYK) model~\cite{Sachdev:1992fk, KitaevTalk2}, or more directly the random matrix theories~\cite{Saad:2019lba,Stanford:2019vob}. Alternatively, an interesting question is whether there exist other configurations whose inclusion into the gravitational path integral would capture properties of a single boundary theory that are washed out after averaging over the ensemble.
This is closely related to the belief that solving the factorization problem will shed light on the microscopic structure of quantum gravity such as the microstates or the states behind the horizon of the black hole; these fine structures are not universal so they can not be captured by the ensemble averaged quantities \cite{Stanford:2020wkf,Almheiri:2021jwq}. 
In \cite{Saad:2021uzi}, the factorization problem is carefully studied in a toy model introduced in~\cite{Marolf:2020xie}, where it is shown that the (approximate) factorization can be restored if other half-wormhole contributions are included. In the dual field theory analysis,  these half-wormhole contributions are identified with non-self-averaging saddle points in the ensemble averaged theories. This idea is explicitly realized in a 0-dimensional ``one-time'' SYK model in~\cite{Saad:2021rcu}, followed by further analyses in different models~\cite{Mukhametzhanov:2021nea,Garcia-Garcia:2021squ, Choudhury:2021nal, Mukhametzhanov:2021hdi,Okuyama:2021eju,Goto:2021mbt,Blommaert:2021fob,Goto:2021wfs}. An explicit connection between the gravity computation in~\cite{Saad:2021uzi} and the field theory computation in~\cite{Saad:2021rcu} is proposed in~\cite{Peng:2021vhs}.

The construction of half-wormhole in~\cite{Saad:2021rcu} is based on the $G,\Sigma$ effective action of the model that comes from the Gaussian statistics of the random coupling. Furthermore, a prescription to identify the half-wormhole contribution is proposed and verified for the 0-dimensional SYK model and GUE matrix model in~\cite{Mukhametzhanov:2021hdi}. This raised a question of whether half-wormhole contributions also exist in different ensemble theories, such as those with random variables from a Poisson distribution~\cite{Peng:2020rno} or a uniform distribution on the moduli space~\cite{Maloney:2020nni,Afkhami-Jeddi:2020ezh,	Cotler:2020ugk,Perez:2020klz,	Benjamin:2021wzr,Dong:2021wot,Collier:2022emf,Chandra:2022bqq}, and whether these contributions share the same general properties as those discussed in~\cite{Saad:2021rcu} and~\cite{Mukhametzhanov:2021hdi}.

In this paper we study the half-wormhole-like contributions that characterize the distinct behaviors of each individual theory in an ensemble of theories, and test the approximation schemes of the half-wormholes in various models. Our main findings are summarized as follows.

\subsection{Summary of our main results}

\begin{itemize}
	\item[\ding{51}] To understand the nature of the half-wormhole contributions in the 1-time SYK model, an approximation scheme is proposed in~\cite{Mukhametzhanov:2021hdi}. Since the proposal does not rely on specific details of the SYK model, such as  the collective $G$ and $\Sigma$ variables, it is interesting to understand if there is a similar approximation that applies to more general ensemble averaged theories. In this paper, we first consider various statistical models with a single or multiple random variables. We compute a variety of different quantities, such as simple observables, power-sum observables and product observables, before and after the statistical average. We propose an approximation formula  for the half-wormhole like contributions in general statistical models, which generalizes the one in~\cite{Mukhametzhanov:2021hdi}, and show their validity explicitly. We find the validity of the ``wormhole/half-wormhole" approximation crucially depend on the large-$N$ factorization property of the observables we consider. The large-$N$ constraints such as traces and determinants play crucial roles in the validity of this approximation.

	\item[\ding{51}] We review the 0-dimensional SYK model introduced in~\cite{Saad:2021rcu} and fill in technical details of some calculations. In particular, in the saddle point analysis of various quantities, such as $\langle \Phi(\sigma)^2\rangle$ and others, we find new non-trivial saddle points whose on-shell values, including the 1-loop corrections, are of the same order as the the trivial saddle that is accounted for the half-wormhole. We then carry out explicit  Lefschetz-thimbles analyses to conclude that the contributions from these non-trivial saddle points should not be included in the path integral, which supports the previous results in~\cite{Saad:2021rcu}. We also extend some of the computations to two-loop order and again find our results support previous conclusions in~\cite{Saad:2021rcu}.

	\item[\ding{51}]We generalize the 0-dimensional SYK model so that the random coupling $J_{i_1\dots i_q}$ can be drawn from more general distributions, with non-vanishing mean or higher order cumulants. 
	
	When $J_{i_1\dots i_q}$ has a non-vanishing mean value, we find new half-wormhole saddle of $z$ in additional to the linked half-wormhole saddle of $z^2$. We introduce new collective variables $G,\Sigma$ to compute $\langle z\rangle$ and identify the contributions from the half-wormhole saddle. We further consider the half-wormhole proposal in this context. We find that depending on the relative ratio between the different cumulants, different ``multiple-linked-wormholes" could be dominant. In particular, in very special limits approximate factorization could hold automatically and no other ``half-wormholes" saddles are needed.
	
	In models with non-vanishing higher cumulants of the random coupling, e.g. $\langle J_{i_1,\dots i_q}^4\rangle \neq 0$, we find a similar conclusion that the saddle point contributes. Equivalently, the bulk configurations that dominate the path integral depends crucially on the ratios of the various cumulants and the result is not universal. 
	
	In addition, we do a preliminary analysis of models whose random couplings $J_{i_1,\dots i_q}$ are drawn from a discrete distribution, the Poisson distribution, where more complicated saddle points can be found.

	\item[\ding{51}] We do a similar analysis explicitly to the Brownian SYK model, and identify the wormhole and half-wormhole saddles at late time. The results are computed from both an explicit integration and a saddle point analysis, and we find a perfect agreement between them. We test the approximation of the partition function by its mean value and the half wormhole saddle, and further show that this approximation is good by demonstrating that the error of this approximation is small. Interestingly, like in the 0-dimensional model we also find non-trivial saddles for $\langle \Phi(\sigma)^2\rangle$ and they should be excluded by a similar Lefschetz thimble analysis.

	\item[\ding{51}]We further investigate  modified 0d and 1d SYK model whose random couplings have non-vanishing mean values that are written in terms of products of some background Majorana fermions~\cite{Goto:2021wfs}. We compute explicitly the wormhole and a new type of saddle point, the ``unlinked half-wormholes", that contribute to the partition function. We show these unlink half-wormholes are closely related to the disconnected saddles due to the non-vanishing mean value of the random coupling.
	\end{itemize}

\section{Statistical models}
\label{sm}

In this section we consider statistical models, which can be considered as toy models of the Random Matrix Theories, to test the idea of half-wormholes in ensemble theories with random variables drawn from different distributions.

\subsection{Models of a single random variable}\label{single}

\begin{figure}
	\centering
	\includegraphics[width=0.35\linewidth]{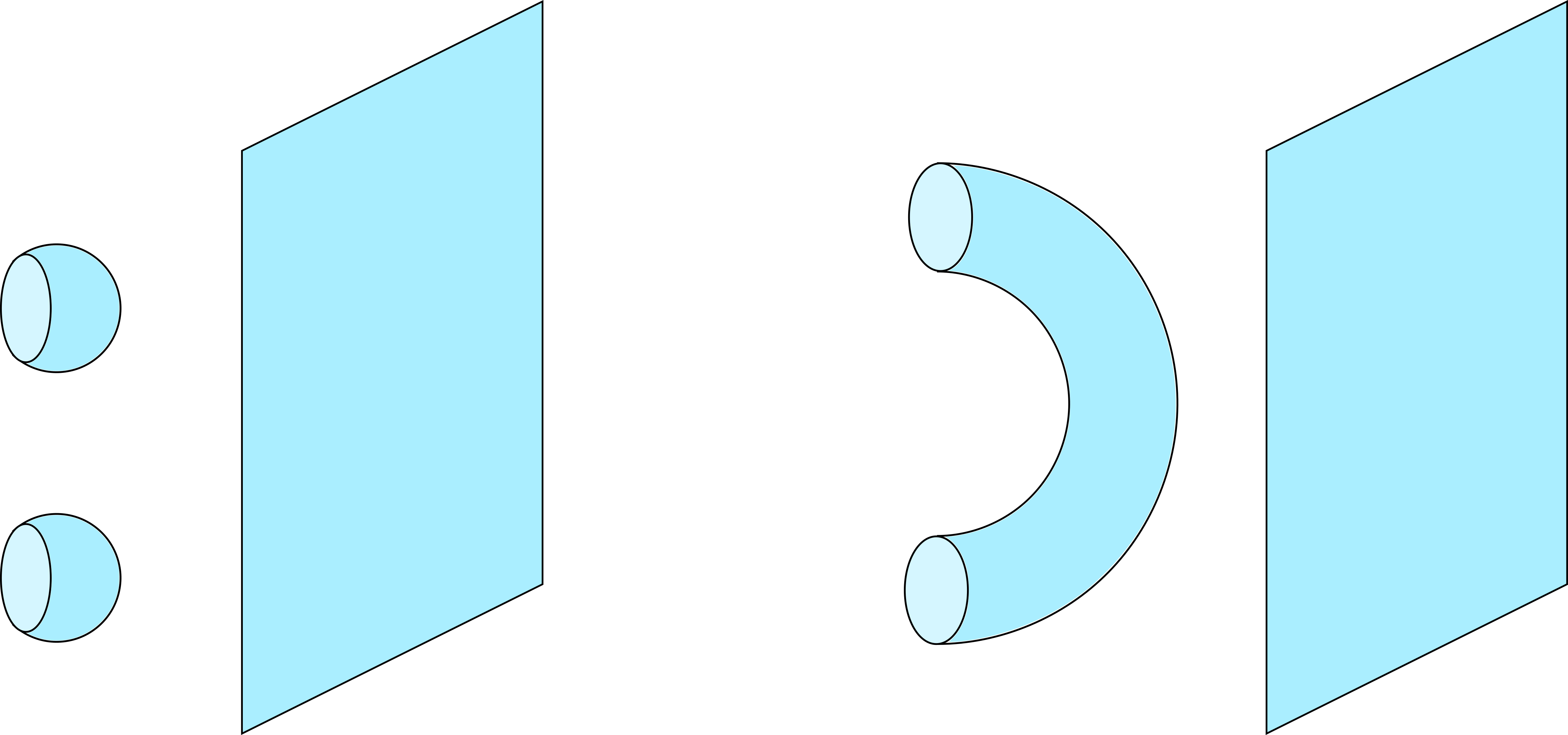}
	\hspace{1.8cm}\includegraphics[width=0.35\linewidth]{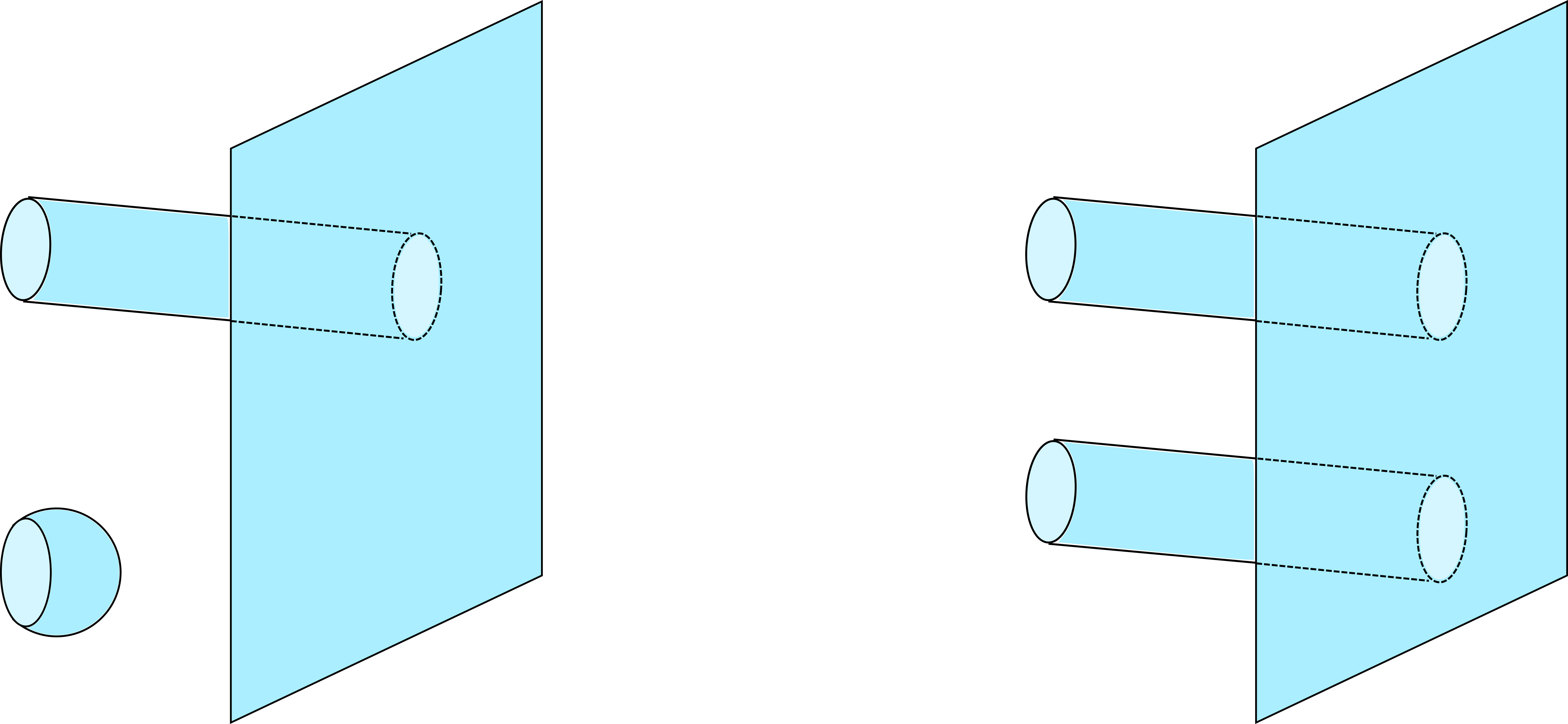}
	\caption{ Each $x$ denotes a circular boundary and the bracket $\langle \cdot\rangle$ denotes a bulk amplitude. The first two diagrams denote $\langle x^2\rangle \langle e^{\im k x}\rangle$ and the last two diagrams denote the ``connected" parts of the correlation function $2\langle x\rangle \langle x e^{\im k x}\rangle_c+\langle x^2 e^{\im k x}\rangle_c $.}
	\label{fac}
\end{figure}

Let $X$ be a random variable with a PDF $P(X)$ that satisfies the inequality
\bea
\langle X^2\rangle\geq \langle X\rangle^2\,,
\eea
that is valid for all conventional probability distributions. 
To identify the ``half-wormhole contributions'' in this model, we consider the unaveraged observable $X$,$X^2$ etc., and rewrite
\bal 
X^n&= \int dx\, \delta(x-X)\frac{x^n P(x)}{P(X)}=\int dx \int \frac{dk}{2\pi} \, e^{\im k (x-X)}\frac{x^n P(x)}{P(X)}=\int \frac{dk}{2\pi}  \frac{e^{-\im k X}}{P(X)} \langle x^n e^{\im k x}\rangle\label{y2}\,,
\eal 
where as usual the angle bracket denotes the average of $x$ with the probability distribution $P(x)$
\bea 
\langle \mathcal{O} e^{\im k x}\rangle=\int dx \, \mathcal{O} e^{\im kx}P(x)\ .
\eea
Such expectation values can further be decomposed into the connected and disconnected parts, for example
\bea 
&&\langle x e^{\im k x}\rangle=\langle x\rangle\langle e^{\im k x}\rangle+\langle x e^{\im kx}\rangle_{\text{c}}\,,\label{x2d1}\\
&&\langle x^2 e^{\im k x}\rangle=\langle x^2\rangle \langle e^{\im k x}\rangle+2\langle x\rangle\langle xe^{\im k x}\rangle_c+\langle x^2e^{\im k x}\rangle_{\text{c}},\label{x2d}\\
&&\langle x^3 e^{\im k x}\rangle=\langle x^3\rangle \langle e^{\im k x}\rangle+3\langle x^2\rangle\langle xe^{\im k x}\rangle_c+3\langle x\rangle\langle x^2e^{\im k x}\rangle_c+\langle x^3e^{\im k x}\rangle_{\text{c}}\,,\label{x2d3}\\
&&\dots \nonumber
\eea 
where the subscript $c$ denotes ``connected'' or ``cumulant'' which can be defined recursively as
\bea 
&&\langle x e^{\im kx}\rangle_{\text{c}}=\langle x e^{\im k x}\rangle-\langle x\rangle\langle e^{\im k x}\rangle,\\
&&\langle x^2e^{\im k x}\rangle_{\text{c}}=\langle x^2 e^{\im k x}\rangle-\langle x^2\rangle \langle e^{\im k x}\rangle-2\langle x\rangle\langle xe^{\im k x}\rangle_{\text{c}},\\
&&\dots \nonumber
\eea 
There is a diagrammatic way to understand this result that closely resembles the 2-dimensional topological gravity model which is introduced in \cite{Marolf:2020xie}. 
Formally writing
\bea 
\langle \mathcal{O}e^{\im k x}\rangle=\langle \mathcal{O}|e^{\im k x}\rangle ,
\eea 
we can interpret the state $|e^{\im k x}\rangle$ as a ``spacetime" D-brane$_{\im k}$ state that is similar to that introduced in~\cite{Marolf:2020xie}. Then the relation~\eqref{x2d} can be understood as in Figure~\ref{fac} where the meaning of the subscript $c$ is transparent.

We would like to get an estimation of the difference between any quantity $X^n$ and its ensemble average $\<X^n\>$, which requires a simple evaluation of $\<x^n e^{ikx}\>$. Motivated by the diagrams in Figure~\ref{fac} and a similar proposal in~\cite{Mukhametzhanov:2021hdi}, we propose the following approximation
\bea \label{apx}
{\langle x^2 e^{\im kx}\rangle_c }\approx \langle x e^{\im kx}\rangle_{\text{c}}\frac{1}{\langle e^{\im kx}\rangle }  \langle xe^{\im kx}\rangle_{\text{c}}\,,
\eea 
which has a diagrammatic interpretation as a recursive computation of configurations with a higher number of contractions to the spacetime brane from gluing the fundamental building blocks $ \langle xe^{\im kx}\rangle_{\text{c}}$ with the ``propagator'' $ \langle e^{\im kx}\rangle^{-1}$.
	
	Equivalently, this relation can be presented as 
	\bea 
	\frac{\langle x^2 e^{\im kx}\rangle }{\langle e^{\im kx}\rangle}&\approx &\langle x^2\rangle-\langle x\rangle^2+ \frac{\langle x e^{\im kx}\rangle \langle xe^{\im kx}\rangle}{\langle e^{\im kx}\rangle \langle e^{\im kx}\rangle} \\
	&= & \langle x^2\rangle+2\langle x\rangle \frac{\langle x e^{\im k x}\rangle_{\text{c}}}{\langle e^{\im kx}\rangle}+\frac{\langle x e^{\im kx}\rangle_{\text{c}} \langle xe^{\im kx}\rangle_{\text{c}}}{\langle e^{\im kx}\rangle \langle e^{\im kx}\rangle}\ .
	\eea 
	Making use of the fact that the quantity $\langle e^{\im kx}\rangle\equiv \varphi(k)$ is  the characteristic function of the probability distribution whose inverse Fourier transformation is the PDF
	\bea 
	\frac{1}{2\pi }\int \varphi(k) e^{-\im k X} dk=P(X)\,,
	\eea
	the relation~\eqref{apx} is equivalent to
	\bea
	X^2&\approx&\langle X^2\rangle-\langle X\rangle^2+\Phi\,, \quad 
	\Phi= \frac{1}{2\pi}\int dk \frac{e^{-\im k X}}{P(X)} \langle e^{\im kx}\rangle \(\frac{\langle x e^{\im kx}\rangle }{\langle e^{\im kx}\rangle}\)^2. \label{Phi0}
	\eea
	A more instructive form of this approximation is
	\bal
	X^2\approx\langle X^2\rangle+\tilde\Phi\,, \quad 
	\tilde\Phi= \frac{1}{2\pi}\int dk \frac{e^{-\im k X}}{P(X)} \langle e^{\im kx}\rangle \(\frac{\langle x e^{\im kx}\rangle^2_c }{\langle e^{\im kx}\rangle^2}
	+\frac{2\<x\>\langle x e^{\im kx}\rangle_c }{\langle e^{\im kx}\rangle}\)\,, \label{Phi1}
	\eal
	where $\<\tilde\F\>=0$. 
	We will call the connected piece $\<X^2\>_c\equiv \langle X^2\rangle-\langle X\rangle^2$ the ``wormhole'' contribution and $\F$ the ``half-wormhole'' contribution although it's mean value is non-vanishing.

	As a simple example,  the Gaussian distribution $\mathcal{N}(\mu,t^2+\m^2)$ has the non-vanishing cumulants
	\bea
	c_1=\mu,\quad c_2=t^2,
	\eea
	such that 
	\bea
	\frac{\langle x e^{\im kx}\rangle }{\langle e^{\im kx}\rangle}=\mu+\im k t^2,\quad  \(\frac{\langle x e^{\im kx}\rangle }{\langle e^{\im kx}\rangle}\)^2=\mu^2-k^2 t^4+2\im k \mu t^2.
	\eea
	Substituting the above into \eqref{Phi0} gives
	\bea
	\Phi=X^2-t^2,
	\eea
	which means that for Gaussian distribution the approximation \eqref{Phi0} is actually exact. 
	Clearly, this approximation cannot be exact for an arbitrarily general probability distribution. For example, for exponential distribution $\mathcal{E}(\lambda)$ the half-wormhole part is given by
	\bea
	\Phi=\frac{X^2}{2}\ ,\qquad x \geq 0\,,
	\eea
	and we quantify the error by its ratio to the variance of $X^2$
	\bea 
	\text{Error}=X^2-\langle X^2\rangle+\langle X\rangle^2-\Phi\,,\qquad 
	\rho=\frac{\langle\text{Error}^2\rangle}{\langle X^4\rangle}=\frac{5}{24}\ .\label{error}
	\eea
	
	In fact, the error of the approximation~\eqref{apx} or~\eqref{Phi0} can be derived explicitly for any general distribution. 
	Denoting the cumulants of the probability distribution as $c_n$, namely
	\bal
	\log\langle e^{\im kx}\rangle\equiv \log\varphi(k)=\sum_{n=0}^\infty c_n \frac{(ik)^n}{n!}\,,
	\eal
	we find\footnote{Notice that $\<\cdot\>_c$ is not a linear functional, so we don't expect similar relations for $\<x^n e^{ikx}\>$. }
	\bal
	(-\im \partial_k)\log\langle e^{\im k x}\rangle=\frac{\langle x e^{\im kx}\rangle }{\langle e^{\im kx}\rangle}=\frac{\langle x e^{\im kx}\rangle_c }{\langle e^{\im kx}\rangle}+\<x\>=\sum_{n=0}^\infty c_{n+1}\frac{(\im k)^n}{n!}\,, \label{1pc0}
	\eal
which means
\bal
\frac{\langle x e^{\im kx}\rangle_c }{\langle e^{\im kx}\rangle}=\sum_{n=1}^\infty c_{n+1}\frac{(\im k)^n}{n!}\ . \label{1pc}
\eal
	Similarly,
	\bal 
	(-\im \partial_k)^2\log\langle e^{\im k x}\rangle=\frac{\langle x^2 e^{\im k x}\rangle}{\langle e^{\im k x}\rangle}-\frac{\langle x e^{\im kx}\rangle \langle xe^{\im kx}\rangle}{\langle e^{\im kx}\rangle \langle e^{\im kx}\rangle}=\sum_{n=0}^\infty c_{n+2}\frac{(\im k)^n}{n!}\,,
	\eal 
	which means
	\bal 
	\frac{\langle x^2 e^{\im kx}\rangle_c }{\langle e^{\im kx}\rangle}- \frac{\langle x e^{\im kx}\rangle_{\text{c}} \langle xe^{\im kx}\rangle_{\text{c}}}{\langle e^{\im kx}\rangle \langle e^{\im kx}\rangle}=\sum_{n=1}^\infty c_{n+2}\frac{(\im k)^n}{n!}\ .
	\eal 
	The  approximation~\eqref{apx}  is thus originated from neglecting all higher $c_k$ with $k>2$.
	
	This  implies that indeed the approximation~\eqref{apx} or~\eqref{Phi0} is exact when the distribution is Gaussian, namely $c_n=0$ for $n>2$.
	
	Similarly we can consider the approximation of $X^n$. We first derive the approximation of  the connected correlators in the presence of spacetime brane.
Taking the higher order derivative of the cumulant generating functions, for example when $n=3$, we get
	\bal 
	(-\im \partial_k)^3\log\langle e^{\im k x}\rangle=\frac{\langle x^3 e^{\im k x}\rangle}{\langle e^{\im k x}\rangle}-3\frac{\langle x^{2} e^{\im kx}\rangle \langle xe^{\im kx}\rangle}{\langle e^{\im kx}\rangle \langle e^{\im kx}\rangle}+2\left(\frac{\langle x e^{\im kx}\rangle }{\langle e^{\im kx}\rangle }\right)^{3}\ .\label{d3log}
	\eal 
	Separating out connected and disconnected parts, we get
	\bal 
	(-\im \partial_k)^3\log\langle e^{\im k x}\rangle=\frac{\langle x^3 e^{\im k x}\rangle_{c}}{\langle e^{\im k x}\rangle}-3\frac{\langle x^{2} e^{\im kx}\rangle_{c} \langle xe^{\im kx}\rangle_{c}}{\langle e^{\im kx}\rangle \langle e^{\im kx}\rangle}+2\left(\frac{\langle x e^{\im kx}\rangle_{c} }{\langle e^{\im kx}\rangle }\right)^{3}+ \langle x^{3}\rangle_c\,,
	\eal 
	where
	\bal
	\langle x^{3}\rangle_c=\langle x^{3}\rangle -3\langle x^{2}\rangle \langle x\rangle +2\langle x\rangle^{3}\,,
	\eal
is the connected correlator that equals to $c_3$. Therefore we arrive at
	\bea
	&&\frac{\langle x^3 e^{\im k x}\rangle_{c}}{\langle e^{\im k x}\rangle}-3\frac{\langle x^{2} e^{\im kx}\rangle_{c} \langle xe^{\im kx}\rangle_{c}}{\langle e^{\im kx}\rangle \langle e^{\im kx}\rangle}+2\left(\frac{\langle x e^{\im kx}\rangle_{c} }{\langle e^{\im kx}\rangle }\right)^{3}=\sum_{n=1}^\infty c_{n+3}\frac{(\im k)^n}{n!}\ .
	\eea  
This means up to the third cumulant we have approximately 
	\bal 
	\frac{\langle x^3 e^{\im k x}\rangle_{c}}{\langle e^{\im k x}\rangle}\approx  3\frac{\langle x^{2} e^{\im kx}\rangle_{c} \langle xe^{\im kx}\rangle_{c}}{\langle e^{\im kx}\rangle \langle e^{\im kx}\rangle}-2\left(\frac{\langle x e^{\im kx}\rangle_{c} }{\langle e^{\im kx}\rangle }\right)^{3}\,,\label{app3}
	\eal 
	and the error of this approximation is due to neglecting  all $c_{k}$ with $k>3$.
	It is clear from  this computation that the error of  this approximation can be determined by \eqref{Phi0}.
	If the accuracy requirement is only up to the second moment, it up to quadratic fluctuations, we can use the approximation~\eqref{apx} again to get
	\bal
\frac{\langle x^3 e^{\im k x}\rangle_{c}}{\langle e^{\im k x}\rangle}\approx\left(\frac{\langle x e^{\im kx}\rangle_{c} }{\langle e^{\im kx}\rangle }\right)^{3}\ ,\label{app2}
	\eal
	which becomes exact when the distribution is Gaussian. In fact, we can derive similar relations by taking higher order derivatives in~\eqref{d3log} to get relations among higher order $\langle x^i e^{\im k x}\rangle_c$'s. If again we need accuracy up to quadratic order one can prove by induction
\bal
\frac{\langle x^n e^{\im k x}\rangle_{c}}{\langle e^{\im k x}\rangle}\approx\left(\frac{\langle x e^{\im kx}\rangle_{c} }{\langle e^{\im kx}\rangle }\right)^{n}\ .\label{appn}
\eal

	We can then approximate the un-average $X^3$ to a required accuracy. In practice, we rewrite the definition of $X^n$ according to~\eqref{y2}, then expand the $\langle x^n e^{\im k x}\rangle$ in~\eqref{y2} in terms of the connected correlators $\langle x^i e^{\im k x}\rangle_c$    according to e.g. \eqref{x2d1}-\eqref{x2d3}. Then depending on the accuracy requirement, we use relations analogous to either~\eqref{app3} or~\eqref{app2}, \eqref{appn}, to write down the approximation and the error of the final approximation is the composition of the errors the different approximations of $\<x^n e^{ikx}\>$. The  general expression of the approximation of $X^n$ and the corresponding errors are complicated. But we will present some general procedures that work for any distribution once an accuracy goal is given.    
	
\subsubsection{Recursion relations for approximations to arbitrary accuracy}
Define $\Phi_n=\frac{1}{2\pi}\int\frac{e^{\im k X}}{P(X)}\langle e^{\im k x}\rangle^{1-n} \langle x e^{\im k x}\rangle^{n}$, we have
\bal
X^m \Phi_n &= \frac{1}{2\pi}\int\frac{ X^m e^{\im k X}}{P(X)}\langle e^{\im k x}\rangle^{1-n} \langle x e^{\im k x}\rangle^{n}
=\frac{1}{2\pi}\int\frac{  e^{\im k X}}{P(X)}\left(-i\pa_k\right)^m\left(\langle e^{\im k x}\rangle^{1-n} \langle x e^{\im k x}\rangle^{n}\right)\ .
\eal 
Evaluating the derivative gives a result involving $\langle x^i e^{\im k x}\rangle$ with $1\leq i \leq m+1$. Rewriting them in terms of  $\langle x^i e^{\im k x}\rangle_c$ with the help of e.g. \eqref{x2d1}-\eqref{x2d3}. Then use the approximation either~\eqref{app3} or~\eqref{app2}, \eqref{appn} according to the required accuracy. Then rewrite the $\langle x^i e^{\im k x}\rangle_c$ in the approximated results back in terms of $\langle x^i e^{\im k x}\rangle$, and the result will be a relation among $\Phi_i$ with $1\leq i \leq m+1$. Making use of the fact that $\Phi_1 = X$ and recursively carrying out the above procedure to evaluate $X^{n-1}\Phi_1$, we get the approximation of $X^n$ to the desired accuracy.

For example, if we require accuracy to the second order, we simply consider
\bal
X \Phi_n 
&=\frac{1}{2\pi}\int\frac{  e^{\im k X}}{P(X)}\langle e^{\im k x}\rangle^{-n} \left(n\langle x^2 e^{\im k x}\rangle \langle x e^{\im k x}\rangle^{n-1}\langle e^{\im k x}\rangle+(1-n)\langle x e^{\im k x}\rangle^{n+1}\right)\ .
\eal 
Following the above procedure to rewrite $\langle x^2 e^{\im k x}\rangle$, we arrive at
\bal
X \Phi_n =n\left(\<x^2\>-\<x\>^2\right)\Phi_{n-1}+\Phi_{n+1}\ .
\eal
For example, we can evaluate
\bal
X^3=X^2\Phi_1=3 \left(\mu_2-\mu_1^2\right) \Phi_1+\Phi_3\,,\label{x3g}
\eal
where we keep only accuracy up to the quadratic order, so $\m_3$ does not appear independently; it is simply replaced by 
\bal
\mu_3=3\m_1\m_2-2\m_1^3\ .\label{mu3r}
\eal

\subsubsection{Explicit relations for Gaussian approximation}	
	If we only want Gaussian  approximations of $X^n$, we can get an explicit  approximation formula. 
	First let introduce some convenient notations
	\bea 
	&&\phi_n=\frac{\langle x^n e^{\im k x}\rangle }{\langle e^{\im k x} \rangle },\quad \phi^c_n=\frac{\langle x^n e^{\im k x}\rangle_c }{\langle e^{\im k x} \rangle },\\
	&&\langle x^n\rangle=\mu_n,\quad \langle x^n\rangle_{\text{cumulant}}=c_n.
	\eea 
	The cumulant $c_m$ can be expressed as a polynomial of moments
	\bea \label{cmp}
	c_m=P_m(\mu_m,\mu_{m-1},\dots,\mu_1).
	\eea 
	Some examples are
	\bea 
	c_1=\mu_1,\quad c_2=\mu_2-\mu_1^2,\quad c_3=\mu_3-3\mu_1\mu_2+2\mu_1^3,\dots
	\eea 
	Note that the coefficient of $\mu_m$ is 1.
	Of course the relations can be inverted
	\bea\label{qm}
	\mu_m=Q_m(c_m,c_{m-1},\dots c_1).
	\eea 
	Similar to \eqref{x2d1},\eqref{x2d} and \eqref{x2d3}, $\phi_n$ can be decomposed as
	\bea \label{phie}
	\phi_m=\tilde{P}_m(\phi_m^c,\dots,\phi_0^c),
	\eea 
	for example
	\bea 
	\phi_1=\phi_1^c+\mu_1\phi_0^c,\quad \phi_2=\phi_2^c+2\mu_1 \phi_1^c+\mu_2,\dots.
	\eea 
	Since $\log \langle e^{\im k x}\rangle$ is the generating function of $c_n$ we have \footnote{The simplest way to see this is to set $k=0$, then it reduces to \eqref{cmp} and to notice that the coefficients of the polynomial $P_m$ do not depend on $k$.}
	\bea 
	(-\im \partial_k)^m\log \langle e^{\im k x}\rangle=P_m(\phi_m,\phi_{m-1},\dots, \phi_1)=\sum_{n=0}c_{n+m}\frac{(\im k)^n}{n!}.
	\eea 
	Using \eqref{phie} and \eqref{qm} the left-hand side can be expanded as a polynomial of $c_i$ with coefficients to be functions of $\phi_i^c$:
	\bea 
P_m(\f_m,\f_{m-1},\dots,\f_{1})=	P_m(\tilde{P}_m(\f_i^c),\tilde{P}_{m-1}(\f_i^c),\dots,\tilde{P}_{1}(\f_i^c))\ .
	\eea 
	For example
	\bea 
	P_2&=&\phi_2-\phi_1^2 =\tilde{P}_2-2{\tilde{P}_1}^2\\
	&=&\phi_2^c+2\mu_1 \phi_1^c+\mu_2-{\phi_1^c}^2-\mu_1^2{\phi_0^c}^2-2\mu_1\phi_1^c\phi_0^c\\
	&=&\phi_2^c-{\phi_1^c}^2+c_1(2\phi_1^c-2\phi_1^c\phi_0^c)+c_2.
	\eea 
	Therefore we end up with
	\bea
	P_m=M_m+ c_1 M_{m-1}^{(1)}+(c_1^2 M_{m-2}^{(1)}+c_2 M_{m-2}^{(2)})+\dots+c_m=\sum_{n=0}c_{n+m}\frac{(\im k)^n}{n!}\,,
	\eea 
	where each $M_i^{(k)}$ is a function of the $\f^c_i$'s. Since the subscript $i$ of $\f^c_i$ and $M_i$ both indicate the power of $x$, it is clear that 
	\bal
	\sum_{a} i_a = m\,,  \qquad \forall \(\prod_{a} \f_{i_a}^c\) \in M_m \,,\label{ias}
	\eal
	where $\prod_{a} \f_{i_a}^c$ is any term in $M_m$.
	Notice that these relations are true for arbitrary $k$, $m$ and distributions, then the non-trivial solution is only 
	\bea 
\quad M_n^{(p)}=0\,,\qquad 	M_m=P_m(\phi^c_m,\phi^c_{m-1},\dots, \phi^c_1)=\sum_{n=1}c_{n+m}\frac{(\im k)^n}{n!}\ .\label{mp}
	\eea 
The Gaussian approximation means $c_m=0$ for all $m>2$. This requires
\bal
P_m(\phi^c_m,\phi^c_{m-1},\dots, \phi^c_1)\approx 0\,, \quad \forall m>1\ .\label{pm0}
\eal
At $m=2$ this relation means
\bal
P_2(\phi^c_2,\phi^c_{1})\approx 0\,,
\eal
which combines with~\eqref{ias} means $\phi^c_2 =\a \(\phi^c_{1}\)^2$ and
\bal
P_2(\a \phi^c_2,\phi^c_{1})=0\ .\label{p2a}
\eal
To fix the normalization $\a$, we notice that since the above relations \eqref{cmp} -\eqref{pm0}, in particular the functional form of $P$, are true for arbitrary distribution,   
we can choose the delta function distribution such that $c_n=0,\forall n\geq 2$ and $\mu_m=\mu_1^m$, we can get the identity
	\bea 
	P_m(\mu^m,\mu^{m-1},\dots \mu)=0,\quad m\geq 2,
	\eea 
	thus combining this with \eqref{p2a} we conclude $\a=1$ and 
	\bal
	\phi^c_2 \approx \(\phi^c_{1}\)^2\,,
	\eal
	where $\approx$ is due to the Gaussian approximation. This is nothing but the approximation~\eqref{apx}.
	Iterating this procedure successively for  different $m$, we reach to
	\bea 
    \phi_m^c\approx {(\phi_1^c)}^m\,,
	\eea 
	in the Gaussian approximation.
	Then we can approximate $X^m$ as
	\bea 
	X^m&=&\frac{1}{2\pi}\int\frac{e^{\im k X}}{P(X)}\langle e^{\im k x}\rangle \phi_m= \frac{1}{2\pi}\int\frac{e^{\im k X}}{P(X)}\langle e^{\im k x}\rangle \tilde{P}_m(\phi_m^c,\dots,\phi_1^c,1)\\
	&\approx&\frac{1}{2\pi}\int\frac{e^{\im k X}}{P(X)}\langle e^{\im k x}\rangle \tilde{P}_m((\phi_1^c)^m,(\phi_1^c)^{m-1},\dots,\phi_1^c,1) \\
	&=& \sum_{i=0}^m {m\choose i}\mu_i \Phi_{m-i}\,,\label{app2}
	\eea
	where $\Phi_i=\frac{1}{2\pi}\int\d k\frac{e^{\im k X}}{P(X)}\langle e^{\im k x}\rangle (\phi_1^c)^{i}$, and it may be understood as generalized wormholes which we will report somewhere else. 
	
	It is easy to check that the result~\eqref{app2} agrees with~\eqref{x3g} once the relation~\eqref{mu3r} is used. 
	
	\subsection{Models with multiple independent identical random variables}
	In statistical models with a single random variable, the various moments are all observables that we can compute. On the other hand, we would like to consider other interesting observables. We therefore proceed to consider operators in statistical models with multiple independent identical random variables. 	
	
	One class of operators in these models is the light operators that are simply linear combinations of the random variables $X_i$.
	We conjecture that if $Y(X_i)$ is some function of a large number $N$ independent random variables $X_i$ such that $Y$ is approximately Gaussian, then the approximation 
	\bea
	&&Y^2\approx \langle Y^2\rangle-\langle Y\rangle^2+\Phi,\label{appro}\\
	&&\Phi(X)= \frac{1}{(2\pi)^N}\int \prod_i \(dk_i \frac{e^{-\im  k_i X_i}}{P(X_i)}\)  \langle e^{\im \sum_ik_i x_i}\rangle \(\frac{\langle Y(x) e^{\im \sum_ik_ix_i}\rangle }{\langle e^{\im\sum_ik_ix_i}\rangle}\)^2. \label{Phi}
	\eea
	is good in the sense that 
	\bal
	\rho \equiv \frac{ \langle \text{Error}^2\rangle}{\langle Y^2\rangle^2}\,,
	\eal 
	is suppressed by $1/N$.
	
Like~\eqref{Phi1} we can rewrite it into 
	\bal
	&Y^2\approx \langle Y^2\rangle+\tilde\Phi,\label{appro2}\\
	&\tilde\Phi(X)= \frac{1}{(2\pi)^N}\int \prod_i \(dk_i \frac{e^{-\im  k_i X_i}}{P(X_i)}\)  \langle e^{\im \sum_ik_i x_i}\rangle \(\frac{\langle Y(x) e^{\im \sum_ik_ix_i}\rangle_c^2 }{\langle e^{\im\sum_ik_ix_i}\rangle^2}+\frac{2\<Y\>\langle Y(x) e^{\im \sum_ik_ix_i}\rangle_c }{\langle e^{\im\sum_ik_ix_i}\rangle}\). \label{Phi2}
	\eal

	\subsubsection{Simple observables}
	
The fundamental logic in this section is that by the central limit theorem (CLT), summing over a large number of i.i.d random variables gives a random variable that approximately obey a Gaussian distribution. 
Explicitly, if $X_i$ is from a normal distribution $\cn(\m,\s^2)$, then the mean of $N$ such i.i.d's 
\bea
\tilde{Y}=\frac{1}{N}\sum_{i=1}^N X_i\,,
\eea
is approximately a Gaussian random variable from $\cn(\m,\s^2/N)$ when $N$ is large enough.

In this paper, it turns out  that it is more convenient to define
\bal
Y=	\sum_{i=1}^N X_i\,,
\eal
so that the connection to the SYK model is more transparent.
Then $Y$ is a Gaussian random variable with probability distribution $\cn(N\m,N\s^2)$ when $N$ is large. In particular, we expect
\bal\label{expect}
\langle Y^4\rangle \approx 3\langle Y^2\rangle^2-2\<Y\>^4\,,\quad \<Y^2\>\approx N\(\<X^2\>-\<X\>^2\)+N^2 \<X\>^2\,, \quad \<Y\>\approx N\<X\>\  .
\eal
They can be checked by a direct calculation
\bea
&&\langle Y^2\rangle=N \langle X^2\rangle+N(N-1)\langle X\rangle^2,\\
&& \langle Y^4\rangle=N\langle X^4\rangle+N(N-1)\( 4\langle X^3\rangle \langle X\rangle+ 3\langle X^2\rangle^2\)\nn \\
&&\quad +6N(N-1)(N-2)\langle X^2\rangle \langle X\rangle^2+N(N-1)(N-2)(N-3)\langle X\rangle^4\ .
\eea
Because all the $X_i$ are independent so that it is straightforward to obtain
\bea \label{Yexp}
\frac{\langle Y e^{\im k_i x_i}\rangle}{\langle e^{\im k_i x_i}\rangle}=\sum_i\frac{\langle x_i e^{\im k_i x_i}\rangle}{\langle e^{\im k_i x_i}\rangle}\equiv \sum_i k_i[1].
\eea
Next we can rewrite the square of~\eqref{Yexp} into the diagonal terms and off-diagonal terms
\bea \label{Y2exp}
\(\frac{\langle Y e^{\im k_i x_i}\rangle}{\langle e^{\im k_i x_i}\rangle}\)^2=\sum_i k_i[1]^2+\sum_{i\neq j}k_i[1]k_j[1].
\eea
To compute the off-diagonal contributions to the half-wormhole, we observe that
\bea
&&\frac{1}{(2\pi)^N} \int \prod_{i}\( dk_i \frac{e^{-\im k_i X_i}}{P(X_i)}\)\langle e^{\im\sum_i k_i x_i}\rangle k_i[1]k_j[1] \\ 
&&=\frac{1}{(2\pi)^2}\int dk_i dk_j \frac{e^{-\im k_i X_i-\im k_j X_j}}{P(X_i)P(X_j)} \langle x_i e^{\im k_i x_i}\rangle \langle x_j e^{\im k_j x_j}\rangle =X_i X_j\ .
\eea
In terms of $\wh{k_i[n]^m}$ which are defined in \eqref{tildek} the half-wormhole can be written as
\bea 
\Phi=\sum_i \widehat{k_i[1]^2}+\sum_{i\neq j}X_i X_j,
\eea
and the error is given by
\bal 
\text{Error}&=\sum_{i}\( X_i^2-\widehat{k_i[1]^2}-t^2\),\quad t^2=\langle X_i^2\rangle-\langle X_i\rangle^2\,,\label{er2a}\\
\langle  \text{Error}^2\rangle&=\sum_{i,j}\langle(X_i^2-\widehat{k_i[1]^2})(X_j^2-\widehat{k_j[1]^2})\rangle+N^2 t^4-2N t^2 \sum_i (X_i^2-\widehat{k_i[1]^2})\ .\label{er2}
\eal
Recalling that $\langle Y^2\rangle\sim  N t^2$ so to prove the conjecture~\eqref{appro} we need to show that the $\co(N^2)$ term in~\eqref{er2} vanish. 
A direct calculation gives
\bea
 \langle \widehat {k_i[1]^2}\rangle_{X_i}&=&\int dX_i P(X_i) k_i[1]^2 \langle e^{\im k_ix_i}\rangle_{x_i} \frac{e^{-\im k_i X_i}}{P(X_i)}dk_i \\
 &=&\int dk_i \delta(-k_i)\langle e^{\im k_ix_i}\rangle_{x_i} k_i[1]^2=\langle X_i\rangle^2\ . \label{k12}
\eea
This means
\bea 
\langle(X_i^2-\widehat{k_i[1]^2}\rangle=\langle X^2_i\rangle-\langle X_i\rangle^2=t^2\ . \quad \Leftrightarrow \quad  \langle \widehat {k_i[1]^2}\rangle_{X_i} \approx \langle X_i\rangle^2\ .\label{xkt}
\eea
In particular, a consequence of this  relation is that although all the 3 terms in~\eqref{er2} are of order $\co(N^2)$, the sum of them cancelled exactly since~\eqref{xkt} does not depend on $i$. 
This then shows that $\langle  \text{Error}^2\rangle \ll \<Y^2\>$ and hence the approximation~\eqref{appro} is valid.

We can derive this result in a more illuminating fashion.
First using \eqref{1pc} $k_i[1]$ can be expressed as
\bea 
k_i[1]=\sum_{n=0} \frac{(-\im k)^n}{n!} c_{n+1}.
\eea
Then using the fact that the inverse Fourier transformation of the characteristic function is the PDF we find 
\bea 
&&\langle \widehat {k_i[1]^2}\rangle_{X_i}=\int dX_i P(X_i) \sum_{n,m=0}\frac{c_{n+1}c_{m+1}}{n!m!}(-\im k_i)^{n+m} \langle e^{\im k_ix_i}\rangle_{x_i} \frac{e^{-\im k_i X_i}}{P(X_i)}dk_i \\
&&\quad =\int dX_i \sum_{n,m=0}\frac{c_{n+1}c_{m+1}}{n!m!}(\pa_{X_i})^{n+m} P(X_i)  =c_1^2=\langle X_i\rangle^2\ .
\eea

\subsubsection{Power-sum observables}
In this section, we consider another class of more general observables 
\bea 
Y=\sum_i f(X_i),\quad Y^2=\sum_{i,j} f(X_i)f(X_j),
\eea
where $X_i$ are still independent identical random variables with PDF $P_{X_i}$ and $f$ is some smooth function so that $f(X_i)$ are also independent and identical random variables with a new PDF $P_{f}$:
\bea
\int dX F[f(X)]P_{X}=\int df F(f) P_f.
\eea
The CLT is still valid but the proposal may not because naively it depends on the function $f$. By smooth function we mean $f(X_i)$ is not singular anywhere such that it can be Taylor expanded
\bea
f(X_i)=\sum_n  a_n X_i^n\,,
\eea
whose expansion coefficients satisfy
\bal
a_n \approx 0 \,, \qquad \forall n>n_0\,, \quad n_0\ll N\ . 
\eal
Accordingly \eqref{Yexp} and \eqref{Y2exp} become
\bea 
&&\frac{\langle Y e^{\im k_i x_i}\rangle}{\langle e^{\im k_i x_i}\rangle}=\sum_i \sum_n a_n k_i[n].\\
&&\(\frac{\langle Y e^{\im k_i x_i}\rangle}{\langle e^{\im k_i x_i}\rangle}\)^2=\sum_{i,j}\(\sum_n a_n k_i[n]\sum_m a_m k_j[m]\).
\eea
So the error is given by
\bal 
\text{Error}&=\sum_i\(f^2(X_i)-t^2-\sum_{n,m}a_na_m \wh{k_i[n]k_i[m]}\) \,,\label{E2a}\\
 \langle\text{Error}^2\rangle&=\langle \sum_{i,j}(f^2(X_i)-\sum_{n,m}a_na_m \wh{k_i[n]k_i[m]} )(f^2(X_j)-\sum_{n,m}a_na_m \wh{k_j[n]k_j[m]} )\rangle \nn\\  &\quad +N^2t^4-2Nt^2\sum_i(f^2(X_i)-\sum_{n,m}a_na_m \wh{k_i[n]k_i[m]} )\,,\label{E2}
\eal
where $t^2=\langle f^2(X_i)\rangle-\langle f(X_i)\rangle^2$.
Similar to the calculation of~\eqref{k12}, one can find
\bea 
\langle \sum_{n,m}a_na_m \wh{k_i[n]k_i[m]} \rangle=\langle f(X_i)\rangle^2,
\eea
which means the leading order terms, ie of order $N^2$, in~\eqref{E2} is  
\bea \label{E2N2}
2\(\langle (f^2(X_j)-\sum_{n,m}a_n a_m \wh{k_j[n]k_j[m]} )\rangle^2-t^4\) N^2 =0\,,
\eea
As a result, the error is small and indeed the approximation~\eqref{appro} is reasonable in this case too. We also show some explicit examples in the Appendix \eqref{exam}. More generally, following the same procedure one can show that the half wormhole proposal is correct for the following family of functions
\bea
Y_k= \sum_{i}^N \(f(X_{i_1},X_{i_2},\dots,X_{i_k})\),\label{Genf}
\eea
where $X_{i_p}$ are independent and identical random variables.

\subsubsection{Product observables}\label{nontrace}
Previously  the function $Y$ we considered are a summation of (polynomials of) independent random variables. The proposal works very well for all the probability distributions.  However in the original construction of half wormhole introduced in~\cite{Saad:2021rcu}, the function $Y$ is a determinant observables which  are ``heavy'' in the traditional field theory language
\bea \label{0SYK}
Y=\text{PF}(J)=\sum'_{A_1<A_2 < \dots<A_p}\text{sgn}(A) {J}_{A_1}{J}_{A_2}\dots {J}_{A_p},
\eea
where the function $\text{PF}(J)$ is called the hyperpfaffian \cite{Barvinok} which is a tensorial generalization of pfaffian and $J_{A_i}$ are random variables. To mimic this construction let us consider a similar model:
\bea 
Y=\sum_{i_1\neq i_2 \neq \dots\neq i_q}^N X_{i_1}X_{i_2}\dots X_{i_q}. \label{GenY}
\eea
$\bullet$ {\bf $q=2$ Gaussian distribution }\\
The simplest case is $q=2$:
\bea
&& Y=\sum_{i\neq j}X_i X_j, \label{Xij2}\\
&& Y^2=\sum_{i\neq j\neq p\neq q }X_i X_j X_p X_q+4\sum_{i\neq j\neq p}X_i^2 X_j X_p+2\sum_{i\neq j}X_i^2 X_j^2\ .\label{Xij2s}
\eea
It is straightforward to get
\bea
&&\langle Y^2\rangle=N(N-1)\(2t^4+4(N-1)\mu^2t^2+N(N-1)\mu^4 \),\label{ay22}\\
&&\langle X_i\rangle=\mu,\quad\langle X^2\rangle-\langle X\rangle^2=t^2.
\eea
So in general $\langle Y^2\rangle$ will scale as $N^4$ if $\m\neq 0$, while if $\mu=0$ it scales as $N^2$. 

One example of the $\mu=0$ case is the Gaussian distribution $\mathcal{N}(\mu=0,t^2)$. We then verifies 
\bea
&&\langle Y\rangle=0,\quad \langle Y^2\rangle=2t^4N(N-1)
\eea
and 
\bal 
\Phi=\sum_{i\neq j\neq p\neq q }X_i X_j X_p X_q+4\sum_{i\neq j\neq p}(X_i^2-t^2) X_j X_p+2\sum_{i\neq j}(X_i^2-t^2) (X_j^2-t^2)\ .
\eal
Therefore we obtain
\bea
&&\text{Error}=-2t^4N(N-1)+4t^2(N-2)\sum_{i\neq j}X_i X_j+4(N-1)t^2\sum_i X_i^2-2t^4N(N-1),\nn \\
&&\langle (\text{Error}/4)^2\rangle=(2+1+1-2)N^4t^8+\# N^3+\dots \label{errf1}
\eea
the leading term does not vanish so the approximation
\bal
Y^2\approx \<Y^2\>-\<Y\>^2+\Phi\,,
\eal
is not good. 

However, for more general Gaussian distributions $\mathcal{N}(\mu,t^2)$ similar calculation gives
\bea
\langle Y\rangle=N(N-1)\mu^2,\quad \langle Y^2\rangle-\langle Y\rangle^2\equiv\tilde{t}^2=2t^2N(N-1)(t^2+2(N-1)\mu^2),
\eea
and
\bea
\text{Error}=-\tilde{t}^2+4t^2(N-2)\sum_{i\neq j}X_i X_j+4(N-1)t^2\sum_i X_i^2-2t^4N(N-1),
\eea
now we find that 
\bea
 \langle \text{Error}^2\rangle= 32 (3 t^2\mu^2+\mu^4)N^5+32(t^4-12 t^2 \mu^2-4\mu^4)N^4+\dots
\eea
and
\bea
\frac{ \langle \text{Error}^2\rangle}{\langle Y^2\rangle^2}= \frac{2(\mu^4+3t^2\mu^2-3t^4)}{(2t^4-4t^2\mu^2)^2 N}+\dots.
\eea
Notice that the error is always small, even when $\mu \rightarrow 0$, and the proposal is valid. This is because when $\mu\neq 0$, the moments of $Y$ behave as
\bea 
&&\langle Y\rangle\approx N^2 \mu,\quad \langle Y^2\rangle \approx N^4 \mu^2,\quad \langle Y^4\rangle \approx N^8\mu^4,
\eea 
as expected from \eqref{expect}. It is thus clear that the $\m\to 0 $ limit is not smooth.

It seems $\m\neq 0$ is fundamentally better than the $\m=0$ case in the sense that the approximation~\eqref{appro} is good. But as we will discuss shortly in section~\ref{exp1} this is not the case and the crucial point is that it is more appropriate to compare the error with the connected contributions and left out the disconnected contributions.

$\bullet$ {\bf General $q$}\\
	Next we consider general distributions.  We show some details of the computation for exponential distribution and Poisson distribution in the Appendix \eqref{exam2}. Here we only give a more abstract derivation.  In terms of \eqref{tildek} the half wormhole \eqref{Phi} can be written as
\bea
\Phi&=&\sum_{i\neq j\neq p\neq q }\wh{k_i[1]}\wh{ k_j[1]}\wh{ k_p[1]}\wh{ k_q[1]}+4\sum_{i\neq j\neq p}\wh{k_i[1]^2 }\wh{k_j[1] }\wh{k_p[1]}+2\sum_{i\neq j}\wh{k_i[1]^2}\wh{ k_j[1]^2}\\
&=&\sum_{i\neq j\neq p\neq q }X_i X_j X_p X_q+4\sum_{i\neq j\neq p}\wh{k_i[1]^2} X_j X_p+2\sum_{i\neq j}\wh{k_i[1]^2}\wh{ k_j[1]^2}. \label{HWg2}
\eea
Therefore the error of the proposal is 
\bea\label{errorij}
\text{Error}= 4\sum_{i\neq j\neq q}(X_i^2-\wh{k_i[1]^2})X_jX_p+2\sum_{i\neq j}(X_i^2X_j^2-\wh{k_i[1]^2}\wh{k_j[1]^2})-\tilde{t}^2.
\eea
The maximal power of $N$ in $\langle \text{Error}^2\rangle$ will be $6$. 

When $\mu \neq 0$, $\langle Y^2\rangle^2\sim N^8$. So in this case the error is small and the approximation is good. 

When $\mu=0$, $\langle Y^2\rangle^2\sim N^4$. The terms of $N^4$ in $\langle \text{Error}^2\rangle$ come from
\bea
\langle\text{Error}^2\rangle&=&\langle  \sum_{i\neq j\neq p\neq q}\{16\times 2(X_i^2-\wh{k_i[1]^2})(X_j^2-\wh{k_j[1]^2})X_p^2 X_q^2 \nn \\
&+& 4(X_i^2X_j^2-\wh{k_i[1]^2}\wh{k_j[1]^2})(X_p^2X_q^2-\wh{k_p[1]^2}\wh{k_q[1]^2})\} \nn \\
&+& 4 t^{16} N^4-8t^4N^2 \sum_{i\neq j}(X_i^2X_j^2-\wh{k_i[1]^2}\wh{k_j[1]^2})\rangle+\dots \\
&=&N^4t^{16} \(32 + 4+4-8\)+\#N^3\dots=32N^4t^{16}+\#N^3\dots \label{fail1}
\eea
which is not vanishing so the error is large and we cannot approximate $Y^2$ by $ \<Y^2\> + \Phi$ probably for the same reason as the $q=2$ case.
One could ask that when $\langle X_{i}^{2}-\widehat{k_{i}[1]^{2}}\rangle=0$, the approximation might be fine, but it requires $t^{2}=0$ which we do not consider at the moment.

$\bullet$ {\bf General distributions}

Now we consider the general case \eqref{GenY}:
\bea 
&&Y=\sum_{i_1\neq i_2 \neq \dots\neq i_q}^N X_{i_1}X_{i_2}\dots X_{i_q} ,\\
&&Y^2=\sum_{k=0}^q \frac{(q!/(q-k)!)^2}{k!}\sum_{j_1\neq j_2\dots j_k\neq i_{1}\dots \neq i_{2q-2k}} X_{j_{1}}^2\dots X_{j_{k}}^2 X_{i_{1}}\dots X_{i_{2q-2k}}.
\eea
If $N\gg q$ then the average $\langle Y^2\rangle$ will have the following scaling behavior in the large $N$ limit
\bea
 \langle Y^2\rangle \sim \begin{cases}  
N^{2q}\mu^{2q} & \quad\mu\neq 0
\\
N^q q! t^{2q} & \quad \mu=0
\end{cases}
\eea
Similar to \eqref{HWg2}, one can find that the half wormhole contribution $\Phi$ can be written as
\bea
\Phi&=&  \sum_{k=0}^q \frac{(q!/(q-k)!)^2}{k!}\sum_{j_1\neq j_2\dots j_k\neq i_{1}\dots \neq i_{2q-2k}} \wh{k_{j_{1}}[1]^2}\dots \wh{k_{j_{k}}[1]^2} X_{i_{1}}\dots X_{i_{2q-2k}},
\eea
so that the error is 
\bea
 \text{Error}&=& \sum_{k=1}^q \frac{(q!/(q-k)!)^2}{k!}\sum_{\substack{j_1\neq j_2\dots j_k\neq \\ i_{1}\dots \neq i_{2q-2k}}}(X_{j_{1}}^2\dots X_{j_{k}}^2-\wh{k_{j_{1}}[1]^2}\dots \wh{k_{j_{k}}[1]^2})X_{i_{1}}\dots X_{i_{2q-2k}} \nn \\
 &-&\langle Y^2\rangle+\langle Y\rangle^2.
\eea
When $\mu\neq0$, the leading contribution to $\langle\text{Error}^2\rangle$ scales as $N^{2q-2}$ so the approximation~\eqref{appro} is correct. 

However when $\mu=0$, the leading contributions to $\langle\text{Error}^2\rangle$ are
\bal
\langle \text{Error}^2\rangle&=E_1+E_2+\# N^{2q-1},\\
E_1&=\langle  \sum_{k=1}^q  \(\frac{(q!/(q-k)!)^2}{k!}\)^2 (2q-2k)!\\
&\quad \times\sum_{\substack{j_1\neq j_2\dots \neq j_{2k}\\
\neq i_1\neq \dots \neq i_{2q-2k}}}(X_{j_1}^2-\wh{k_{j_{1}}[1]^2})\dots (X_{j_k}^2-\wh{k_{j_{2k}}[1]^2})X_{i_1}^2\dots X_{i_{2q-2k}}^2 \rangle\nn\\
\quad &=N^{2q}t^{4q}(2 q)! \left(\, _3F_2\left(-q,-q,-q;1,\frac{1}{2}-q;\frac{1}{4}\right)-1\right)\neq 0,\\
E_2&=\langle \( q! \sum_{i_1\neq i_2 \neq \dots\neq i_q}^N (X_{i_1}^2\dots X_{i_q}^2-\wh{k_{i_1}[1]^2}\dots \wh{k_{i_q}[1]^2})-q! N^q \)^2\rangle=0\ .
\eal
So the error is large as in the previous case \eqref{fail1} and the approximation~\eqref{appro} is not good. 

In our toy model \eqref{GenY} we did not include the ``diagonal" terms while from our analysis above we have shown in the large $N$ limit it is the ``off-diagonal" term that dominates. So our conclusions for \eqref{GenY} are also valid for the following general function
\bea
Y=\sum_{i_1,i_2,\dots,i_q=1}^N X_{i_1}X_{i_2}\dots X_{i_q}. \label{GenY2}.
\eea
As a simple demonstration, let us still consider the simplest case with $q=2$:
\bea
&&Y=\sum_{i,j}X_i X_j,\\
&&Y^2=\sum_{i}X_i^4+4\sum_{i\neq j}X_i^3 X_j+3\sum_{i\neq j}X_i^2 X_j^2 \nonumber \\
&&\qquad+6\sum_{i\neq j\neq k}X_i X_j X_k^2+\sum_{i\neq j\neq m\neq n}X_i X_jX_m X_n.
\eea
Comparing  
\bea
\langle Y^2\rangle=&& N^4 \kappa_1^4+4 N^2 \kappa_3\kappa_1+3N^2 \kappa_2^2+6N^3 \kappa_2\kappa_1^2+N\kappa_4,\\
&&\kappa_1=\langle X\rangle=\mu,\quad \kappa_2=\langle X^2\rangle-\langle X\rangle^2=t^2,\\
&&\kappa_3=\langle X^3\rangle-3\langle X\rangle \langle X^2\rangle+2\langle X^3\rangle,\\
&&\kappa_4=\langle X^4\rangle-3\langle X^2\rangle^2-4\langle X\rangle \langle X^3\rangle+12\langle X\rangle^2\langle X^2\rangle-6\langle X\rangle^4,
\eea
with \eqref{ay22} one find that if $t\neq 0$, the scaling behavior of $\langle Y^2\rangle$ is same as before. The half wormhole contribution $\Phi$ can be work out similarly:
\bea
\Phi&=&\sum_i\wh{k_i[2]^2}+\sum_{i\neq j} \wh{k_i[2]}\wh{k_j[2]} +\sum_{i\neq j\neq m\neq n} X_iX_jX_mX_n+4\sum_{i\neq j\neq m}\wh{k_i[1]^2}X_jX_m \nn \\
&+&2\sum_{i\neq j}\wh{k_i[1]^2}\wh{k_j[1]^2}+2 \sum_{i\neq j\neq k}\wh{k_i[2]} X_j X_k+4\sum_{i\neq j}\wh{k_i[2]k_i[1]} X_j
\eea
Then the error is given by
\bea\label{errorii}
\text{Error}&=&4\sum_{i\neq j\neq k}(X_i^2-\wh{k_i[1]^2})X_j X_k+2\sum_{i\neq j\neq k}(X_i^2-\wh{k_i[2]})X_j X_k+2\sum_{i\neq j}(X_i^2 X_j^2-\wh{k_i[1]^2}\wh{k_j[1]^2})\nn\\
&+&\sum_{i\neq j}(X_i^2 X_j^2-\wh{k_i[2]}\wh{k_j[2]})+4\sum_{i\neq j}(X_i^3-\wh{k_i[2]k_i[1]})X_j+\sum_i(X_i^4-\wh{k_i[2]^2})\rangle-\tilde{t}^2. \nonumber \\
&=&4\sum_{i\neq j\neq k}(X_i^2-\wh{k_i[1]^2})X_j X_k+2\sum_{i\neq j}(X_i^2 X_j^2-\wh{k_i[1]^2}\wh{k_j[1]^2})\nn\\
&+&4\sum_{i\neq j}(X_i^3-\wh{k_i[2]k_i[1]})X_j+\sum_i(X_i^4-\wh{k_i[2]^2})\rangle-\tilde{t}^2, \label{extr2}
\eea
where we have used the identity
\bea 
\wh{k[2]}=\int \d k \frac{e^{-\im k X}}{P(X)} \langle x^2 e^{\im k x}\rangle=X^2.
\eea 
Comparing with \eqref{errorij}, there are two extra terms in \eqref{errorii}, but they will never contribute\footnote{If $\mu\neq 0$ they maximally contribute to $N^5$ and when $\mu=0$ they maximally contribute to $N^3$.} to the leading power of $N$ when $t\neq0$. So again it seems the approximation~\eqref{appro} is good when $\m\neq 0$ but not good when $\m=0$. We will explain in the next section how to understand these results and modify the proposal~\eqref{appro}.

\subsection{Large-$N$ constraints and half-wormhole approximation}

In the previous sections we consider a few different examples. To summarize, the half-wormhole conjecture~\eqref{appro} and~\eqref{Phi} is valid for a large families of statistical models. However, for some examples discussed in section~\ref{nontrace} this approximation is not good.

\subsubsection{Why and how to modify the approximation proposal}\label{exp1}
The failed examples indicate that the proposed $\Phi$ does not capture all semi-classical components in the observable $Y^2$ to be approximated.

As discussed previously, the approximation~\eqref{appro} should come from the approximation~\eqref{apx}. The relation~\eqref{apx} indeed fails for the case where the approximation~\eqref{appro} is not good in section~\eqref{nontrace}. To see this explicitly, we consider the simplest example~\eqref{Xij2} where   
\bal
Y^2&=\sum_{i\neq j\neq p\neq q }X_i X_j X_p X_q+4\sum_{i\neq p\neq q}X_i^2 X_p X_q+2\sum_{i\neq j}X_i^2 X_j^2\,,
\eal
which means we need to consider the following terms in the approximation $\F$
\bal
\<X_i X_j X_p X_q e^{i \sum_{a} k_a X_a }\>\,,\qquad \< X_j^2 X_p X_q e^{i \sum_{a} k_a X_a }\>,,\qquad \< X_i^2 X_j^2 e^{i \sum_{a} k_a X_a }\>\ .\label{list}
\eal
However, in the proposal~\eqref{appro} the $\F$ term contains only $\<Ye^{ik_a x_a}\>^2$, which means only terms like
\bal
\<X_i X_j  e^{i \sum_{a} k_a X_a }\> \<X_p X_q e^{i \sum_{a} k_a X_a }\>\,, \qquad i\neq j\,, p \neq q\,, \label{list2}
\eal
contribute. 
Therefore to check why the proposal~\eqref{appro} that fails, we want to understand what is ``missing" in~\eqref{list2} comparing with the correct answer involving~\eqref{list}.

Because the $x_i$'s are identical independent random variables, the cumulant $c_n$ for each $x_i$ are the same and the moment generating function is just a product of the moment generating functions of each $x_i$. Therefore we can reduce the problem of finding a good approximation of the above product terms to each flavor of $x_i$ and find the approximation for each of them. This should give a good approximation for each term. \footnote{Although this would obscure the interpretation of $Y$ as an independent function, we still choose to proceed this way in order to check how the approximation~\eqref{appro} fails. }

Recall the approximation is to replace $\frac{\<X_i^n e^{ikx}\>_c}{\<e^{ikx}\>}$ by $\(\frac{\<X_i  e^{ikx}\>_c}{\<e^{ikx}\>}\)^n$ for $n>1$, ie~\eqref{appn}, therefore only the last two terms in~\eqref{list} are affected by the approximation. In particular, the first term in~\eqref{list} gives the same contribution as the term~\eqref{list2} that leads to the inaccurate approximation~\eqref{appro}. So the non-vanishing contributions from the last two terms in the leading order of $1/N$ should then be responsible for the failure of the approximation~\eqref{appro} in this example. As discussed above, a good approximation to the $x_j$ factor of $\< X_j^2 X_p X_q e^{i \sum_{a} k_a X_a }\>$ should be 
\bal
\frac{\< X_j^2  e^{i  k_j X_j }\>}{\<   e^{i  k_j X_j }\>}\approx \<X_j^2\>+2\<X_j\>\frac{\< X_j  e^{i  k_j X_j }\>_c}{\<   e^{i  k_j X_j }\>} +\(\frac{\< X_j  e^{i  k_j X_j }\>_c}{\<   e^{i  k_j X_j }\>}\)^2\ .
\eal
The contribution to the half-wormhole $\F$ from this term $\< X_j^2 X_p X_q e^{i \sum_{a} k_a X_a }\>$ is thus
\bal
(t^2+\m^2) X_p X_q+2 \m X_j X_p X_q+\F_2^j X_p X_q\ .\label{con1}
\eal
Similarly, the $\< X_p^2 X_q^2 e^{i \sum_{a} k_a X_a }\>$ type terms gives a contribution 
\bal
&(t^2+\m^2)^2+4\m^2  X_p X_q +\F_2^p \F_2^q +4 \m (t^2+\m^2) \(X_p+X_q\) \nn\\
&\quad +(t^2+\m^2) \(\F_2^p+\F_2^q\)+4\m  \(X_p \F_2^q +X_q \F_2^p\)\ .\label{con2}
\eal
Now we should sum over $j,p,q$ to get all the contributions to the computation of $\<Y^2\>$ and further to Error$^2$. 

To understand the structure of the contribution to Error$^2$, we denote
\bal
\text{Error} = (Y^2-\Phi+\<Y\>^2)-\<Y^2\>=M-\<Y^2\>\,, \qquad   \<M\>=\<Y^2\>\,,  
\eal
then if we switch the notation of $\<\text{Error}^2\>$ to a slightly more indicative one $\<\text{Error}_1\text{Error}_2\>$, we have
\bal
\<\text{Error}_1\text{Error}_2\>=\<M_1M_2\>+\<Y^2\>^2-2\<Y^2\>\<M\>=\<M_1M_2\>-\<Y^2\>^2\ .\label{keyapp}
\eal
Therefore we find if $\<M_1M_2\>\approx \<M_1\>\<M_2\>$ to the leading order, then  the error is small and the approximation~\eqref{appro} is good. 

This is precisely how the previous proposal~\eqref{appro} failed. For example, in the Error~\eqref{errf1}, it is precisely the contraction among the two factors $\sum_{i\neq j} X_iX_j$ that gives another factor of $2N^4 t^8$ in $\langle (\text{Error}/4)^2\rangle$ and prevent it from vanishing. On the other hand, if we check the results~\eqref{con1} and~\eqref{con2} we find, to the leading order of $N$, the term that have non-trivial contribution to Error$^2$ is
\bal
4(N-2)(t^2+\m^2) X_p X_q\,,
\eal
that comes from summing the first terms in~\eqref{con1} over $j$; the other terms are either suppressed by $1/N$ or do not give nontrivial contraction between the two copies of Error as discussed above. Then we immediately notice that this is precisely the term, with $\m=0$ in this case, that is missing in $\F$ to remove the ``problematic" term in the Error that we just discussed. Therefore, once we use the correct approximation with all terms in~\eqref{list}, the error should be small and the approximation should be good.
The other examples in section~\eqref{nontrace} could also be modified in a similar way so that the errors become small. 

Further notice that one of the upshot of the approximation~\eqref{appro} is, as pointed out in~\cite{Mukhametzhanov:2021hdi}, that we can safely ignore the direct correlation between the two $Y$'s (or $z$'s in the context of~\cite{Mukhametzhanov:2021hdi}) and the two terms are ``linked" through the correlation with $e^{i k x}$. What we found in the previous section are however cases where these direct correlations cannot be ignored. The new ingredient of the  approximation~\eqref{Phi2} we will present shortly is precisely a partial correlation between the $Y$'s directly, not just through the  $e^{i k x}$ factors. In this sense the saddles in the general models discussed in section~\ref{nontrace}  are hyper-linked half-wormholes with extra partially direct connections. 

With this we propose a modified approximation 
\bal
&Y^2\approx \langle Y^2\rangle+\tilde\Phi,\label{appro2}\\
&\tilde\Phi(X)= \frac{1}{(2\pi)^N}\int \prod_i \(dk_i \frac{e^{-\im  k_i X_i}}{P(X_i)}\)  \langle e^{\im \sum_ik_i x_i}\rangle \frac{\left[ Y(x)^2 e^{\im \sum_ik_ix_i}\right] }{\langle e^{\im\sum_ik_ix_i}\rangle}\,, \label{Phi2}
\eal
where $\left[ Y(x)^2 e^{\im \sum_ik_ix_i}\right]$ denotes all possible terms contains at least one contraction between $Y^2$ and the spacetime brane $e^{ikx}$. 

\begin{figure}
	\centering
	\includegraphics[width=0.8\linewidth]{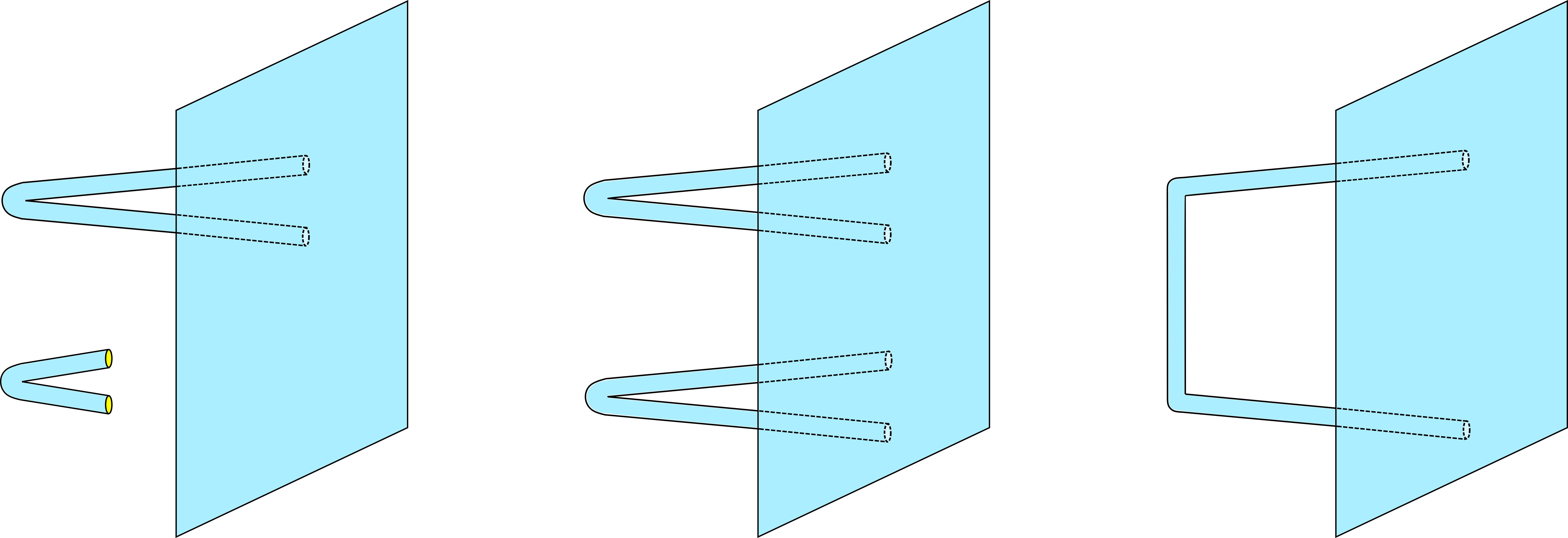}
	\caption{A pictorial illustration of the 3 terms in~\eqref{3terms} respectively. Each vertex on the left is a factor of $Y(x)$, brane on the right denotes $e^{i k x}$, and each bracket $\langle \cdot e^{ikx}\rangle_c$ corresponds to a component of the bulk amplitude that connects the brane with a set of vertices. Notice that the first diagrams should be considered as two diagrams each has a vertex connected to the brane.}
	\label{conn2}
\end{figure}

In the example~\eqref{Xij2}, each term in $Y$ contains two $X_i$ legs, therefore we have 
\bal
\frac{\left[ Y(x)^2 e^{\im \sum_ik_ix_i}\right] }{\langle e^{\im\sum_ik_ix_i}\rangle}=\frac{2\<Y\>\langle Y(x) e^{\im \sum_ik_ix_i}\rangle_c }{\langle e^{\im\sum_ik_ix_i}\rangle}+\frac{\langle Y(x) e^{\im \sum_ik_ix_i}\rangle_c^2 }{\langle e^{\im\sum_ik_ix_i}\rangle^2}+\frac{\langle Y(x)^2 e^{\im \sum_ik_ix_i}\rangle_c }{\langle e^{\im\sum_ik_ix_i}\rangle}\,,\label{3terms}
\eal
where the different terms correspond to one contraction to the brane, two separate contractions to the brane and a pair of connected contractions to the brane. The~${}_c$ here means the contribution cannot be made disconnected if we only cut on the brane. Among these terms the last one is precisely the one missed in the previous proposal~\eqref{Phi}. A demonstration of these terms are shown in Figure~\ref{conn2}. We notice that this approximation is closely related to the relation between $\hat{Z}$ and $\hat{W}$ discussed in~\cite{Peng:2021vhs}, see e.g. Figure.~9 there.

From this analysis, it is more obvious to understand why the Errors are all small when $\m \neq 0$ in section~\ref{nontrace}. When disconnected contributions exist, the leading order contributions of the Error$^2$ always come from the disconnected component and hence the Error is guaranteed to be small. However, this is not very meaningful since as in most of the large-$N$ theories studied in the literature, we isolate away the disconnected contributions and always focus on the connected contributions.

\subsubsection{Why the proposal works for the Pfaffian in the SYK model}

From the above discussion, it seems that for a generic operators with complicated product structure, the original proposal~\eqref{appro} almost surely fails. However, we know from explicit computations in~\cite{Saad:2021rcu, Mukhametzhanov:2021nea} that the approximation works well for the hyperpfaffian of the random couplings which is also related to the partition function of the SYK model. 

We believe the reason for this is the large-$N$ factorizations properties due to large-$N$ constraints. By this we mean when the operators are defined to have extra structures, for example as a trace or a determinant over the $N$ flavors, such extra structure remains to affect the computation of the Error. When this is true, which indeed is our case, then the contractions between the two copies of Error are necessarily suppressed by the large-$N$ factors; either $1/N$ when the structure is trace as in~\eqref{er2a} or higher powers of $1/N$ when the structure is a determinant. Therefore all contractions between the two copies of Errors are suppressed and at the leading order the result factorizes and hence the original proposal~\eqref{appro} works.      
\footnote{A related fact is that when the approximation is no longer good the relation between  the 4$^{\text{th}}$ moment $\<Y^4\>$ of the observable~\eqref{Xij2} and the second moment $\<Y^2\>$ deviates significantly from the Gaussian distribution. In Gaussian distribution, this contribution is $3\<Y^2\> \subset \<Y^4\>$, on the other hand, for the observable $Y$ in~\eqref{Xij2} we get
\bea 
\langle Y^4\rangle &=&8\sum_{i\neq j}\langle X_i^4 X_j^4\rangle+60\sum_{i\neq j\neq p\neq q }\langle X_i^2 X_j^2 X_p^2 X_q^2\rangle +48 \sum_{i\neq j\neq p}\langle X_i^4 X_j^2 X_p^2\rangle \\
&\approx &60 N^4 t^8 \neq 3\langle Y^2\rangle^2-2\langle Y\rangle^4\approx 12 N^4 t^8\ .
\eea 
But at the moment we have not succeeded in making a causal relation between this fact and the fact that the Error is small. The explanation in the main text does better in doing so.}

A somewhat ad hoc reason for the need of traces or determinant in the definition of the operator to make the discussion about (half-)wormhole meaningful is the following. There is no ``spacetime" in our statistical models, so we cannot use any locality property to identify a function of the random variables as a single operator; the most we can do is to use a trace or determinant structure to identify a group of random variables as an operator. If there is no such trace/determinant constraints, it is equally legitimate to regard the result as computing  correlations of a large number of the fundamental random variables and the (half-)wormhole interpretation is not necessarily relevant.

A different interpretation of the importance of the existence of such trace or determinant structure could be considered as some emergent global symmetry among the random variables (probably when appropriately analytically continued). By this we simply mean if we treat the random variables $X_i$ as ``fields", then the action, ie the probability distribution, and the operators we considered in the computation all have $SO(N)$ symmetry among them. Then the invariant tensors of $SO(N)$ directly lead to the trace or determinant structures we just described.  
It is interesting to make this point more clear, and we plan to come back to this question somewhere else.

We did not find a general proof of the above assertion~\eqref{Phi2} or~\eqref{3terms}, but as a check we can, according to our assertion, modify the definition of the function $Y$ and put in by hand some constraints, mimicking a trace structure. Then we find with this constraints the approximation~\eqref{appro} is indeed valid. 
For instance we could introduce a restriction in the sum
\begin{align}
    Y=\sum_{i+j=M}X_{i}X_{j},\quad N<M<2N,\quad i\neq j\,,
\end{align}
where $N$ is the total number of $X$'s and $M$ is an integer. Without loss of generality we assume $M$ is even in the following, and the computation for odd $M$ is the same. 
Following the previous computations, we get
\begin{align}
    Y^{2}&=2\sum_{i+j=M}X_{i}^{2}X_{j}^{2}+\sum_{i\neq j\neq m\neq n}X_{i}X_{j}X_{m}X_{n}\,,
\end{align}
and
\begin{align}
    \langle Y\rangle =K\mu^{2},\quad \langle Y^{2}\rangle=2K(t^{2}+\mu^{2})^{2}+K(K-2)\mu^{4},\quad K=2N-M\ .
\end{align}
Taking $X_i$ from the same Gaussian distribution in the previous cases we get the expression for the error
\begin{align}
    \text{Error}=4t^{2}\sum_{i+j=M}X_{i}^{2}-4Kt^{4}-4Kt^{2}\mu^{2}.
\end{align}
It is straightforwardly to show that the expectation values 
\begin{align}
    \langle \text{Error}\rangle =0\,, \qquad 
    \langle (\text{Error}/4)^{2}\rangle =2Kt^{8}+4Kt^{6}\mu^{2}\ .
\end{align}
Clearly in this case $ \langle (\text{Error}/4)^{2}\rangle$ is $1/N$ suppressed compared to $\langle Y^{2}\rangle^{2}$ independent on the value of $\m$. Hence the approximation~\eqref{appro} is always valid in the presence of  this extra constraint. 
Similar restrictions could be imposed to models with general $q$. It turns out that again the computation is quite similar 
and we expect the approximation to be valid in these cases too. 

\section{SYK at one time point: $\langle J_a\rangle = 0$}
\label{SYK0}
In this section, we study the half-wormhole contributions in some 0d SYK model that can be considered as the usual 0+1d SYK model on a single instant of time. This section is largely a review of previous results in~\cite{Saad:2021rcu,Mukhametzhanov:2021nea,Mukhametzhanov:2021hdi}; we provide more details of various saddle point  results and carry out Lefschetz thimble analysis of some computations when needed. 

\subsection{SYK model with one time point}
Let us first revisit the analysis of the 0-dimensional SYK model introduced in \cite{Saad:2021rcu}. We are interested in the following Grassmann integral
\bea \label{z0}
z=\int d^N \psi \exp (\im ^{q/2}\sum J_{i_1\dots i_q}\psi_{i_1\dots i_q})\,,
\eea
where $\psi_{i_1\dots i_q}=\psi_{a_1}\psi_{a_2}\dots \psi_{a_q}$ and $\psi_i$ are Grassmann numbers. The number $z$ can be understood as the partition function of $0+0$ dimensional analogue of SYK model.
The random couplings $J_{i_1\dots i_q}$ is drawn from a Gaussian distribution 
\bea 
\langle J_{i_1\dots i_q}\rangle=0,\quad \langle J_{i_1\dots i_q}J_{j_1\dots j_q}\rangle=t^2\delta_{i_1j_1}\dots \delta_{i_qj_q},\quad t^2=\frac{(q-1)!}{N^{q-1}}\ .\label{gauss}
\eea 
We sometimes use the collective indies $A,B$ to simplify the notation
\bea 
A=\{a_1<\dots < a_q\}\,,\qquad J_A\psi_A\equiv J_{a_1\dots a_q}\psi_{a_1\dots a_q}\ .
\eea 
Integrating out the Grassmann numbers directly gives \eqref{0SYK}\footnote{Here we choose the measure of Grassmann integral to be $\int d^N \psi \psi_{1\dots N}=\im^{-N/2}$.}:
\bea 
z=\int d^N \psi \exp(\im ^{q/2}J_A \psi_A)=\sum'_{A_1<\dots<A_p} \text{sgn}(A)J_{A_1}\dots J_{A_p}\,,\quad p=N/q\,,\label{pfa}
\eea 
where the expression \eqref{pfa} is nothing but the hyperpfaffian $\text{Pf}(J)$. Since $\langle z\rangle=0$ due to~\eqref{gauss}, we focus on $z^2$ and $\langle z^2\rangle$
\bea 
&& z^2=z_L z_R=\int \d^N \psi^L \d^N \psi^R \exp\left\{\im^{q/2}\sum_A J_A\(\psi_A^L+\psi_A^R\)\right\}\,, \label{0z2a0}\\
&&\langle z^2\rangle=\int \d^{2N} \psi \exp\left\{\frac{N }{q}\(\frac{1}{N} \sum_{i=1}^N \psi_i^L \psi_i^R\)^q\right\}\,,\label{0z2a}
\eea 
where we have assumed that $q$ and $N$ are even. The exact values of \eqref{0z2a} can be computed by introducing the standard $G,\Sigma$ variables
\bea 
\langle z^2\rangle &=& \int \d^{2N}\psi \int_{\mathbb{R}}\d G\delta\(G-\frac{1}{N}\sum_{i=1}^N \psi_i^L \psi_i^R\)\exp\(\frac{N}{q}G^q\) \\
&=&\int_{\mathbb{R}}\d G\int_{\im \mathbb{R}}\frac{\d \Sigma}{2\pi \im /N}\exp \left\{N \(\log(\Sigma)-\Sigma G+\frac{1}{q} G^q\)\right\} \\
&=&  N^{-N}\int_{\mathbb{R}}\d G\exp\(\frac{N}{q}G^q\) (-\pa_G)^N \delta(G)\\
&=&\frac{N!(N/q)^{N/q}}{N^N (N/q)!} =e^{-(1-\frac{1}{q})N}\sqrt{q}\(1+\frac{1-q}{12N}+\mathcal{O}(\frac{1}{N^2})\)\,,\label{e2loop}
\eea 
where in the last step we expand around $N\to \infty$ to the next-to-leading order.

Next we consider the non-averaged quantity~\eqref{0z2a0}. Following~\cite{Saad:2021rcu}, we rewrite
\bea 
&& z^2=\int_R \d\sigma \Psi(\s) \Phi(\s)\,,\quad \Psi(\sigma)=\int \frac{dg}{2\pi/N}\exp[N(-\im \sigma g-1/q g^q)]\,,\label{psi}
\eea 
where the coupling dependent piece $\Phi$ is
\bea 
\Phi(\s)=\int \d^{2N}\psi \exp\left\{ \im e^{-\frac{\im \pi}{q}}  \sigma \psi_i^L\psi_i^R+\im^{q/2}J_A(\psi_A^L+\psi_A^R)-\frac{N}{q}\left(\frac{1}{N}\psi_i^L\psi_i^R\right)^q \right \}\, .
\eea 
Its averaged value is 
\bea 
\langle \Phi(\sigma)\rangle=(\im e^{-\frac{\im \pi}{q}}\sigma)^N\ .
\eea

As suggested in \cite{Saad:2021rcu}, to understand the relation between each individual result and the averaged result, we could figure out in what region of the $\sigma$-plane $\Phi$ is self-averaging. This is reflected in the quantity 
$\<\(\F(\s)-\<\F(\s)\>\)^2\>$. Therefore we compare $\langle \Phi(\sigma)\rangle^2$ with $\langle \Phi(\sigma)^2\rangle $ 
\bea \label{aphi2gs}
\langle \Phi(\s)^2\rangle=\int_R\frac{\d^4 \sigma_{AB} \d^4 g_{AB}}{(2\pi /N)^4}e^{N\left[\log(-e^{-\frac{2\im \pi}{q}}(\s^2+\s_{14}\s_{23}-\s_{13}\s_{24})) -\im \s_{AB}g_{AB}-\frac{1}{q}g_{AB}^q\right]}\,,
\eea 
where we relabel $L=1,L'=3,R=2,R'=4$ and $(AB)=(13),(14),(23),(24)$. The integral can be done exactly~\cite{Saad:2021rcu} following a similar computation we used to get~\eqref{e2loop}
\bea 
\langle \Phi(\s)^2\rangle=(-e^{-\frac{2\im \pi}{q}})^N\hspace{-4mm}\sum_{n_1+n_2+n_3=\frac{N}{q},n_i\geq 0}\hspace{-2mm}\frac{N!}{N^{2q(n_2+n_3)}}\(\frac{N}{q}\)^{2(n_2+n_3)}\frac{\sigma^{2qn_1}(qn_2)!(qn_3)!}{(qn_1)!(n_2!)^2(n_3!)^2}\,,
\eea 
which can be organized into a polynomial in $\sigma$
\bea 
\langle \Phi(\s)^2\rangle &=&(-e^{-\frac{2\im \pi}{q}})^N\(\sigma^{2N}+\frac{2N!q!}{(N-q)! q^2 N^{2q-2}}\sigma^{2N-2q}+\dots+e^{2N\frac{1-q}{q}}2q\) \\
&\sim &(-e^{-\frac{2\im \pi}{q}})^N\(\sigma^{2N}+\frac{2(q-1)!}{q N^{q-2}}
\sigma^{2N-2q}+\dots+e^{2N\frac{1-q}{q}}2q\)\,, \label{aphi2}
\eea 
where the phase factor is trivial whenever $q$ divides $N$.

\subsection{The saddle points analysis}

The above results can be reproduced by saddle point approximation in large~$N$ limit. 

\subsubsection{The averaged $\langle z^2\rangle$}
To obtain the same  result~\eqref{e2loop} from saddle point approximation, we first we rotate the contour 
\bea 
\Sigma=\im e^{-\im \frac{\pi}{q}}\s,\quad G=e^{\im \frac{\pi}{q}} g\,,
\eea 
to get
\bal \label{z2t1}
\langle z^2\rangle =\int_R \d g \int_R \frac{\d \sigma}{2\pi/N}\exp\left\{N\(\log(\im e^{-\frac{\im \pi}{q}}\sigma)-\im \sigma g-\frac{1}{q}g^q\)    \right\} \equiv \int_R \d g \int_R \frac{\d \sigma}{2\pi/N}e^{NS}\,,
\eal
so that the integral converges. The saddle point equations are
\bea \label{whs}
-\im \sigma -g^{q-1}=0\,,\quad g^q=-1\,,\quad \rightarrow\quad  g=e^{\frac{(2m+1)\im \pi}{q}}\,,\quad m=0,\dots,q-1\ .
\eea 
All of them give the same on-shell action
\bea 
\langle z^2\rangle_s=\frac{N}{2\pi} e^{-(1-\frac{1}{q})N}\ .
\eea 
To match with the exact result \eqref{e2loop} we need to consider  fluctuations around the saddle points. For simplicity let us take $q=4$ and focus on one of the saddle points
\bea 
\sigma_s=g_s=-(-1)^{\frac{3}{4}},\quad \langle z^2\rangle_{s}=\frac{N}{2\pi}e^{-\frac{3}{4}N}.
\eea
Expanding the exponent around this saddle  
\bea 
\sigma=\sigma_s+x,\quad g=g_s+y
\eea 
to the second order
\bea 
S_2\sim-\frac{3}{4}+\frac{3\im x^2}{2}-\im xy-\frac{\im y^2}{2}+[(-1)^{3/4}x^3+\frac{(-1)^{3/4}}{3}y^3]+\frac{y^4-x^4}{4}\,,
\eea 
and evaluating  the integral directly gives the fluctuation that combines with the saddle contribution to 
\bea 
\langle z^2\rangle_{\text{saddle}+\text{loop}}= e^{-\frac{3}{4}N}\frac{1}{2}\(1-\frac{1}{4N}\)\ .
\eea 
Adding contributions from all 4 saddles we arrive at
\bea 
\langle z^2\rangle_{\text{saddle}+\text{loop}}=2e^{-\frac{3}{4}N}\(1-\frac{1}{4N}\)\,,
\eea 
that agrees with \eqref{e2loop} at the two-loop order.

\subsubsection{The unaveraged $z^2$: the wormhole saddle}

The result~\eqref{aphi2} can be reproduced from a saddle point analysis in the large-$N$ limit. 
The saddle point equations are
\bea 
g_{AB}^{q-1}=-\im \s_{AB}\,,\quad -\im g_{13}=\frac{\s_{24}}{f},\quad \im g_{14}=\frac{\s_{23}}{f},\quad \im g_{23}=\frac{\s_{14}}{f},\quad -\im g_{24}=\frac{\s_{13}}{f}\,,\label{eom}
\eea 
where $f\equiv\s_{14}\s_{23}-\s_{13}\s_{24}+\s^2$. The trivial solution $\sigma_{AB}=g_{AB}=0$ leads to 
\bea 
\langle \Phi(\sigma)^2\rangle_{\text{trivial}+1\text{loop}}=\langle \Phi(\sigma)\rangle^2\,,
\eea 
which says the trivial saddle always agrees with the first term in \eqref{aphi2}. 

Next let us consider non-trivial solutions with $\s_{AB}\neq 0$.  From the equations of motion we obtain 
\bea \label{identities}
&&x^{q-2}=y^{q-2},\quad (x^{q-1}-y^{q-1}+\s^2)^{2}=x^{q-2}=y^{q-2}\,,\\
&&g_{13}^q=g_{24}^q,\quad g_{23}^q=g_{14}^q
\eea 
where
\bea 
x=g_{13}g_{24},\quad y=g_{14}g_{23}\ .
\eea
It is easy to check that solutions of the above equation satisfies $x=ye^{\frac{2m\pi \im}{q-2}}$, and for each choice of $m$ there are $2q^2$ solutions of $g_{ab}$. For simplicity let us again focus on the $q=4$ case such that there are only two classes $x=\pm y$.

$\bullet$ When $x=y$ we find another 32 non-trivial saddles. The on-shell action of all of them are the same
\bea 
\langle \Phi(\s)^2\rangle_{\text{non-trivial}}^+=N^4\langle \Phi(\sigma)\rangle^2=\langle \Phi(\sigma)^2\rangle_{\text{trivial}}\,,
\eea 
where the factor $N^4$ comes from the measure of \eqref{aphi2gs}. However the 1-loop fluctuations around them are different
\bea 
&& \text{trivial saddle}: \frac{1}{N^4}\,, \quad \text{non-trivial saddles}: \frac{1}{8 N^4}\ .
\eea 
We notice that including the 1-loop effect, the trivial saddle is larger and it reproduces the large $N$ behavior of the exact result. On the other hand, the non-trivial saddle contributions are also comparable; so it is possible that we should also take into account of their contributions as well. However, if we add all the trivial and non-trivial saddle-point values, the result will  obviously exceed the exact value~\eqref{aphi2}. 
In fact, by a simple Lefschetz-thimble analysis, see e.g.~\cite{Witten:2010cx}, which is reviewed In Appendix~\ref{Lefschetz}, we conclude that these non-trivial saddles should not be included.

\begin{figure}
	\centering
	\includegraphics[width=5 cm]{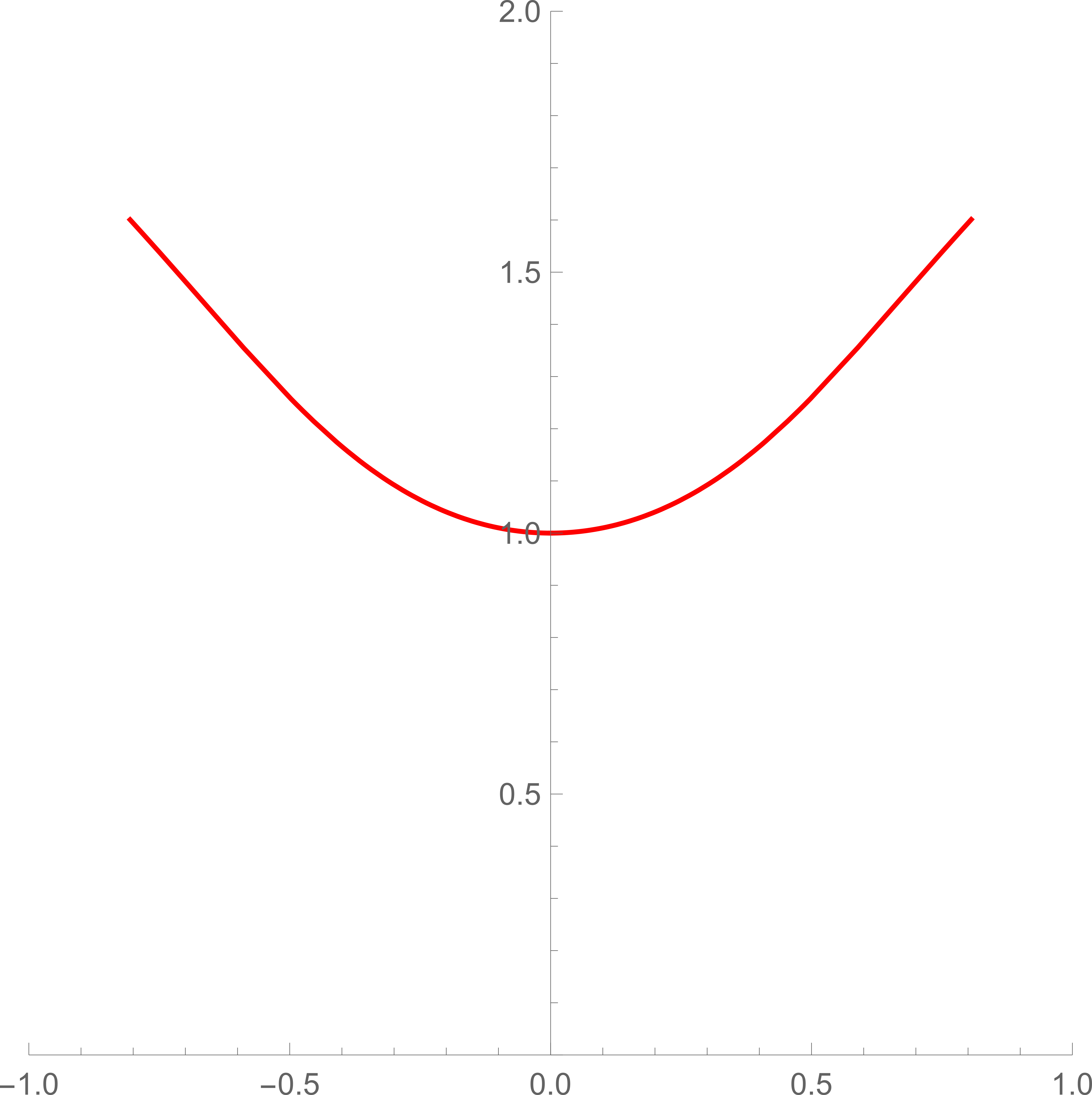}
    \includegraphics[width=5 cm]{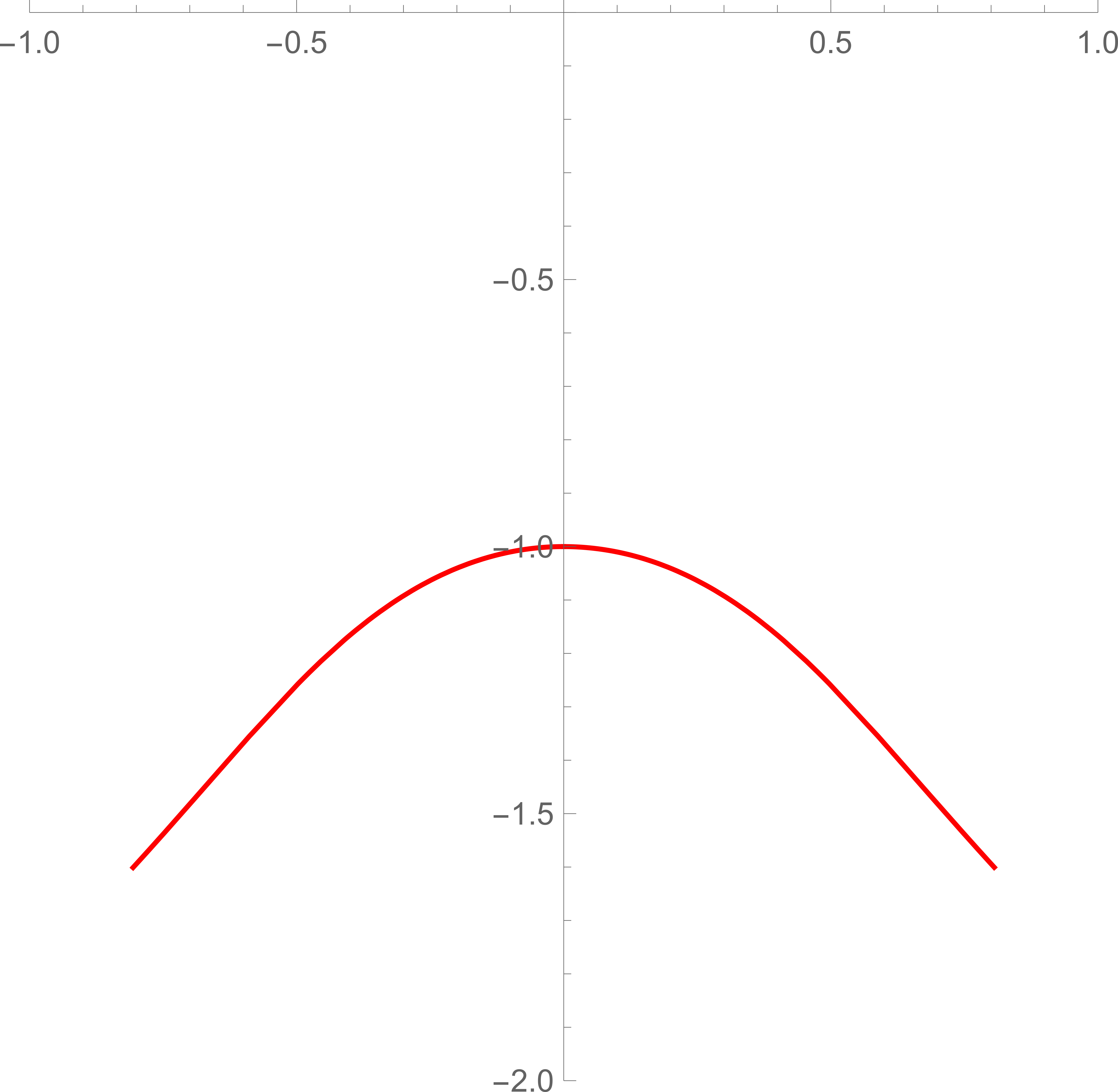}
	\caption{Anti-thimble on the $\sigma_{13}$ plane (left) and the $\sigma_{24}$ plane (right).}
	\label{antithimble1}
\end{figure}

In particular, we choose a Morse function to be the real part of the action~\eqref{aphi2gs}  
\begin{align}
h\equiv \Re(S)=&\sum_{abj} \left(-\frac{g_{abj}^{4}}{4}+\frac{3 g_{ab1}^{2}g_{ab2}^{2}}{2}+g_{ab1}\sigma_{ab2}+g_{ab2}\sigma_{ab1}\right)\nonumber\\
	&\quad +\frac{1}{2}\log\left((\sigma_{142}\sigma_{231}+\sigma_{141}\sigma_{232}-\sigma_{132}\sigma_{241}-\sigma_{131}\sigma_{242})^{2}\right.\nonumber\\
	&\left.\quad+(1+\sigma_{141}\sigma_{231}-\sigma_{142}\sigma_{232}-\sigma_{131}\sigma_{241}+\sigma_{132}\sigma_{242})^{2}\right)\,,
\end{align}
where we have chosen $q=4$ for simplicity and $\sigma=1$ since we are interested in the case $\s \neq 0$\footnote{The $\s=0$ case is analyzed in~\cite{Saad:2021rcu} }. The $g_{abi}$ and $\s_{abj}$ are the real and imaginary parts of the field $g_{ab}$ and  $\s_{ab}$
\begin{align}
	g_{ab}=g_{ab1}+\im g_{ab2},\quad \sigma_{ab}=\sigma_{ab1}+\im\sigma_{ab2}\ .
\end{align}
The downward flow equations of the Morse function are
\begin{align}
	\frac{dg_{abj}}{dt}=-\frac{\partial h}{\partial g_{abj}},\quad \frac{d\sigma_{abj}}{dt}=-\frac{\partial h}{\partial \sigma_{abj}}\ .
\end{align}
The end point of each anti-thimble is one of the saddles at $g_{abj}^{c}$ and $g_{abj}^{c}$, which leads to the following boundary conditions of the flow equation
\begin{align}
	\lim_{t\to +\infty}g_{abj}=g_{abj}^{c},\quad \lim_{t\to +\infty}\sigma_{abj}=\sigma_{abj}^{c}\ .
\end{align}
We can then solve the flow equation and obtain the Lefschetz anti-thimbles going through each saddle point and if they intersect with the original integration contour the saddle point contributes to the integral. 

For example in Figure~\ref{antithimble1} we illustrate examples of the anti-thimbles of the saddle point 
\begin{align}
	g_{13}=1,\quad g_{24}=-1,\quad g_{14}=(-1)^{3/4},\quad g_{23}=(-1)^{1/4},\\
	\sigma_{13}=\im,\quad \sigma_{24}=-\im,\quad \sigma_{14}=(-1)^{3/4},\quad \sigma_{23}=-(-1)^{1/4}\,,
\end{align}
that do not intersect with the original integration contour, namely the real axis. This means the contribution of this saddle should not be included to the integral.

\begin{figure}
	\centering
	\includegraphics[width=5 cm]{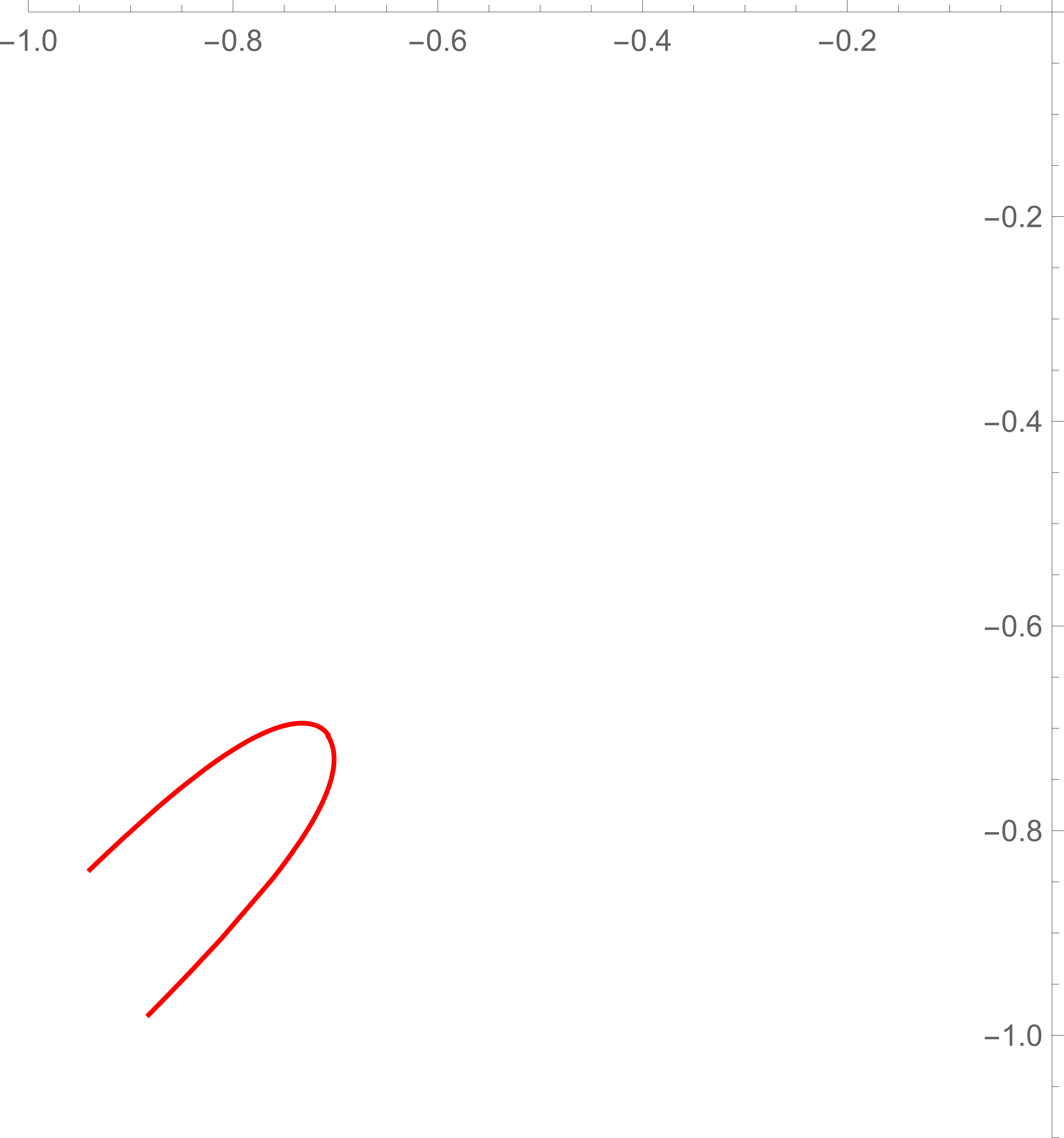}
	\hspace{3cm}\includegraphics[width=5 cm]{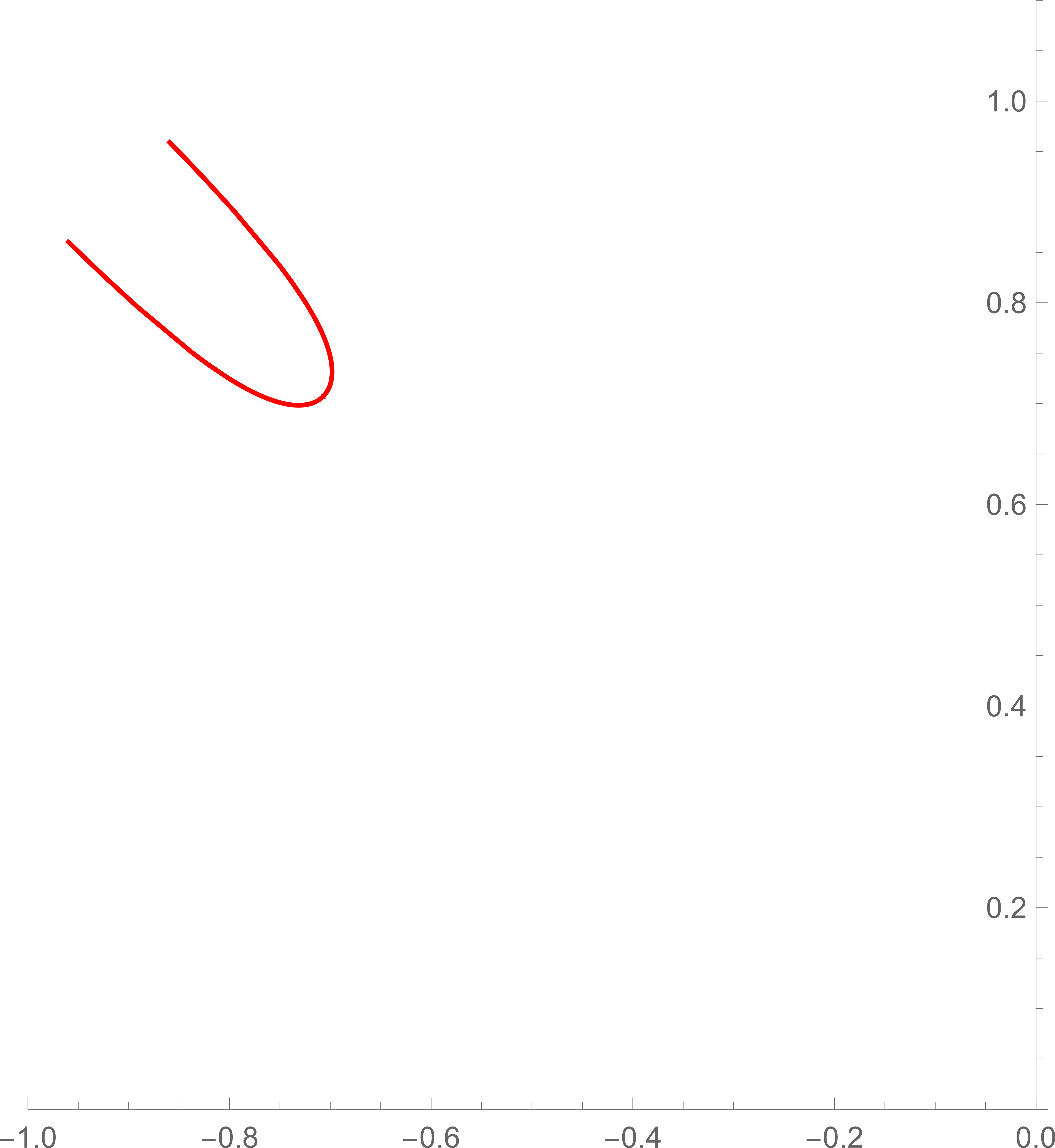}
	\caption{Anti-thimble on the $g_{13}$ plane (left) and the $g_{24}$ plane (right). }
	\label{antithimble2}
\end{figure}

Examples of anti-thimbles of another saddle point 
\begin{align}
	g_{13}=-(-1)^{1/4},\quad g_{24}=(-1)^{3/4},\quad g_{14}=-1,\quad g_{23}=-1,\\
	\sigma_{13}=(-1)^{1/4},\quad \sigma_{23}=(-1)^{3/4},\quad \sigma_{14}=-\im,\quad \sigma_{23}=-\im\,,
\end{align}
is shown in Figure~\ref{antithimble2}. Again they do not intersect with the real axis so the contribution from this saddle should not be included either.

We can run this analysis over all the nontrivial saddles and find none of them contribute to the integral. As a result, the path integral can be approximated entirely by the trivial saddle.

\begin{figure}
	\centering
	\includegraphics[scale=0.3]{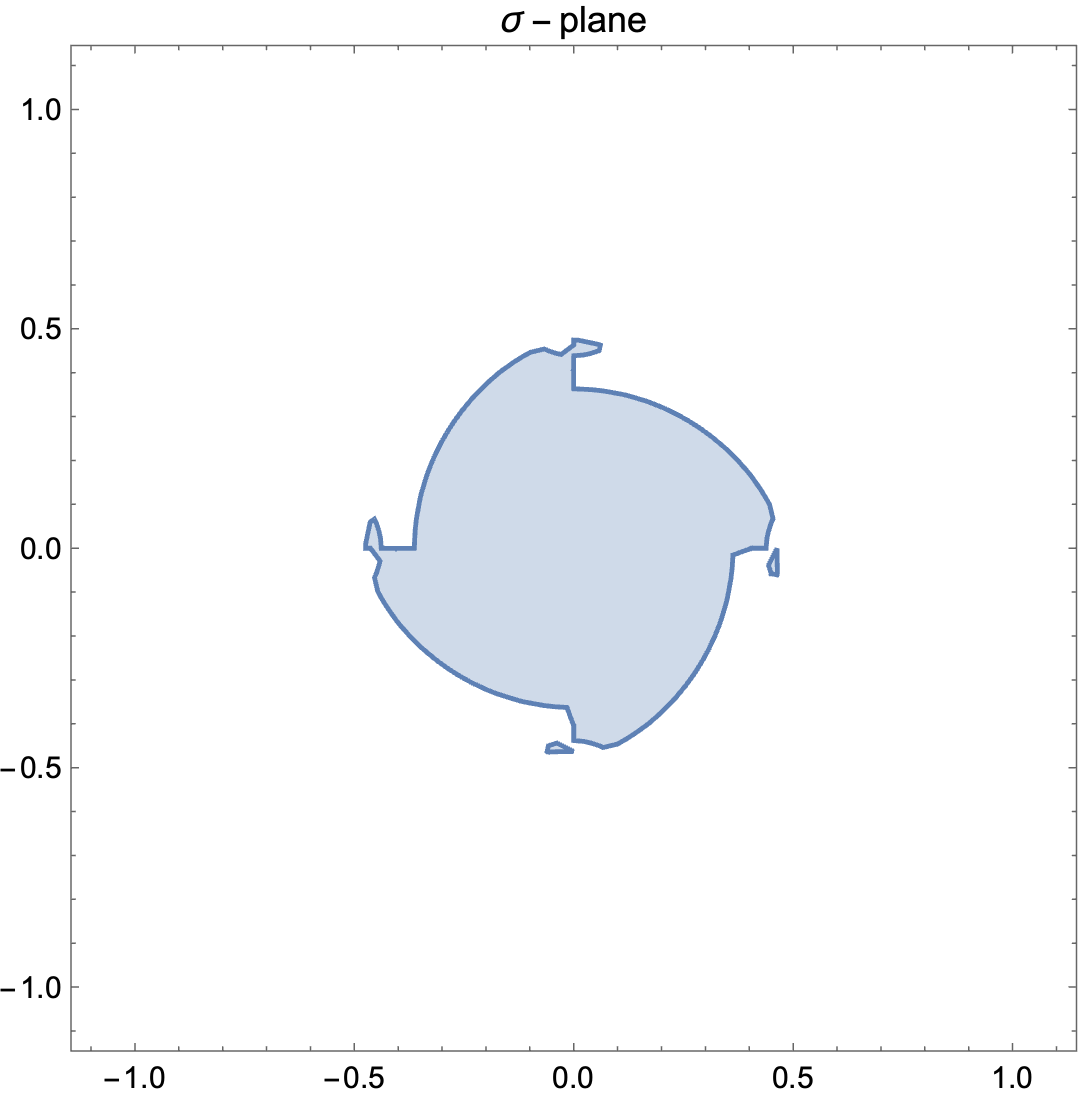}
	\caption{The shaded region is where a non-trivial saddle in~\eqref{nontri1} dominates over the trivial saddle. The plot for the other two non-trivial saddles can be obtained from this plot by simple rotations.}
	\label{plot1}
\end{figure}
 
$\bullet$ When $x=-y$, there are also nontrivial saddle points and a similar analysis of Lefschetz thimbles demonstrate that they do not contribute to the integral. 

Actually, there is a quicker way to arrive at the same conclusion. We find that the on-shell actions corresponding to these saddle points are
\bea 
\(\frac{\sigma^2}{2}\)^{\frac{N}{3}}e^{-N\pm \frac{3}{2}2^{\frac{1}{3}}N e^{\frac{2\im m \pi}{3}}\sigma^{\frac{4}{3}}},\quad m=0,\pm 1,\quad \sigma \rightarrow \infty\ .\label{nontri1}
\eea 
However these saddle points should be  saddle points of the entire multi-dimensional integral including the integral over $\s$. As a result this saddle should also satisfy the fall-off condition of the $\s$ integral, otherwise they will not contribute to the $\s$ integral. Therefore we should only consider the decaying saddle points namely
\bea \label{sub1}
\(\frac{\sigma^2}{2}\)^{\frac{N}{3}}e^{-N +\frac{3}{2}2^{\frac{1}{3}}N e^{\pm \frac{2\im  \pi}{3}}\sigma^{\frac{4}{3}}},\quad \(\frac{\sigma^2}{2}\)^{\frac{N}{3}}e^{-N -\frac{3}{2}2^{\frac{1}{3}}N \sigma^{\frac{4}{3}}}\ .
\eea 
We plot the region where these non-trivial saddle dominates over the trivial saddle in Figure~\ref{plot1}, and it is easy to observe from the figure that the wormhole saddle~\eqref{whs} of $\langle z^2\rangle$,  located at $|\sigma|=1$,  is in the region where the trivial saddle dominates.

Another family of solutions to the equation of motion~\eqref{eom} has $x=0$ or $y=0$.  
On shell actions on these saddles behave as
\bea \label{sub2}
\sigma^{\frac{2N}{3}}e^{-N +\frac{3}{2}N e^{\pm \frac{2\im  \pi}{3}}\sigma^{\frac{4}{3}}},\quad \sigma^{\frac{2N}{3}}e^{-N -\frac{3}{2}N \sigma^{\frac{4}{3}}}\,,
\eea
whose dominant regions are similar to Figure~\ref{plot1} and they are sub-leading comparing with the trivial saddle. 

Putting all the result together we confirm that the trivial saddle point dominate in the $g_{ab}$ and $\s_{ab}$ integral and the wormhole saddle~\eqref{whs} is self-averaging.

\subsubsection{The unaveraged $z^2$: the linked half-wormhole saddles}
The trivial saddle point discussed in the previous section gives vanishing contribution at $\s \sim 0$, so we expect other saddle points dominate the path integral here. In~\cite{Saad:2021rcu} they are referred to as the (linked) half-wormhole saddles. Here we provide some further details of the saddle contribute at $\s \sim 0$ and show that it agrees with the exact result in~\eqref{aphi2}, ie 
\bal
\<\F(0)^2\>_{\text{ext}} \sim 2q e^{-\frac{3}{2}N}\ . \label{extphi0}
\eal
We can apply the same analysis, except that now we evaluate at $\s \sim 0$, as in the previous section. 
As expected, the trivial saddle gives 
\bal
e^{N\log(\s)} \sim 0\ .
\eal
The subleading non-trivial saddles~\eqref{sub1} and~\eqref{sub2} discussed in the previous section has on-shell values
\bal
\frac{e^{-\frac{3}{2}N}}{2^{N/2}},\quad e^{-\frac{3}{2}N}\ ,
\eal
respectively when $\s=0$. So \eqref{sub2} dominates.
Adding them up precisely gives the exact solution~\eqref{extphi0}
\bal
2q e^{-\frac{3}{2}N}\,,
\eal

The general lesson is that the linked half wormhole saddle points are always in the integral, and furthermore they are also always saddles. It's only that they are, for most of the time, hidden behind the leading saddles. They can only be exposed in regions where the leading saddle decreases faster, namely the  $\s\sim 0$ region in this case.

\section{SYK at one time point: $\langle J_a\rangle \neq 0$}

In the following, we will generalize the study of half-wormhole along several directions. The main question we want to address is how the distribution of the random coupling affects the wormhole and half-wormhole saddles.

First let us consider the case where the random coupling is drawn from a general Gaussian distribution $\cn(u,t^2)$\footnote{When we write $J_A$, we have in mind that the index set $A$ is automatically sorted, and all $J$'s with other permutations of $A$ picks up signs accordingly.} 
\bea\label{para}
\langle J_A\rangle=J_A^0=u,\quad \langle J_A^2\rangle-\langle J_A\rangle^2=\t^2 \frac{(q-1)!}{N^{q-1}}\equiv t^2\,,
\eea
in particular, the mean value of the random coupling could be non-vanishing.

The ensemble averaged quantities can be computed directly 
by first averaging over the couplings and then integrating out the fermions
\bea
 \langle z\rangle &=&\text{PF}(J^0)\,,\label{y1}\\
 \langle z^2\rangle &=& \int d^{2N}\psi \exp\(\im^q t^2 \sum_A \psi_A^L \psi_A^R +\im^{q/2}J_A^0(\psi_A^L+\psi_A^R)\)\label{y2ab}\\
 &=&\sum'_{A,B}\text{sgn}(A)\text{sgn}(B)\(J_{A_1}^0J_{B_1}^0+\delta_{A_1 B_1}t^2)\)\dots \(J_{A_p}^0J_{B_p}^0+\delta_{A_p B_p}t^2)\) \ .\label{y2a}
\eea 
\subsection{Half-wormhole saddle in $z$}
\begin{figure}
\begin{center}
	\includegraphics[width=0.5\linewidth]{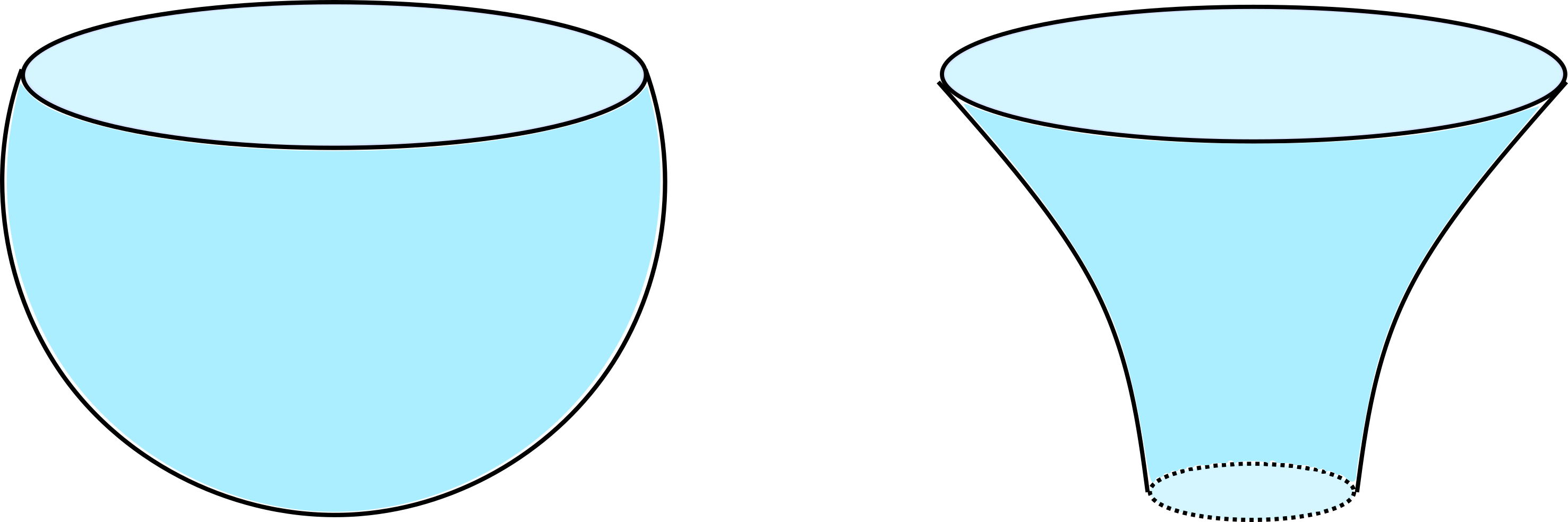}
  \caption{The single half-wormhole saddle of $z$.} \label{halfz}
  \end{center}
\end{figure}

Since $\langle z\rangle\neq 0$, we expect a disk saddle point in the path integral presentation of $z$ that gives the contribution of $\<z\>$. Moreover, like linked half-wormhole contribution to $z^2$ in the model with $u=0$, it is possible that there are also single half-wormhole saddles contributing to $z$, \footnote{This single half-wormhole saddle is related to the half-wormhole saddle of JT gravity introduced in \cite{Blommaert:2021fob}. } as shown in Figure.~\ref{halfz}. We will show in the following that such saddles indeed exist and together with their contribute $\Theta_1$ the following approximation is good
\bea \label{hz}
z\approx \langle z\rangle+\Theta_1\ .
\eea 
Let us clarify the notation we use in this paper,  we call the non-self-averaged component in $z$ as ``single half-wormhole" or simply ``half-wormhole", and we refer to the non-self-averaged saddle in $z^2$ as ``linked half-wormhole". 
To demonstrate~\eqref{hz} explicitly, recall that the partition function is given by
\bea \label{z11}
z=\int \d^N \psi \exp \(\im^{q/2}\sum J_{i_1\dots i_q}\psi_{i_1\dots i_q}\)\ .
\eea 
The ensemble averaged quantity $\<z\>$ does not vanish 
\bea \label{z1a}
\langle z\rangle=\int \d^N \psi \exp (\im^{q/2}\sum J^{(0)}_{i_1\dots i_q}\psi_{i_1\dots i_q})=u^p\frac{(pq/2)!}{p!((q/2)!)^p}\equiv m_p u^p\,, \quad pq =N\ .
\eea 
In the following we present a heuristic but simple proof of this result. A more rigorous but technical proof is presented in Appendix \ref{proof}. For simplicity let us first consider the $q=4$ case
\bea 
\langle z\rangle_{}=\int d^{N}\psi \, e^{-u \sum_{A}\psi_A},\quad A=\{a_1<\dots<a_4\}\ .
\eea 
We introduce the collective variable $G$ 
\bea 
G=\frac{1}{N}\sum_{1\leq i<j\leq N}\psi_i\psi_j,\quad G^2=\frac{2!}{N^2}\sum_A \psi_A\,,\label{gij}
\eea 
then $\langle z\rangle$ can be rewritten as
\bea \label{z4ij}
\langle z\rangle_{}=\int_{\mathbb{R}}\d G \int_{\im \mathbb{R}}\frac{d\Sigma}{2\pi \im/N} \d^N \psi\, e^{-\frac{u}{2 }N^2G^2} e^{- \Sigma (NG-\sum_{i<j}\psi_i\psi_j)}\ .
\eea 
Now we can integrate the out the fermions to get
\bea 
\int d^N \psi\, e^{\Sigma \sum_{i<j}\psi_i \psi_j}=(\Sigma)^{N/2} m_p\, |_{(q=2)}=\Sigma^{N/2}\ .
\eea 
Then \eqref{z4ij} becomes
\bea \label{trick1}
\langle z\rangle_{q=4}&=&\int_{\mathbb{R}}\d G \int_{\im \mathbb{R}}\frac{d\Sigma}{2\pi \im/N}\Sigma^{N/2} e^{- \frac{ u N^2G^2}{2}}e^{-N \Sigma G}\, \nn \\
&=&N^{-N/2}(\partial_G)^{N/2}e^{-\frac{ u N^2 G^2}{2}}\, |_{G=0} \, =\( \frac{u}{2}\)^{N/4} \frac{(N/2)!}{(N/4)!}=m_p u^p|_{q=4}\, .
\eea 
For general $q$, the proof is similar with the modification 
\bea 
\sum_A\psi_A=\frac{N^{q/2}}{(q/2)!}G^{q/2}\, .
\eea 
In summary,  we have generalized the $G,\Sigma$ trick and derived an effective action to compute $\langle z\rangle$:
\bea \label{gsz}
\langle z\rangle_{}=\int_{\mathbb{R}}\d G \int_{\im \mathbb{R}}\frac{d\Sigma}{2\pi \im/N}\Sigma^{N/2} e^{u \im^{q/2}\frac{N^{q/2}}{(q/2)! }G^{q/2}} e^{- N\Sigma G}\, .
\eea 
It would be convenient to rotate the integral contour as
\bea 
\Sigma\rightarrow \im  e^{-\im \frac{2\pi}{q}} \s,\quad G\rightarrow e^{\im \frac{2\pi}{q}} g
\eea 
such that we obtain a ``standard" action:
\bea \label{z1t1}
\langle z\rangle_{}=\int_{\mathbb{R}}\frac{dg d\s }{2\pi/N }\exp\left\{\frac{N}{2}\(\log(\im e^{-\frac{2\pi\im }{q}}\s) -2\im \s g-\frac{2\mu}{q}g^{q/2} \)\right\},
\eea 
where we define
\bea \label{para2}
\mu\equiv \im^{q/2}u \frac{2{N}^{q/2-1}}{(q/2-1)!},\quad \leftrightarrow \quad  u=(-\im)^{q/2}\mu\frac{(q/2-1)!}{2 N^{q/2-1}}.
\eea 
Rescaling $\mu$ to 1,  the saddle point equations are then 
\bea \label{whs2}
\frac{1}{\s }-2\im g=0,\quad -2\im\sigma-\m g^{q/2-1}=0,\quad \rightarrow \quad \m g^{q/2}=-1\ .
\eea 
Comparing \eqref{z1t1}  with \eqref{z2t1}  it is easy to find that to reproduce the exact result \eqref{z1a} we have to added the contributions from all the $q/2$ saddles.

\begin{figure}
  \includegraphics[scale=0.25]{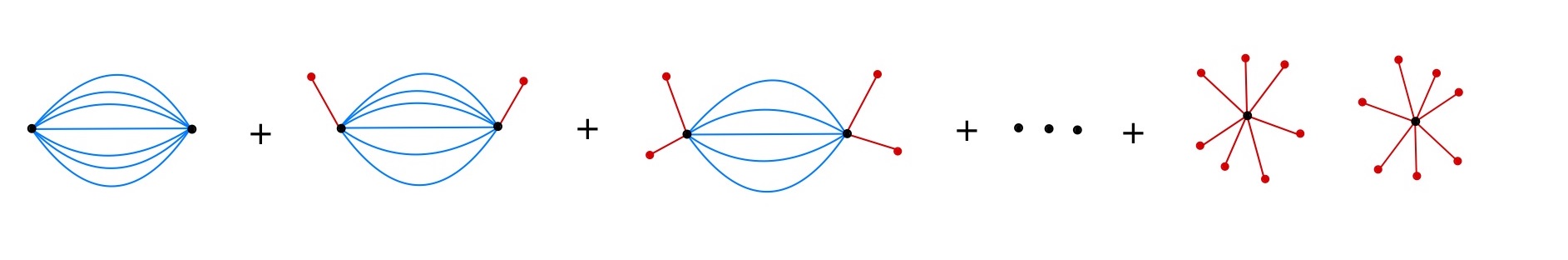}
  \caption{Feynman diagrams for $\langle z^2\rangle, \langle \Theta_1^2\rangle,\langle z\Theta_1\rangle$. Each black dot represents a $z$ or $\Theta_1$, each red dot and the attached  line represents a contraction with the $J_A^0$ source, and each blue line is a contraction of a pair of $J_A$.} \label{z2figure}
\end{figure}

Having found the suitable saddle contributions to the averaged partition function $\<z\>$, we proceed to analyze the difference between the non-averaged quantity and the mean value $z-\<z\>$. We start with inserting the identity
\bea \label{hidentity}
1=\int_{-\infty}^\infty dG_h\int_{-\im \infty}^{\im \infty} \frac{Nd\Sigma_h}{2\pi \im }e^{-\Sigma_h(NG_h-\sum_{i<j}\psi_i \psi_j)+\frac{N\mu}{q}\(G_h^{q/2}-\(\frac{1}{N}\sum_{i<j}\psi_i \psi_j\)^{q/2}\) }\,,\nn \\
\eea 
into the non-averaged partition function $z$.
To make the integral well defined, we again rotate the contour by $\Sigma_h=\im e^{-2\im \pi/q}\sigma_h,G_h=e^{2\im \pi/q}g_h$, then $z$ can be cast into the form 
\bea 
z=\int_{-\infty}^\infty \frac{N\d \sigma_h}{2\pi} \Psi(\sigma_h)\hat{\Theta}(\sigma_h)\,,
\eea 
where the first factor is similar to \eqref{psi}
\bea 
\Psi(\sigma_h)=\int_{\mathbb{R}}\frac{\d g_h}{2\pi/N} \exp[N(-\im \sigma_h g_h-\frac{\mu}{q} g_h^{q/2})]\,,
\eea 
and the second factor is
\bea \label{thetah}
\hat{\Theta}(\sigma_h)=\int \d^N \psi \exp[\im  e^{-\frac{2\im \pi}{q}}\sigma_h \sum_{i<j}\psi_i \psi_j+\im^{q/2} J_A\psi_A-\im^{q/2}u \sum_A\psi_A ]\ .
\eea 
Averaging over the coupling, we get back to the computation in~\eqref{z1t1} where $\s_h =\frac{1}{2i} \left(\m^{-2/q} e^{4\p i (n+\frac{1}{2})/q}\right)$. We expect a separate saddle point to appear in this integral which leads to the difference $z-\<z\>$. 
The $\Psi(\s_h)$ is peaked at $\s_h=0$, so we look for dominant contributions around $\s_h \approx 0$, which is
\bea \label{theta0syk}
\Theta_1=\hat{\Theta}(0)=\text{Pf}(J-J^0)=\sum'_{A}\text{sgn}(A)(J_{A_1}-J_{A_1}^{0})\dots (J_{A_p}-J_{A_p}^{0})\ .
\eea 
It is clear that its average vanishes $\langle \Theta_1\rangle=0$. 
Then we propose the approximation
\bea 
z\approx \langle z\rangle+\Theta_1\ .
\eea 
which is~\eqref{hz}. 
According to the power of $J^0_A=u$, we can further expand
\bal
\Theta_1&= \sum_{k=0}^{p} \Theta_1^{(k)} u^k\ .
\eal
To verify this approximation, we define the error function
\bea \label{error1}
\text{Error}=z-\langle z\rangle-\Theta_1\ .
\eea  
A direct calculation gives
\bea 
\langle \text{Error}^2\rangle=\langle z^2\rangle-\langle z\rangle^2+\langle \Theta^2\rangle-2\langle z\Theta\rangle
\eea 
The quantities $\langle z^2\rangle, \langle \Theta^2\rangle,\langle z\Theta\rangle$ can be computed with the Feynman diagrams as shown in Fig. \ref{z2figure}.
Recall that value of $\langle z\rangle$ is given by the star diagram that is one connected component of the last term in Fig.~\ref{z2figure}
\bea 
\langle z\rangle=\frac{(pq/2)!}{p!((q/2)!)^p}\mu^p\equiv m_p\mu^p\,,\label{zvev}
\eea 
The value of $\langle z^2\rangle$ can  be computed either from summing over the diagrams, 
\bea \label{z2our}
\langle z^2\rangle=\sum_{k=0}^p c_k m_{p-k}^2t^{2k}u^{2p-2k}\equiv \sum_k z_2^{(k)}\,,
\eea 
where
\bal
&&c_k=\frac{1}{k!}{ N\choose q }{ N-q\choose q }\dots {N-(k-1)q\choose q}=\frac{N!}{k! (q!)^k (N-kq)!}\,, \label{ck}
\eal
or by introducing the collective variables 
\bea 
G_{LR}=\frac{1}{N}\sum_i \psi_i^L \psi_i^R,\quad G_L=\frac{1}{N}\sum_{i<j}\psi_i^L \psi_j^L,\quad G_R=\frac{1}{N}\sum_{i<j}\psi_i^R \psi_j^R\,,
\eea 
and doing the path integral
\bea 
\langle z^2\rangle &=& \int_R \d^3 G_i \int_{\im \mathbb{R}}d^3 \Sigma_i\, e^{\frac{N}{q}(\tau^2 G_{LR}^q+\mu G_L^{q/2}+\mu G_{R}^{q/2})- N(\Sigma_i G_i)}\int \d^{2N}\psi e^{\frac{1}{2}{\Psi}M{\Psi}},\\
&=&\int_R \d^3 G_i \int_{\im \mathbb{R}}d^3 \Sigma_i\,e^{\frac{N}{q}(\tau^2 G_{LR}^q+\mu G_L^{q/2}+\mu G_{R}^{q/2})- N(\Sigma_i G_i)}\sqrt{\text{det}[\S_L\S_RA^2+\S_{LR}^2]}\nonumber\\
&=&\int_R \d^3 G_i \int_{\im \mathbb{R}}d^3 \Sigma_i\, e^{\frac{N}{q}(\tau^2 G_{LR}^q+\mu G_L^{q/2}+\mu G_{R}^{q/2})- N(\Sigma_i G_i)}\text{det}[\im \sqrt{\S_L\S_R} A+ \S_{LR}]\nonumber\\ 
&=& \int_R \d^3 G_i \int_{\im \mathbb{R}}d^3 \Sigma_i\, e^{\frac{N}{q}(\tau^2 G_{LR}^q+\mu G_L^{q/2}+\mu G_{R}^{q/2})- N(\Sigma_i G_i)}\frac{1}{2}\((\Sigma_{LR}+\im \sqrt{\S_L\S_R})^N+(\Sigma_{LR}-\im \sqrt{\S_L\S_R})^N\)\nonumber \\
&=&\int_R \d^3 G_i \int_{\im \mathbb{R}}d^3 \Sigma_i\, \sum_{m=0}^{N/2} {N \choose 2m}( \Sigma_{LR})^{2m}(\im^2 \Sigma_L\Sigma_R)^{\frac{N}{2}-m} e^{\frac{N}{q}(\tau^2 G_{LR}^q+\mu G_L^{q/2}+\mu G_{R}^{q/2})}e^{- N(\Sigma_i G_i)}\nonumber\,, \label{exz2s}
\eea 
where we have defined
\bea 
&&\Psi=\(\psi_1^L,\dots,\psi_{N}^L,\psi_1^R,\dots,\psi_{N}^R\),\quad M=\begin{pmatrix}
	\Sigma_L A& \Sigma_{LR} I_{N}\\
	-\Sigma_{LR} I_{N}& \Sigma_R A\\
\end{pmatrix},\\
&& A=-A^T,\quad A_{ij}=1,\quad \forall i<j .
\eea 
Using the same tricks as \eqref{trick1}, \eqref{exz2s} can be evaluated exactly as
\bea 
\langle z^2\rangle &=&N^{-N}\sum_{k=0}^p{N\choose kq}(\partial_{G_{LR}})^{kq}(\im^2\partial_{G_L}\partial_{G_R})^{\frac{N-kq}{2}}e^{\frac{N}{q}(\tau^2 G_{LR}^q+\mu G_L^{q/2}+\mu G_{R}^{q/2})}|_{G_i=0} \\
&=&N^{-N}\sum_{k=0}^p\im^{N-kq}{N\choose kq}\frac{(kq)!}{k!}\(\frac{N\tau^2}{q}\)^k\left[\frac{(\frac{q(p-k)}{2})!}{(p-k)!}\right]^2\(\frac{N \mu}{q}\)^{2p-2k}\\
&=&\sum_{k=0}^p c_k m_{p-k}^2t^{2k}u^{2p-2k},
\eea 
which agrees with \eqref{z2our} as it should be.

Furthermore, from this result we find $z_2^{(0)}=\langle z \rangle^2$ which is given by the last diagram in Fig.~\ref{z2figure} and $z_2^{(p)}=\langle z^2\rangle_{\mu=0}$ which is given by the first diagram in Fig.~\ref{z2figure}. The expression of $\Theta_1$ \eqref{theta0syk} implies that $\langle \Theta_1^2\rangle=\langle \Theta_1 z\rangle=z_2^{(p)}$, therefore we find
\bea 
&&\langle \text{Error}^2\rangle=\sum_{k=1}^{p-1}c_k m_{p-k}^2t^{2k}u^{2p-2k}\equiv \sum_{k=1}^{p-1} z_2^{(k)}\, , \label{summation0syk}
\eea 
where $m_p$ is defined in~\eqref{z1a}.
In the large-$N$ limit, some of the terms in the summation~\eqref{z2our} dominate. If $z_2^{(p)}$ or $z_2^{(0)}$ dominates then the error is small.

However the dominant term is not always given by  a fixed $z_2^{(k)}$. A simple argument is the following. To find the dominant term we can compute the ratio\footnote{Recall that $p=N/q$.}
\bea 
&&r_k=\frac{z_2^{(k)}}{z_2^{(k-1)}}=\frac{t^2 (-k+p+1) (-4 k+4 p+1) (-4 k+4 p+3)}{3 u ^2 (2 k (p-k)+k)}\, ,\\
&&r_p=\frac{t^2}{p u^2},\quad r_{1}\sim \frac{p^2 t^2}{ u^2}\,,
\eea 
here for simplicity we have chosen $q=4$. First we notice that $r_k$ decreases with respect to $k$. Therefore if $r_1 \leq 1$ i.e.
\bea \label{disk}
\frac{u}{t }\geq {p}\,,
\eea 
then the dominant term will be $z_2^{(0)}$. It means that all the wormhole saddles are suppressed. However if $r_p\geq 1$ i.e.
\bea \label{wh}
\frac{u}{t }\leq \frac{1}{\sqrt{p}}
\eea 
then the dominant term will be $z_2^{(p)}$, in other words the effect of $\mu$ can be neglected. For other cases with
\bea 
\frac{1}{\sqrt{p}}<\frac{u}{t }<p, \label{between}
\eea 
by fine tuning the value of $u/t$, every diagram in Fig. \eqref{z2figure} is possible to be dominant. 
For the choices \eqref{para} and \eqref{para2} which lead to reasonable large $N$ behavior  we have
\bea 
\frac{u}{t}\sim \frac{\mu}{\tau} \frac{(q/2-1)!}{\sqrt{(q-1)!}} N^{\frac{1}{2}}\sim \sqrt{p},
\eea 
which exactly lies in the \eqref{between}. It also implies there should be other saddles contributing to \eqref{thetah}.

 On the other hand, the can derive the saddle point equations
\bea 
&& G_{L(R)}^{-1+\frac{q}{2}}=\frac{2}{\mu}\Sigma_{L(R)},\quad G_{LR}^{-1+q}=\frac{1}{\tau^2}\Sigma_{LR},\label{res1}\\
&& G_{L(R)}=\frac{\im \S_{R(L)}}{2\sqrt{\S_L \S_R}}\frac{f_+^{n-1}-f_-^{n-1}}{f_+^n+f_-^n},\quad G_{LR}=\frac{f_+^{n-1}+f_-^{n-1}}{f_+^n+f_-^n}\,,\label{sd2f}
\eea 
where $f_\pm=\Sigma_{LR}\pm \im \sqrt{\S_L\S_R}$.  Again for simplicity we will choose $\tau^2=\mu=1$.
There are always two types of trivial solutions
\bea 
&&\text{wormhole solution}:\quad G_L=G_R=0,\quad G_{LR}=e^{\frac{2\im m\pi }{q}},\\
&&\text{disconnect solution}:\quad G_{LR}=0,\quad G_{L}=e^{\frac{4\im m_L \pi }{q}} ,\quad G_{R}=e^{\frac{4\im m_R \pi }{q}}
\eea 
with on-shell action
\bea 
&&\text{wormhole solution}: \quad \langle z^2\rangle_{\text{wh}}=e^{-N(1-\frac{1}{q})}e^{\frac{2\im m \pi N}{q}}\\
&&\text{disconnect solution}:\quad \langle z^2\rangle_{\text{dis}}={2^{-N}}e^{-N(1-\frac{2}{q})}{e^{\frac{4\im m \pi N}{q}}}.
\eea 
Note that the ratio of these two contribution is
\bea 
\frac{\langle z^2\rangle_{\text{wh}}}{\langle z^2\rangle_{\text{dis}}}=\(2e^{-1/q}\)^N,
\eea 
so when $q\geq 2$ it is the wormhole saddle dominates. 
The general analytic solution is hard to obtain. However in the large $N$ limit we expect that only $f_+$ or $f_-$ will survive. Assuming $f^N_-\rightarrow 0,N\rightarrow \infty$, \eqref{sd2f} get dramatically simplified
\bea 
G_{L(R)}=\frac{\S_{R(L)}}{-2\im \sqrt{\S_R\S_L}}\frac{1}{\S_{LR}+\im \sqrt{\S_L \S_R}},\quad G_{LR}=\frac{1}{\S_{LR}+\im \sqrt{\S_L \S_R}},\label{res2}
\eea 
from which we obtain
\bea 
G_{LR}^q+G_{R}^{q/2}+G_{L}^{q/2}=1,\quad G_{R}^{q/2}=G_{L}^{q/2}.
\eea 
For the case of $q=4$, \eqref{res1} and \eqref{res2} can be solved explicitly and it contributes the on-shell action
\bea 
\langle z^2\rangle_{\text{non-trivial}+}\approx e^{-0.63 N}e^{\frac{2m\im \pi N}{4}} >\langle z^2\rangle_{\text{wh}}= e^{-0.75 N}e^{\frac{2m\im \pi N}{4}}.
\eea 
We also checked that for these solutions $\lim_{N\rightarrow \infty}f_-^N = 0$. Similar saddles can also be found for the case of $f_+^N=0$. Therefore we conclude that in the large $N$ limit the dominate saddles are the non-trivial ones. 

In the regime of \eqref{between}, the ansatz \eqref{theta0syk} of half-wormhole saddle is not adequate. We have  to consider the contribution from the $\sigma_h$ fluctuation to $\Theta$. This can be done by expanding $\hat{\Theta}(\sigma_h)$ with respect $\s_h$, substituting into $z$ and integrating over $\s_h$. Equivalently this can be done by expanding the exact value of $z$ 
\bea 
z&=&\text{PF}(J_A)=\text{PF}(u+J_A-J_A^0)\nonumber\\&=&\sum'_{A}\text{sgn}(A)(u+J_{A_1}-J_{A_1}^{0})\dots (u+J_{A_p}-J_{A_p}^{0})\equiv \sum_{n=0}^p \Theta^{(n)}\,,
\eea 
with respect to $u$.  For examples
\bea 
&&\Theta^{(p-1)}= \sum_A'\text{sgn}(A)(J_{A_1}-J_{A_1}^0)\dots J_{A_i}^0\dots (J_{A_p}-J_{A_p}^0)\, ,\\
&&\Theta^{(0)}=\langle z\rangle \, ,\quad \Theta^{(p)}=\Theta.
\eea 
Then from the Feynman diagrams it is not hard to find  in Fig.  \ref{z2figure} that
\bea 
\langle {\Theta^{(k)}}{\Theta^{(k)}}\rangle=\langle \Theta^{(k)} z\rangle= z_2^{(k)}.
\eea
So if $z_2^{(k)}$ is the dominant term, we can choose the half-wormhole saddle to be $\Theta^{(k)}$. Or we can think of that for each wormhole saddle $z_2^{(k)}$ there is a corresponding half-wormhole saddle $\Theta^{(k)}$ such that
\bea 
z\approx \langle z\rangle+ \Theta^{(k)}.
\eea
We will present a further analysis on this model somewhere else.

\subsection{Linked half-wormhole saddles in $z^2$}

\begin{figure}
\begin{center}
  \includegraphics[scale=0.25]{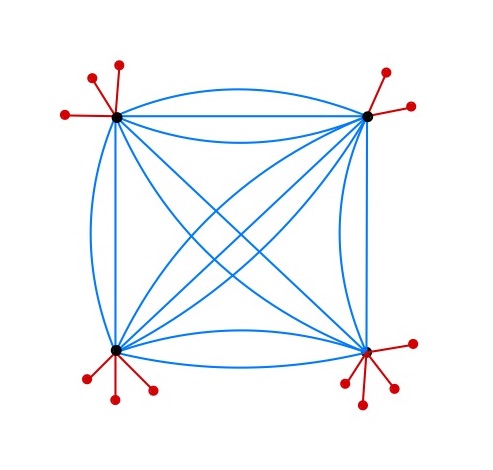}
  \caption{Feynman diagrams for $\langle z^4\rangle$} \label{z4}
  \end{center}
\end{figure}

\begin{figure}
\begin{center}
  \includegraphics[scale=0.25]{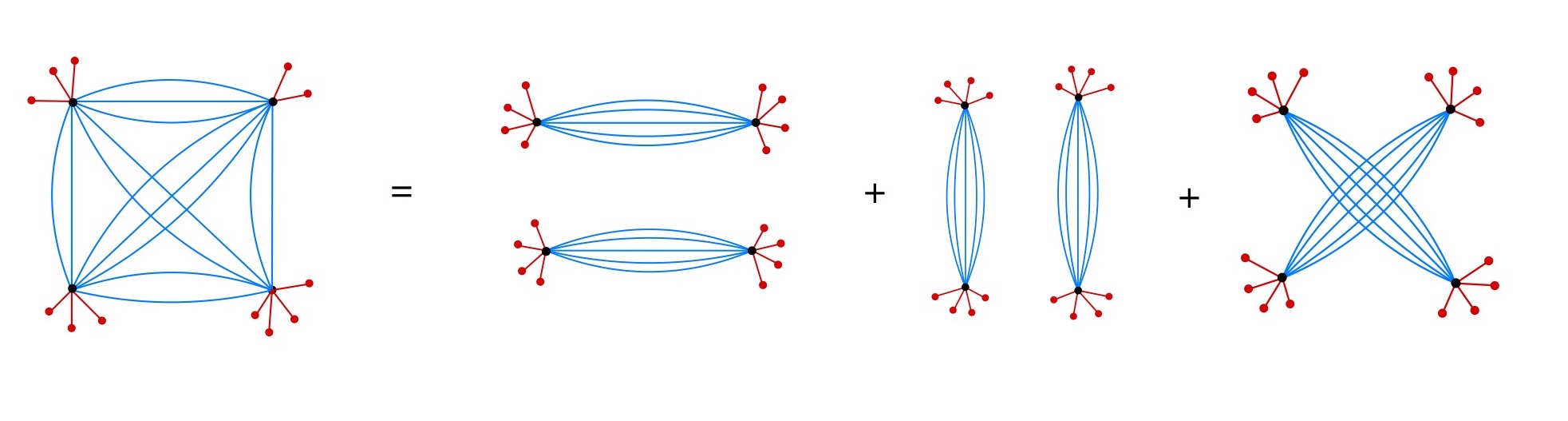}
  \caption{$\langle z^4\rangle \approx 3 {z_2^{(k)}}^2$} \label{z42}
  \end{center}
\end{figure}

In this section we study the linked half-wormhole contribution to $z^2$, and, in particular, we would like to understand the relation with the single half-wormhole saddles in $z$,

To get a general picture, we first compute $\langle z^4\rangle$ from the Feynman diagrams shown in Fig.\ref{z4}. In general it is a cumbersome combinatorial problem but in the large $N$ limit we know that it should be factorized into disconnected diagrams as 
\bea \label{z4sa}
\langle z^4\rangle \approx 3 {z_2^{(k)}}^2\, ,\qquad \langle z^2\rangle\approx z_2^{(k)}\, ,
\eea 
which is shown in Fig.\ref{z42} and here we have assumed that $z_2^{(k)}$ is the dominant wormhole saddles.

This means there are more refined structures of the nontrivial saddles in $z^2$, comparing with the general discussion in~\cite{Saad:2021rcu}.  
Inspired by our analysis of the single half-wormhole for $z$, we insert another two copies of identities \eqref{hidentity} in $z^2$
\bea 
&&z^2=\int \d\sigma_w \d\sigma_{h_L}\d \sigma_{h_R}\Psi(\sigma_w,\sigma_{h_L},\sigma_{h_L})\hat{\Lambda}(\sigma_w,\sigma_{h_L},\sigma_{h_L})\,, \\
&&\Psi(\sigma_w,\sigma_{h_L},\sigma_{h_L})=\Psi(\s_w)\Psi(\s_{h_L})\Psi(\s_{h_R})\, ,\\
&&\hat{\Lambda}(\sigma_w,\sigma_{h_L},\sigma_{h_L})=\int \d^{2N} \psi \exp[\im  e^{- \frac{2\im\pi}{q}}\sigma_{h_L} \sum_{i<j} \psi^L_{ij} +\im  e^{-\frac{2\im \pi}{q}}\sigma_{h_R} \sum_{i<j} \psi^R_{ij} +\im e^{\frac{\im \pi}{q}}\sigma_w \psi_i^L \psi_i^R\nn \\
&&\qquad \qquad \qquad \qquad +\im^{q/2} J_A(\psi_A^L+\psi_A^R)-\im^{q/2}u \sum_A(\psi_A^L+\psi_A^R)-\im^q t^2\psi_A^L\psi_A^R  ],
\eea 
where we have introduced three pairs of $G,\Sigma$ variables
\bea 
&&G_w=\frac{1}{N}\psi_i^L\psi_i^R,\quad G_{h_L}=\frac{1}{N}\sum_{i<j} \psi^L_{ij},\quad G_{h_R}=\frac{1}{N}\sum_{i<j} \psi^R_{ij},
\eea 
and rotated the contour as before. As before, the function $\Psi$ is highly peaked around $\Psi(0,0,0)$ so we expect that there is a half-wormhole saddle point 
\bal 
\Lambda=\hat{\Lambda}(0,0,0)&=\sum'_{A,B}\text{sgn}(A)\text{sgn}(B) \prod_{k=1}^p\((J_{A_k}-J_{A_k}^0)(J_{B_k}-J_{B_k}^0)-\delta_{A_k B_k}t^2\)\,,
\eal 
whose average manifestly vanishes $\langle \L\rangle=0$ and it further satisfies $\langle \L^2\rangle=2 {z_2^{(p)}}^2$.
However because of the large $N$ behavior \eqref{z4sa}, again we have to consider the fluctuations of $\s_h$. It is achieved by  expand $\hat{\L}(0,\sigma_{h_L},\sigma_{h_R})$ with respect to $\s_{h_{L(R)}}$ or equivalently by expanding 
\bal 
\sum'_{A,B}\text{sgn}(A)\text{sgn}(B)\prod_{k=1}^p\((u+J_{A_k}-J_{A_k}^0)(u+J_{B_k}-J_{B_k}^0)-\delta_{A_k B_k}t^2\) \equiv \sum_{n=0}^{p} \Lambda^{(k)}.
\eal 
Some examples are
\bea 
 &&\L^{(p-1)}=\sum_i\sum'_{A,B}\text{sgn}(A)\text{sgn}(B)\((J_{A_1}-J_{A_1}^0)(J_{B_1}-J_{B_1}^0)-\delta_{A_1 B_1}t^2\)\dots \nn\\
&& J_{A_i}^0J_{B_i}^0\dots\((J_{A_p}-J_{A_p}^0)(J_{B_p}-J_{B_p}^0)-\delta_{A_p B_p}t^2\),\quad \L^{(0)}=\langle z\rangle^2,\quad \Lambda^{(p)}=\Lambda\ .\nonumber
\eea 
Then similarly one can find that
\bea 
\langle \Lambda^{(k)}\Lambda^{(k)}\rangle =\langle z \Lambda^{(k)}\rangle= 2 {z_2^{(k)}}^2
\eea 
so that when $z_2^{(k)}$ is the dominant wormhole saddle in the large $N$ limit the
\bea 
z^2\approx \langle z^2\rangle+\Lambda^{(k)}\approx z_2^{(k)}+\Lambda^{(k)}\,,
\eea 
is a good approximation.

\section{SYK at one time point: $\langle J_a\rangle=0,\quad \langle J_a^4\rangle_c\neq 0$}
\begin{figure}[h]
\begin{center}
  \includegraphics[scale=0.35]{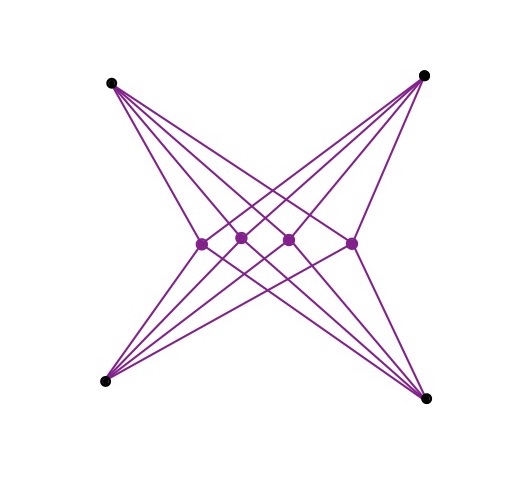}
  \caption{$z_4^{(p)}$} \label{z4v}
  \end{center}
\end{figure}

Another class of interesting distributions of the random coupling is non-Gaussian. In this section we consider a special subset of them that have vanishing mean values, namely
\bal
\langle J_A\rangle=0\,,\qquad 
\langle J_A^2\rangle= t^2\,, \qquad \langle J_A^4\rangle=v^4+3 \langle J_A^2\rangle^2\ .\label{j4}
\eal
It is easy to compute that the partition function of the 0d SYK model with such random couplings are 
\bea
 \langle z\rangle=0,\quad \langle z^2\rangle=\frac{N!}{p!(q!)^p}t^2,\ .
\eea
The higher moments of $J_A$ in~\eqref{j4} contributes nontrivially to $\langle z^4 \rangle$
\bea
\langle z^4\rangle &=&\sum_{A,B,C,D}'\text{sgn}(A)\text{sgn}(B)\text{sgn}(C)\text{sgn}(D)\langle J_{A_1}J_{B_1}J_{C_1}J_{D_1}\dots J_{A_p}J_{B_p}J_{C_p}J_{D_p} \rangle\,,
\eea 
which can be expanded
\bea
&&\langle z^4\rangle=\sum_{k=0}^{p} c_k n_{N-qk} v^{4k} t^{4(p-k)}\equiv \sum_k z_4^{(k)},\nn \\
&&n_{N}=\frac{N!}{(q!)^{2N/q}}\sum_{\substack{n_1+n_2+n_3=N/q\\n_i\geq 0}}\frac{(qn_1)!(qn_2)!(qn_3)!}{(n_1!n_2!n_3!)^2} ,\nn \\
&&c_k n_{N-qk}=\frac{N!}{k!(q!)^{2p-k}}\sum_{\substack{n_1+n_2+n_3=N/q-k\\n_i\geq 0}}\frac{(qn_1)!(qn_2)!(qn_3)!}{(n_1!n_2!n_3!)^2}
\eea
where $c_k$ is the number of ways to choose $k$ $q$-subsets out of $N$ and $n_N$ is the multiplicities coming from the different Wick contractions, i.e.
\bea
 \langle z^4\rangle_{v=0}=n_N t^{4p}.
\eea
To find the dominant term in the large $N$ limit let us define the ratio
\bea 
&&\tilde{r}_k=\frac{z_4^{(k)}}{z_4^{(k-1)}}\sim \frac{v^4}{t^4}\frac{1-k+p}{k}\frac{4! (4p-kp)!}{(4p-4k+4)!},\\
&&\tilde{r}_1\sim \frac{v^4}{t^4}\frac{1}{p^{2}},\quad \tilde{r}_p\sim \frac{\u^4}{t^4}\frac{1}{p} \, ,
\eea 
where we have taken $q=4$ for simplicity. By taking the derivative with respect to $k$ we find that $\tilde{r}_k$ will initially decrease and then increase with increasing $k$ so $\tilde{r}_p$ is the maximal value. If $\tilde{r}_p\leq 1$ i.e.
\bea 
\frac{v^4}{t^4}\leq p\,,
\eea 
then the dominant term will be $z_4^{(0)}$ therefore the contributions of higher moments can be ignored in this limit. Recall that the half-wormhole saddle of $z^2$ when $\langle J_A\rangle=0$ can be written as
\bea\label{phig}
\Phi=\sum'_{A,B}\text{sgn}(A)\text{sgn}(B)\(J_{A_1}J_{B_1}-\delta_{A_1 B_1}t^2\)\dots\(J_{A_p}J_{B_p}-\delta_{A_p B_p}t^2\) \, ,
\eea 
such that
\bea 
\langle \Phi^2\rangle \approx \langle \Phi z^2\rangle \approx 2\langle z^2\rangle^2,
\eea 
and 
\bea 
\langle \text{Error}^2\rangle&=&\langle z^4\rangle-\langle z^2\rangle^2+\langle \Phi^2\rangle -2\langle z^2\Phi^2\rangle \nn \\
&\approx & 3\langle z^2\rangle^2-\langle z^2\rangle^2+2\langle z^2\rangle^2-4\langle z^2\rangle^2=0,
\eea
in the leading order of $N$ as before. However if $\tilde{r}_p>1$, then it will be possible that $z_4^{(p)}$ is the leading term whose corresponding Feynman diagram is shown in Fig.\ref{z4v}.
Therefore there will be no half-wormhole saddle anymore since the (two-mouth) wormhole saddles are not dominant. 

One can consider more general distribution with all the cumulants to be non-vanishing. The analysis and the results will be similar. If $v$ is very large then it is the four-way wormhole saddle that dominate. It is therefore possible to introduce a new ''four-linked-wormhole" saddle as we show in next section. However, if $v$ is relatively small it is still the two-mouth wormhole (with some legs as shown in Fig.\ref{z2figure}) that dominates. We will present a more thorough analysis of these points separately.

\section{SYK at one time point: $\langle J_a\rangle=\langle J_a^2\rangle=\langle J_a^3\rangle=0$}\label{j4}

In this section, we consider a special model where we could focus on the ``multi-linked" wormhole saddle points.
In this model the random coupling only have non-vanishing $4^{\text{th}}$ cumulant
\bea \label{choicej}
\langle J_a\rangle=\langle J_a^2\rangle=\langle J_a^3\rangle=0,\quad \langle J_a^4\rangle=v^4\ .
\eea
Such a distribution could also be considered as an extremal limit of other distributions.

\subsection{Averaged quantities: $\langle z^4\rangle$ and $\langle z^8\rangle$}
Due to our special choice \eqref{choicej} the first non-vanishing averaged quantity is 
\bea 
\langle z^4\rangle &=& \int \d^{4N}\psi \exp\(v^4\sum_{A_1<\dots < A_q}\psi_{A_1}^1\psi_{A_1}^2\psi_{A_1}^3\psi_{A_1}^4\dots \psi_{A_q}^1\psi_{A_q}^2\psi_{A_q}^3\psi_{A_q}^4  \) \nn \\
&=&\int \d^{4N}\psi \exp\(\frac{v^4}{q!} (\sum_{i}^N\psi_{i}^1\psi_{i}^2\psi_{i}^3\psi_{i}^4)^q\)\, .
\eea 
Then we can introduce the $G,\Sigma$ trick
\bea 
\langle z^4\rangle &=&\int \d^{4N}\psi \int \d G\, \delta(G_4-\sum_{i}^N\psi_{i}^1\psi_{i}^2\psi_{i}^3\psi_{i}^4)\exp \(\frac{v^4}{q!}G_4^q\) \nn \\
&=&\int \d^{4N}\psi\int \d G\frac{\d\Sigma}{2\pi \im}\exp\(-\Sigma(G_4-\sum_{i}^N\psi_{i}^1\psi_{i}^2\psi_{i}^3\psi_{i}^4)\)\exp \(\frac{v^4}{q!}G_4^q\) \nn \\
&=&\int \d G\int\frac{\d\Sigma}{2\pi \im} \exp\( N \log\Sigma-\Sigma G_4+\frac{v^4}{q!}G^q\) \nn \\
&=&(\pa_{G_4})^N \exp \(\frac{\u^4}{q!}G_4^q\)\, |_{G_4=0}=\(\frac{v^4}{q!}\)^{N/q}\frac{N!}{(N/q)!}=v^{4p}\frac{N!}{p!(q!)^p}\, .
\eea 
Alternatively, we can obtain this result by integrating out the fermions first to get the hyperpfaffin, taking the $4^{\text{th}}$  power, and then do the average
\bal \label{z8}
\langle z^4\rangle =\sum_{ABCD}\text{sgn}(A,B,C,D)\langle J_{A_1}J_{B_1}J_{C_1}J_{D_1}\dots J_{A_p}J_{B_p}J_{C_p}J_{D_p}\rangle=v^{4p}\sum_A \, 1=v^{4p}\frac{N!}{p!(q!)^p}\,.
\eal 
The computation of $\langle z^8\rangle$ is more involved 
\bea 
\langle z^8\rangle=\int \d^{8N}\psi \exp\(\frac{v^4}{q!} (\sum_{i}^N\psi_{i}^a\psi_{i}^b\psi_{i}^c\psi_{i}^d)^q\)\,, 
\eea 
where 
\bea 
(a,b,c,d)\in \{1\leq a<b<c<d\leq 8\}\ .
\eea 
In the following we use the collective index $A'$ to label the $4$-element subset. Then we introduce antisymmetric tensors $G_{abcd}=G_{A'}$ and $\Sigma_{abcd}=\Sigma_{A'}$ as the collective field variables such that \eqref{z8} can be expressed as
\bea 
\langle z^8\rangle &=& \int \frac{\d G_{A'}\d \Sigma_{A'}}{(2\pi \im)^{70}}(\text{PF}(\Sigma_{A'}))^N\exp\(-\sum_{A'}[\Sigma_{A'}G_{A'}+\frac{v^4}{q!}G_{A'}^q]\) \nn \\
&=&\(\sum'_{A'_1<A'_2}\text{sgn}(A')\pa_{G_{A'_1}}\pa_{G_{A'_2}}\)^N\exp\(\frac{v^4}{q!}G_{A'}^q\) |_{G_{A'}=0} \nn \\
&\approx& \(\frac{v^4}{q!}\)^{\frac{2N}{q}} \frac{N!^2}{p!^2}\frac{1}{2}{8\choose 4}=35\(\frac{v^4}{q!}\)^{\frac{2N}{q}} \frac{N!^2}{p!^2}\, ,
\eea 
where in the last line we have taken the large $N$ limit. In this limit we have 
\bal
\langle z^8\rangle\approx 35 \langle z^4\rangle^2 \ .
\eal

\subsection{The un-averaged  $z^4$}
Following similar ideas as in the previous sections, we insert a suitable identity to the expression of $z^4$
\bea 
z^4&=&\int \d^{4N}\psi \exp\(\im^{q/2}\sum_{A,i}J_A\psi_A^i\)\int \d G_4 \delta(G_4-\sum_{i}^N\prod_{a=1}^4\psi_{i}^a)\exp \(\frac{v^4}{q!}[G_4^q-(\sum_{i}^N\prod_{a=1}^4\psi_{i}^a)^q]\)\,  \nn \\
\eea 
Rotating the contour as before we can rewrite $z^4$ as
\bea 
z^4=\int \d \sigma \Psi(\sigma)\hat{\Gamma}(\sigma)\, ,
\eea 
where $\Psi(\sigma)$ is same as \eqref{psi} and the second factor is
\bea 
\hat{\Gamma}(\sigma)=\int \d^{4N}\psi \exp\(\im e^{-\frac{\im \pi}{q}}\sigma \prod_a \psi^a_i+\im^{q/2}\sum_{A,a}J_A\psi^a_A-v^4\sum_A \prod_a\psi_A^a\).
\eea 
Therefore we expect the half-wormhole saddle is given by
\bea 
\Gamma=\hat{\Gamma}(0)&=&\sum_{ABCD}\text{sgn}(A,B,C,D)\prod_{k=1}^p (J_{A_k}J_{B_k}J_{C_k}J_{D_k}-\delta_{A_k}^{B_k}\delta_{C_k}^{B_k}\delta_{C_k}^{D_k}v^4) \, ,
\eea 
which satisfies
\bea
&& \langle \Gamma\rangle=0\,,  \qquad \langle \Gamma^2\rangle =\langle \Gamma z^4\rangle\approx 34 \langle z^4\rangle^2 \, ,\\
&& \langle( z^4-\langle z^4\rangle-\Gamma)^2\rangle=\langle z^8\rangle-\langle z^4\rangle^2+\langle \Gamma^2\rangle-2\langle \Gamma z^4\rangle\approx 0\, .
\eea 
We find clearly that the contribution from this four-linked-wormhole saddle is not equal to the square of (two-linked) half-wormhole saddle. Even though we derive it in the 0-SYK toy model, it should exist in other SYK-like theory as long as the $G,\Sigma$ trick can be applied. We will present some more details about these more general discussions somewhere else.

\section{SYK at one time point: Poisson distribution}

Up to now we have only considered random couplings with continuous probability distributions. It is also interesting to consider random couplings that take discrete values such as the Poisson distribution.

In fact the Poisson distribution, whose PDF and moments are given by \eqref{poissonpdf} and \eqref{poissonms}, can be regarded as an opposite extremum to what we have considered above in the sense that all the cumulants are equal $\langle J^n\rangle_{c}=N\lambda$, $\forall n$. From the gravity point of view, it means that all the wormholes with different number of boundaries have the same amplitude. Ensemble theory or theories with random coupling with Poisson distribution have been studied in \cite{Marolf:2020xie,Peng:2020rno,Peng:2021vhs}. If we view the index $i$ of $\psi^i$ as the label of different time points, then the effect of ensemble average is to introduce (``non-local") interaction between different time points. In particular, starting with action \eqref{z0} we can compute the first few moments\footnote{Here we have rescaled $q \rightarrow 2q$, $N\rightarrow 2N$.}
\bal
\langle z\rangle&= \int \d^{2N} \psi \, e^{N\im^{q} \lambda \sum_A \psi^1_A},\\
\langle z^2\rangle&=\int \d^{4N} \psi\, e^{N\im^{q}\lambda\sum_A(\psi_A^1+\psi_A^2)}e^{N\im^{2q} \lambda\sum_{A}\psi_A^1\psi_A^2},\\
\langle z^3\rangle&=\int \d^{6N} \psi\, e^{N\im^{q}\lambda\sum_A(\psi_A^1+\psi_A^2+\psi_A^3)}e^{N\im^{2q} \lambda\sum_{A}(\psi_A^1\psi_A^2+ \psi_A^1\psi_A^3+\psi_A^2\psi_A^3)}e^{N\im^{3q}\lambda \sum_A \psi_A^1\psi_A^2\psi_A^3}\ .
\eal 
 For a generic $k$, we find
\bea 
\langle z^k\rangle=\int \d^{2kN}\psi e^{\lambda \sum_A \sum_{n=1}^k \frac{1}{n!}(\im^{q}\sum^k_{i=1} (\psi_A^i))^n}\ .
\eea 
Formally we can define
\bea 
\cz(\lambda)\equiv\langle z^\infty \rangle=\int \d \psi \exp \left\{N\lambda \sum_A (e^{\im^{q}\sum_{i=1} \psi_A^i}-1)\right\}\,.
\eea 
We can compute these moments by integrating out the fermions directly
\bea 
\langle z^n\rangle=\langle \text{Pf}(J_A)^n\rangle\ .
\eea  
However the ensemble average of $\text{PF}(J_A)^n$ is very complicated. Alternatively, if we only care about the large $N$ behavior we can use the $G,\Sigma$ trick and do a saddle point approximation. For example,  the $G,\Sigma$ expression of $\langle z\rangle$ is similar to \eqref{gsz}
\bea 
\langle z\rangle=\int d\Sigma dG (-\im )^{N}\Sigma^{N} e^{N \im^{{q}}\lambda\frac{  G^{{q}}}{{q}!}}e^{\im N \Sigma G}.\label{za1}
\eea 
The saddle point equations are
\bea 
&& \Sigma G=\im ,\quad \frac{\lambda}{(q-1)!}(\im G)^q =1\,,
\eea 
whose solutions are
\bea 
&&\im G=\(\frac{ (q-1)!}{\lambda}\)^{1/q}e^{\frac{2m\pi \im}{q}},\quad m=1,\dots,q\,.\label{phase}
\eea 
It has been argued in \cite{Saad:2021rcu} these $q$ saddle points should be added together to reproduce the correct large $N$ behavior in a very similar calculation. We expect the same to apply in the current situation\footnote{Here we have dropped the normalization factor $\im^N$.}
\bea 
\langle z\rangle_{\text{Disk}}= e^{-N(1-\frac{1}{q})}\(\frac{N^{q}\lambda}{(q-1)!}\)^p\sum_m e^{\frac{2m\pi \im}{q}}=q e^{-N(1-\frac{1}{q})}\(\frac{N^{q}\lambda}{(q-1)!}\)^p,
\eea 
where $p=N/q$ as before.
Adding the 1-loop factor $1/\sqrt{q}$ we end up with the correct large-$N$ behavior
\bea
\langle z\rangle_{\text{Disk}+1\text{ loop}}=\frac{1}{\sqrt{q}} e^{-N(1-\frac{1}{q})}\(\frac{N^{q}\lambda}{(q-1)!}\)^p .
\eea 
Other moments can be computed similarly. For example, to compute $\langle z^2\rangle$, we need to introduce three collective variables
\bea 
G_1=\sum_{i<j}\im \psi_i^1\psi_j^1,\quad G_2=\sum_{i<j}\im \psi_i^2\psi_j^2,\quad G_{12}=\sum_{i}\psi_i^1\psi_i^2
\eea 
such that
\bea 
\im^q\sum_A\psi_A^1=\frac{G_1^q}{q!},\quad \im^q\sum_A\psi_A^2=\frac{G_2^q}{q!}, \quad \im^{2q}\sum_A \psi_A^1\psi_A^2=\frac{G_{12}^{2q}}{(2q)!}.
\eea 
Imposing these relations with the help of a set of Lagrangian multiplier fields $\S_1$, $\S_2$ and $\S_{12}$, the $\langle z^2\rangle$ can be expressed as
\bea 
\langle z^2\rangle &=& \int [\d^3 G_i d^3 \Sigma_i] e^{N\frac{\lambda}{q!}(G_1^q+G_2^q+\frac{q!}{(2q)!}G_{12}^{2q})}e^{\im N\sum_i(\Sigma_i G_i)}\int \d^{2N}\psi e^{\frac{1}{2}{\Psi}M{\Psi}},\\
&=&\int [\d^3 G_i d^3 \Sigma_i] \sqrt{\text{det}[\Sigma_1\Sigma_2 A^2-\Sigma_{12}^2 I_{2N}]}e^{\frac{N\lambda}{q!}(G_1^q+G_2^q+\frac{q!}{(2q)!}G_{12}^{2q})}e^{\im N \sum_i(\Sigma_i G_i)} \\
&=&\int [\d^3 G_i d^3 \Sigma_i]\im^{2N} \text{det}[\sqrt{\Sigma_1\Sigma_2} A+\Sigma_{12} I_N]e^{N\frac{\lambda}{q!}(G_1^q+G_2^q+\frac{q!}{(2q)!}G_{12}^{2q})}e^{\im N \sum_i(\Sigma_i G_i)} \\
&=&\int [\d^3 G_i d^3 \Sigma_i] \frac{\im^{2N}}{2}\((\Sigma_{12}+\sqrt{\Sigma_1\Sigma_2})^{2N}+(\Sigma_{12}-\sqrt{\Sigma_1\Sigma_2})^{2N} \)e^{N \frac{N\lambda}{q!}(G_1^q+G_2^q+\frac{q!}{(2q)!}G_{12}^{2q})}e^{N \im \sum_i(\Sigma_i G_i)}\nonumber \\
&=&\int [\d^3 G_i d^3 \Sigma_i]\im^{2N} \sum_{k=1}^{N} {2N \choose 2k}\Sigma_{12}^{2N-2k}(\Sigma_1\Sigma_2)^ke^{N\frac{\lambda}{q!}(G_1^q+G_2^q+\frac{q!}{(2q)!}G_{12}^{2q})}e^{N \im \sum_i(\Sigma_i G_i)}\label{exz2}
\eea 
where we have defined
\bea 
&&\Psi=\(\psi_1^1,\dots,\psi_{2N}^1,\psi_1^2,\dots,\psi_{2N}^2\),\quad M=\begin{pmatrix}
	\Sigma_1 A&-\im \Sigma_{12} I_{2N}\\
	\im \Sigma_{12} I_{2N}& \Sigma_2 A\\
\end{pmatrix},\\
&& A=-A^T,\quad A_{ij}=1,\quad \forall i<j .
\eea 
The saddle point equations lead to
\bea 
&& \im\Sigma_i+\frac{\lambda}{(q-1)!}G_i^{q-1}=0,\quad i=1,2,\\  &&\im\Sigma_{12}+\frac{\lambda}{(2q-1)!}G_{12}^{2q-1}=0\,, \quad \sum_i \Sigma_i G_i=2\im\ .
\eea 
This set of equations have multiple solutions. For example, the wormhole saddle is
\bea 
&& G_1=G_2=\Sigma_1=\Sigma_2=0,\quad G_{12}=\(\frac{2 (2q-1)!}{\lambda}\)^{1/2q}e^{\frac{2m\pi \im}{2q}},\\
&&\langle z^2\rangle_{WH+1\text{loop}}=\frac{1}{\sqrt{2q}} e^{-2N(1-\frac{1}{2q})}\( \frac{(2N)^{2q}\lambda}{2(2q-1)!}\)^p \label{whs}
\eea
and the disconnected saddle is 
\bea 
&& G_{12}=\Sigma_{12}=0,\quad G_1=G_2=\(\frac{ (q-1)!}{\lambda}\)^{1/q},\\
&&\langle z^2\rangle_{disc+1\text{loop}}=\frac{1}{q}e^{-2N(1-\frac{1}{q})}\( \frac{N^{q}\lambda}{(q-1)!}\)^{2p}=\langle z\rangle_{\text{Disk}+1\text{loop}}^2. 
\eea 
The ratio of these two saddles is
\bea 
\frac{\langle z^2\rangle_{WH+1\text{loop}}}{\langle z^2\rangle_{disc+1\text{loop}}}=\sqrt{\frac{q}{2}} \(\frac{q!^2 2^{2q} }{e\lambda q (2q)!}\)^p\, .\label{ratio}
\eea 
In the large $N$ or $p=N/q$ limit, the wormhole saddle can dominate only when 
$\lambda < \frac{q!^2 2^{2q} }{e\lambda q (2q)!}\left(\frac{q}{2}\right)^{\frac{1}{2p}}$ which is consistent with our previous results. 

Then a natural question is that in this limit how about other n-boundary wormhole saddles? In the following let us focus on 
a particular $n$-linked-wormhole saddles. When $n=2k$ is even, the situation is similar to the one in section~\ref{j4}:
\bea 
\langle z^{2k}\rangle_{\text{connected}}&=&\int d^{4kN} \psi  \d G\frac{\d\Sigma}{2\pi}\exp\(\im N\Sigma\left(G-\sum_{i}^{2N} \prod_{a=1}^{2k}\psi_{i}^a\right)\)\exp \(N\frac{\lambda}{(2q)!}G^{2q}\)\\
&=&\int \d G\frac{\d\Sigma}{2\pi}(\im \Sigma)^{2N} \exp \(\frac{N\lambda}{{(2q)}!}G^{2q}+\im N \Sigma G\)\,,\label{z2k}
\eea 
where the collective variable $G$ is 
\bea 
G=\sum_{i}^{2N} \prod_{a=1}^{2k}\psi_{i}^a\ .
\eea 
The expression~\eqref{z2k} is of the same form as \eqref{za1} so the saddle point approximation is 
\bea 
\langle z^{2k}\rangle_{2k-WH+1\text{loop}}=\langle z^2\rangle_{2-WH+1\text{loop}}=\frac{1}{\sqrt{2q}} e^{-2N(1-\frac{1}{2q})}\( \frac{(2N)^{2q}\lambda}{2(2q-1)!}\)^p.
\eea 
When $n=2k+1$ is odd, the situation is similar to the one of $n=1$:
\bea
\langle z^{2k+1}\rangle_{\text{connected}}&=&\int d^{(4k+2)N} \psi  \d G\frac{\d\Sigma}{2\pi}\exp\(\im N \Sigma(G-\sum_{i<j}^{2N} \prod_{a=1}^{2k+1}\psi_{i}^a\prod_{a=1}^{2k+1}\psi_{j}^a\)\exp \(\frac{N\lambda}{q!}G^q\)\nn\\
&=&\int \d G\frac{\d\Sigma}{2\pi}(\im \Sigma)^{2N} \exp \(\frac{N\lambda}{{q}!}G^{q}+\im N\Sigma G\),\label{z2k1}
\eea 
where the collective variable $G$ is obviously defined as
\bea 
G=\sum_{i<j}^{2N} \prod_{a=1}^{2k+1}\psi_{i}^a\prod_{a=1}^{2k+1}\psi_{j}^a,
\eea 
therefore the saddle point approximation is 
\bea 
\langle z^{2k+1}\rangle_{2k+1-HW+1\text{loop}}=\langle z\rangle_{\text{Disk}+1\text{loop}}=\frac{1}{\sqrt{q}} e^{-N(1-\frac{1}{q})}\(\frac{N^{q}\lambda}{(q-1)!}\)^p\ .
\eea 
These  higher $n$-linked-wormholes should be compared with the corresponding powers of the disk solution, and furthermore since $\langle z^2\rangle_{2-WH+1\text{loop}}\gg 1$, we conclude that  all these multiple-linked-wormholes are suppressed. In other words, the ensemble of $z$ can be  approximated by a Gaussian when the ratio~\eqref{ratio} is of order 1.

\section{The Brownian SYK model}
\label{SYK1}
In this section, we study the wormhole and half-wormholes saddles in the Brownian SYK model~\cite{Saad:2018bqo}. In the Brownian SYK model, the couplings are only correlated at the same instant of time so that after integrating over the coupling we end up with a local effective action\footnote{See Appendix \eqref{am} for general discussion on averaged model.}. The quantity that is analogous to the partition function but with some information of real time evolution is
\bal
U(T)=\mathbf{T} e^{-\im\int_{0}^{T} dt H(t)}\ . 
\eal 
To check the nature of its fluctuations that is not caused by the phase factor, we consider the norm square of its trace
\bal
\left|\Tr\, U(T)\right|^2\ .\label{tu2}
\eal
This quantity is manifest real in the sense the complex conjugate maps $\Tr\, U(T)$ to $\Tr\, U(T)^*$.
The trace is over the Hilbert space, which has a path integral interpretation
\bal
\Tr\, U(T)&=\int \mathcal{D} \psi_{a} \exp \left\{-\im \int_{0}^{T} d t\left[-\frac{\im}{2} \psi_{a} \partial_{t} \psi_{a}+J_{a_{1} \ldots a_{q}}(t)\im^{\frac{q}{2}} \psi_{a_{1} \ldots a_{q}}\right]\right\}\,,\label{tu}
\eal
where the Lagrangian density is manifestly real.

To compute~\eqref{tu2}, we introduce two replicas of fermions; $\psi^{(L)}$  constitute the fermions in $H$ of $U$ and   $\psi^{(R)}$  in $U^*$. Therefore the complex conjugate should map between $\psi^{(L)}$ and $\psi^{(R)}$. 
One conventional way to define $\psi^{(R)}$ from $\psi^{(L)}$ is
\bal
\psi^{(R)}_a=\left(\psi^{(L)}_a\right)^*\ .\label{cc}
\eal
Then the complex conjugation of~\eqref{tu} is
\bal
\Tr\, U(T)^*&=\int \mathcal{D} \psi_{a}^{(R)} \exp \left\{-\im \int_{0}^{T} d t\left[\frac{\im}{2} \psi^{(R)}_{a} \partial_{t} \psi^{(R)}_{a}-J_{a_{1} \ldots a_{q}}(t)\im^{\frac{q}{2}} \psi^{(R)}_{a_{1} \ldots a_{q}}\right]\right\}\,,\label{tuc1p}
\eal  
We can further do a field redefinition $\psi^{(R)} \to \im\psi^{(R)}$ so that the kinetic term has the ``right" sign\footnote{Here we choose to absorb an extra $i^N$ phase factor into the definition of the path integral measure. There might be $N \bmod 4$ effects that we will discuss separately.}
\bal
\Tr\, U(T)^*&=\int \mathcal{D} \psi_{a}^{(R)} e^{-\im \int_{0}^{T} d t\left[-\frac{\im}{2} \psi^{(R)}_{a} \partial_{t} \psi^{(R)}_{a}-J_{a_{1} \ldots a_{q}}(t)(-\im)^{\frac{q}{2}} \psi^{(R)}_{a_{1} \ldots a_{q}}\right]}\,,\label{tuc1}
\eal  
Combining~\eqref{tu}, with $\psi_a$ replaced by $\psi_a^{(L)}$, and~\eqref{tuc1}, the quantity we would like to compute is 
\bal
|\operatorname{Tr} U(T)|^{2}=\int \mathcal{D} \psi_{a}^{(L)} \mathcal{D} \psi_{a}^{(R)} e^{\im \int_{0}^{T} d t\left[\frac{\im}{2} \psi_{a}^{(j)} \partial_{t} \psi_{a}^{(j)}-J_{a_{1} \ldots a_{q}}(t)\left(\im^{\frac{q}{2}} \psi_{a_{1} \ldots a_{q}}^{(L)}-(-\im)^{\frac{q}{2}}\psi_{a_{1} \ldots a_{q}}^{(R)}\right)\right]}\ .\label{z1}
\eal

A side remark is that the complex conjugation is closely related to time reversal symmetry $\mathcal{T}$, and also because $[\mathcal{T},H]=0$,  we expect~$\Tr\, U(T)^*=\Tr\, U(-T)$. Indeed, we find 
\bal
&\Tr\, U(-T)=\int \mathcal{D} \psi_{a}^{(R)} \exp \left\{-\im \int_{0}^{-T} d t\left[-\frac{\im}{2} \psi^{(R)}_{a} \partial_{t} \psi^{(R)}_{a}+J_{a_{1} \ldots a_{q}}(t)\im^{\frac{q}{2}} \psi^{(R)}_{a_{1} \ldots a_{q}}\right]\right\}\\
&=\int \mathcal{D} \psi_{a}^{(R)} \exp \left\{\im \int_{-T}^0 d t\left[-\frac{\im}{2} \psi^{(R)}_{a} \partial_{t} \psi^{(R)}_{a}+J_{a_{1} \ldots a_{q}}(t)\im^{\frac{q}{2}} \psi^{(R)}_{a_{1} \ldots a_{q}}\right]\right\}\\
&=\int \mathcal{D} \psi_{a}^{(R)} \exp \left\{\im \int_{0}^T d t\left[-\frac{\im}{2} \psi^{(R)}_{a} \partial_{t} \psi^{(R)}_{a}+J_{a_{1} \ldots a_{q}}(t)\im^{\frac{q}{2}} \psi^{(R)}_{a_{1} \ldots a_{q}}\right]\right\}=\Tr\, U(T)^*\,,\label{tut}
\eal
where we simply use $\psi^{(R)}$ to represent a different set of fermions that will be integrated over in the path integral; in particular, we do not think of them as the complex conjugate of the $\psi^{(L)}$. In the last line we assume the system to be  invariant under time translation, and the last equality is clear from~\eqref{tuc1p}. Therefore the quantity we are interested can also be written as $\Tr\, U(T)\Tr\, U(-T)$. \par 

Note that the random couplings satisfy
\bea \label{SYK1j}
\langle J_A\rangle=0,\quad \langle J_A(t)J_B(t')\rangle=\delta(t-t')\delta_{AB}\cj^2,\quad \cj^2=2J\frac{(q-1)!}{N^{q-1}},
\eea 
and the our normalization of one-dimensional Majorana fermions  is
\bea 
\{\psi_i,\psi_j\}=\delta_{ij}\ .
\eea 
To simplify our notation, we simply denote $\left|\Tr\, U(T)\right|^2$ by $\<|z|^2\>$ in the rest computation.

\subsection{$\<|z|^2\>$ in the Brownian SYK model: accurate evaluation}
As argued in \cite{Saad:2018bqo}, we focus on the time independent configurations. Therefore we can directly integrate out the fermions and averaging over the random coupling according to~\eqref{SYK1j}. In the large $N$ limit and for even $q$ this leads to
\bal
\<|z|^2\>&=\int_{\mathbb{R}} \mathcal{D}G_{LR}\int_{i \mathbb{R}}\frac{\mathcal{D}\S_{LR}}{2\p i/(TN)} e^{N\left[\log\left(2\cos \frac{ T\S_{LR}}{2}\right)-\frac{2JT}{q2^q}+i^q \frac{2JT}{q}G^q_{LR}-2\frac{T}{2}\S_{LR}G_{LR}\right]}\ .
\eal
The integration measure is normalized such that if we first to the $G$ integral then the $\S$ integral, we get the result of free fermions $\langle |z|^2\rangle |_{J=0}=2^N$. Notice that the $G_{ij}$ function defined above is real under the complex conjugation~\eqref{cc}.
Making use of the identity
\bea 
\int_{\im\mathbb{R}} \frac{d\Sigma}{2\pi i/(  TN)}e^{- NT \Sigma (G-x)}=\delta(G-x)
\eea 
we get
\bal 
\langle |z|^2\rangle=  e^{-\frac{J N T}{q2^{q-1}}}\sum_{k=0}^N {N\choose k}e^{\frac{JTN}{q2^{q-1}}(\frac{2k}{N}-1)^q}\equiv c_e\,. \label{ce}
\eal 
In the large-$N$ ( and also large-$NT$ ) limit, the dominant contribution are determined from two factors: the combinatoric factor and the exponential. At early time $T\ll 0$, contributions from the different exponential factors are roughly the same, so the dominant term is determined from the largest term in the combinatoric factor 
\bal
{N \choose N/2} = \frac{N!}{\left((N/2)!\right)^2}\sim 2^N\sqrt{\frac{2}{\p N}} \,,
\eal  
which leads to the contribution
\bal
c_s=\sqrt{\frac{2}{\p N}} 2^N e^{-\frac{JTN}{q 2^{q-1}} }\ . \label{smalltdom}
\eal
At late time, the different exponential factors dominant over the combinatoric factors, so the dominant contribution is from the 
maximal exponential factor, which is at $k=0,N$ with  contributions to the sum being
\bal
c_l=2\ .\label{largetdom}
\eal

\begin{figure}
	\centering
	\includegraphics[width=0.7\linewidth]{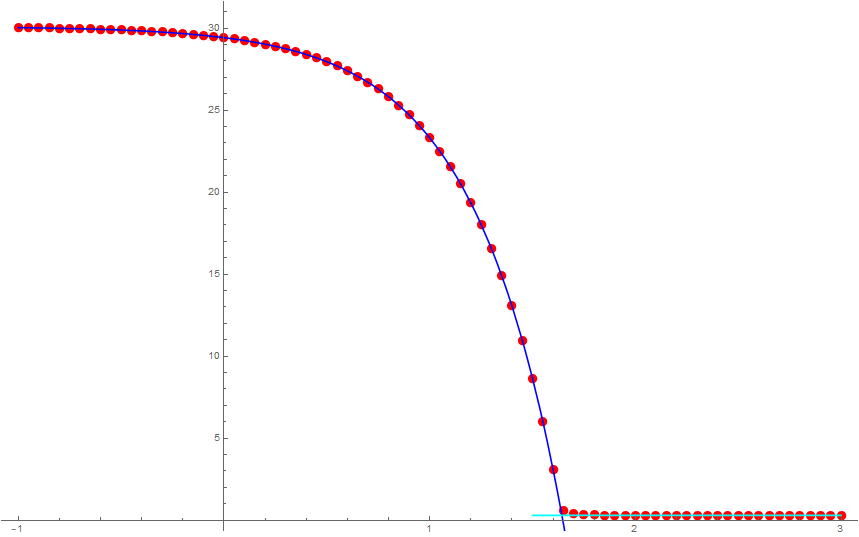}
	\caption{The behavior of $\langle |z|^2\rangle$, where the horizontal axis is $\log_{10}T$ and the vertical axis is $\log_{10}\<|z|^2\>$. The red dots are computed using~\eqref{ce} with  $q=4$, $J=1$ and a cutoff at $n=100$. The blue curve is the contribution from the trivial saddle~\eqref{smalltsaddle}, and the cyan curve is the contribution from the wormhole saddle~\eqref{longtsaddle}.}
	\label{fig:brownian1}
\end{figure}

The behavior of $\langle |z|^2\rangle$ is shown in Figure.~\ref{fig:brownian1}  where the early time exponential decay and the late time constant behavior is manifest.

\subsection{$\<|z|^2\>$ in the Brownian SYK model: large-$N$ saddle point evaluation }
In the following, we perform a saddle point analysis to reproduce these distinct behaviors. 
We deform the integration contour together with  a change of variable
\bal
 G = \im e^{\frac{\im \pi}{q}} g,\quad \S=   e^{-\frac{\im \pi}{q}} \sigma \ .
\eal
The action then reduces to
\bal
\<|z|^2\>&=\int \mathcal{D}g\int\frac{\mathcal{D}\s_{LR}}{2\p i/(TN)} e^{N\left[\log\left(2\cos \frac{T e^{-i \frac{\p}{q}} \s}{2}\right)-\frac{2JT}{q2^q}- \frac{2JT}{q} g^q-T i \s g\right] }\ .
\eal

The equation of motion of the $\s$ field leads to 
\bal
g=\frac{\im}{2} e^{-\frac{\im \pi}{q}}\tan(\frac{T\sigma}{2}  e^{-\frac{\im \pi}{q}})\,,
\eal
while the equation of motion of $g$ is
\bal
\im g\s+2 J g^q=0.
\eal
The two equations indicate a condition that $g$ should satisfy
\bal
g+\frac{1}{2}e^{-\frac{\im \pi}{q}}\tanh({JT} g^{q-1} e^{-\frac{\im \pi}{q}})=0\ .\label{saddle1}
\eal
Solutions to this equation are in general irrational. In the following we solve it with different approximations.

\subsubsection{Saddle point solution: the $q=1$ case  }

Formally, we can consider the $q=1$ case where the saddle point solution can be found explicitly. In particular, when $q=1$, the saddle point equation~\eqref{saddle1} reduces to
\bal
g=-\frac{1}{2}\tanh\({TJ}\)\,,\qquad \s =\im J\ .
\eal
The on-shell action (with the 1-loop correction) is then
\bal
\<|z|^2\>&=e^{{-\frac{2JTN}{2}}} \left(e^{-{J T}}+e^{{J T}}\right)^N\,.
\eal
On the other hand, when $q=1$, the summation expression~\eqref{ce} can be evaluated explicitly to
\bal
\<|z|^2\>=e^{{-{2JTN}{}}} \left(e^{-{J T}}+e^{{J T}}\right)^N\ .
\eal
The exact result agrees with the above saddle point result.

\subsubsection{Saddle point solution: $q>1$ at short time $T \s \ll 1$ }
In the saddle point approach, the effective action in the short time limit  can be expanded into
\bal
S_{\text{2}}=N\left(2T \left(-\frac{J  g^q}{q}-\frac{i g \s}{2}\right)-\frac{1}{8} e^{-\frac{2 i \pi }{q}} \s^2 T^2-\frac{e^{-\frac{4 i \pi }{q}} \s^4 T^4}{192}+\mathcal{O}(T^5)\right)\,,\label{action2}
\eal
where
\bal
S_{\text{2}}=S-N\log(2)+NT\frac{2J }{q 2^{q}}\,,
\eal
is the part of the action that depends on the dynamical fields, in other words, the constant piece in  $S$ has been factored out to define $S_2$.  
Notice that although we have $T\ll 1$ in this limit, we still want the saddle point approximation to be good, this means we want $NT \gg 1$.

Before going to the details, we first discuss the region where this is a valid perturbative analysis. In the above expansion, the only $g$ dependence is in the $\left(-\frac{\im \s g}{2}-\frac{J}{q}g^q\right)$ term, which means the set of saddle point equations always contain the following equation
\bal
2\im J g^{q-1}=\s \ .
\eal 
This relation means the saddle point contribution to the on shell action has the general form
\bal
S_{\text{2}}=N\left(\frac{2\left(-\frac{i}{2}\right)^{\frac{q}{q-1}} (q-1) T J^{\frac{1}{1-q}} \tilde{\s}^{\frac{q}{q-1}}}{q} -\frac{1}{8} e^{-\frac{2 i \pi }{q}} \s^2 T^2-\frac{e^{-\frac{4 i \pi }{q }} \s^4 T^4}{192}+\mathcal{O}(T^3)\right)\ .\label{action2a}
\eal  
Next, we would like to make sure our expansion of the $\log$ term is valid, this requires
\bal
T\s<1\ .
\eal

The remaining terms could switch dominance depending on the value of $\s$ in the saddle point solution. 
\bal
\text{short time: } &T \s^{\frac{q}{q-1}} > T^2\s^2 > T^4\s^4\,, \quad \Leftrightarrow \quad  0<\s < T^{\frac{q-1}{2-q}}\,,\label{tlarge1}\\
\text{intermediate time: } &T^2\s^2 >T \s^{\frac{q}{q-1}} >  T^4\s^4\,, \quad \Leftrightarrow \quad T^{\frac{q-1}{2-q}}<\s < T^{\frac{3 (-1 + q)}{4 - 3 q}} \ .\label{tlarge2}
\eal
In all these cases the 
saddle point equation~\eqref{saddle1} reduces to the approximate form
\bal 
g=\im e^{-\frac{2\im \pi}{q}}\s T\,,\quad \s=\im g^{q-1}J \, .\label{saddlea}
\eal

$\bullet$ {\bf Short time}
 
In this case, the dominant term in the action is 
\bal
S_{\text{2}}=2NT \left(-\frac{J g^q}{q}-\frac{ g \s}{2}\right)\approx 2N\left(-\frac{ \left((-1)^{\frac{1}{q-1}} (q-1)\right) T J^{\frac{1}{1-q}} \s^{\frac{q}{q-1}}}{q2^{\frac{q}{q-1}}}\right)\ .
\eal 
The $\s T$ term in the saddle point equation~\eqref{saddlea} can be dropped and the only solution is 
\bal
g=0\,,\qquad \s=0\,,
\eal
 
\begin{figure}
	\centering
	\includegraphics[width=0.5\linewidth]{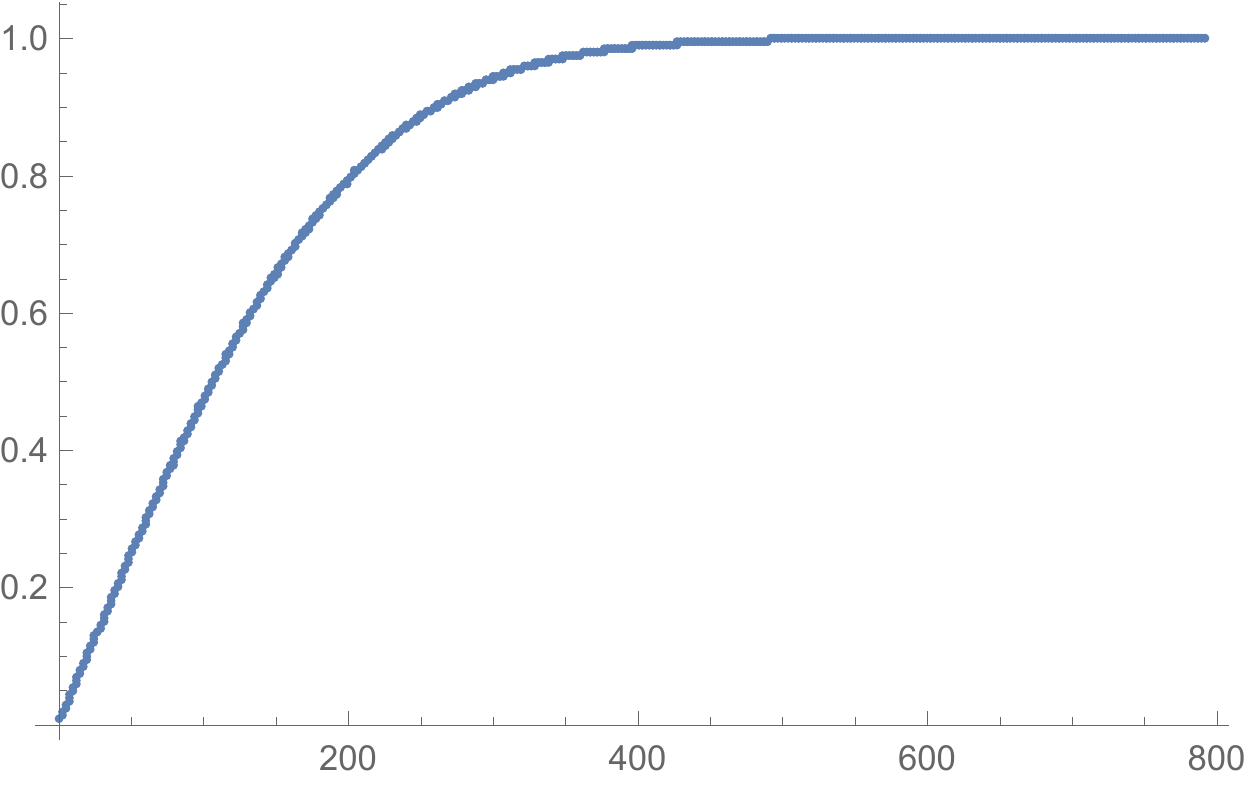}
	\caption{$N=10^5$, $q=4$, $J=1$, $T=0.1$. The vertical axis is $c_e/\left(\sqrt{\frac{\p N}{2}}c_s\right)$ with a cutoff $2\sqrt{\p N}$ in the summation~\eqref{ce}. The horizontal axis is the number of terms kept in the summation.}
	\label{fig:sum1}
\end{figure}

This gives the following saddle contribution to the on-shell action
\bal
 {N T}\exp\left\{N\left[\log\left(2\right)-\frac{2JT}{q 2^q }\right]\right\}=2^{N} e^{-\frac{JTN}{q2^{q-1}}} N T\ .
\eal
Next we need to consider the quadratic fluctuations around this saddle
\bal
g\to g+  \delta g 
\,,\qquad \s \to \s +  
\delta \s \ .
\eal 
This gives the 1-loop factor
\bal
\frac{1}{N T}\,,
\eal
Therefore this saddle point approximation gives 
\bal
\<|z|^2\>_{s+1\text{loop}}&=\exp\left\{N\left[\log\left(2\right)-\frac{JT}{q 2^q }\right]\right\}=2^{N} e^{-\frac{JTN}{q 2^{q-1} }}\ .\label{smalltsaddle}
\eal
We next want to compare this saddle point approximation with the exact result~\eqref{ce}. It is clear that in the small $T$ region the dominant saddle should be~\eqref{smalltdom}. However, it is also clear that the $k\sim \frac{N}{2}$ terms in the sum should also give comparable contributions. Indeed, if we compare the result~\eqref{smalltdom} with~\eqref{smalltsaddle}, we find 
\bal
\<|z|^2\>_{s+1\text{loop}} = \sqrt{\frac{\p N}{2}} c_s\ .
\eal   
On the other hand, as we can check numerically, 
\bal
c_e = \sqrt{\frac{\p N}{2}} c_s\ ,\quad \text{when} \quad J\sim1,\quad T \ll 1\ .\label{ma}
\eal
An example of this numerical check is shown in Figure.~\ref{fig:sum1}. Or we can understand this approximation as the following. When $J$ is of order 1  the exponent $e^{T N(1-\frac{2k}{N})^q}\leq e^{TN}$. So if $T\sim \frac{1}{N}$ this exponent is always of order $1$ so that \eqref{ce} can be  approximated by $e^{-\frac{JNT}{q 2^{q-1}}}\sum_k {N\choose k}=e^{-\frac{JNT}{q 2^{q-1}}} 2^N$.  Actually $T\sim 1/N$ is only a sufficient condition for  \eqref{ma} to hold; as can be observed from the numerical data $T$ can be much larger than $1/N$.

Therefore, indeed we find our saddle point approximation agrees very well with the exact result.

$\bullet$ {\bf Intermediate time }

In this region both the $\s^2$ and $\s^{\frac{q}{q-1}}$ terms in the action are roughly of the same order, so we need to solve the approximated saddle point equation~\eqref{saddlea}. There are two solutions
\bal
\s=0\,,\qquad g=0\,,\label{saddle0}
\eal
and
\bal
g=\left(-e^{2 \pi  \im m-\frac{2 \im \pi }{q}}\frac{J T}{2}\right)^{\frac{1}{2-q}}\,,\quad \s= 2\im J g^{q-1} \ .\label{saddleqm2}
\eal
and both saddles should in principle be taken into account. Notice that the expression seems to blow up at $T\to 0$, but we have fixed the range to be $\sim T_0$ so the expression remains finite.
The saddle point contribution from the non-trivial solution~\eqref{saddleqm2} is proportional to
\bal
( N T)2^{N}e^{-\frac{2JNT}{q 2^q}}\exp\(\(\frac{1}{2}-\frac{1}{q}\)g^q\)\,,
\eal
The loop correction around each of the saddle is
\bal
\frac{1}{ N T \sqrt{q-2}}\ .
\eal
The full contribution is 
\bal\label{saddleqm3}
2^Ne^{-\frac{2JNT}{q 2^q}}\frac{1}{\sqrt{q-2}}\exp\(N\frac{2-q}{2q}e^{\frac{2 qm\im\pi}{2-q}}\(\frac{JT}{2}\)^{\frac{2}{2-q}}\)
\eal
However, it is easy to check numerically that the contributions to the on-shell action from these saddles~\eqref{saddleqm3} never dominate when $q>2$. Therefore the trivial saddle point  always has larger contribution and dominate the path integral in this range of time.  

\subsubsection{Long time $T \s \gg 1$ }

At long time, we can replace the $\cosh$ function by an exponential function. There are two choices, which leads to two different solutions 
\bal
\log\left(2\cos (\frac{T\sigma}{2}  e^{-\frac{\im \pi}{q}}) \right)& \sim \frac{\im }{2}Te^{-\frac{\im \pi}{q}}\sigma\,,\qquad  \Re(\im e^{-\im \frac{\p}{q}} \s)>0\label{case1}\\
\log\left(2\cos (\frac{T\sigma}{2}  e^{-\frac{\im \pi}{q}}) \right)& \sim -\frac{\im}{2} Te^{-\frac{\im \pi}{q}}\sigma\,,\qquad   \Re(\im e^{-\im \frac{\p}{q}} \s)<0\ . \label{case2}
\eal  
The solution of the saddle point equation in this case is time independent, 
\bea\label{nont}
g=\pm \frac{1}{2}e^{-\frac{\im \pi}{q}}\,,\sigma=2 \im J g^{q-1}\ .
\eea
For even $q$ the on-shell actions of these saddles, including the 1-loop corrections, are also time independent  
\bal 
\langle |z|^2\rangle_{\text{WH}+1\text{1oop}}=2\times 1=2\,,\label{longtsaddle}
\eal 
where the factor of  2 comes from adding up the contributions from the two saddles~\eqref{nont} and the result reproduces~\eqref{largetdom}. In addition, the contribution from the trivial saddle  vanishes at late time, so the non-trivial saddles~\eqref{case1} and~\eqref{case2} dominate.
Since $g\neq0$, these saddle points are identified with wormhole saddles.

\subsection{$\langle |z|^4\rangle$ in the Brownian SYK model}
One of our goal in this section is to find possible half-wormhole saddles and study their relation to the wormhole saddle.
To achieve this, it is helpful to first consider $ |z|^4\equiv z_1z_2z_3 z_4 $:
\bea 
\int  \mathcal{D}^{4N} \psi e^{\im \int_{0}^{T} d t\left[\frac{\im}{2} \psi_{a}^{(i)} \partial_{t} \psi_{a}^{(i)}-J_{a_{1} \ldots a_{q}}(t)\left(\im^{\frac{q}{2}} \psi_{a_{1} \ldots a_{q}}^{(1)}-(-\im)^{\frac{q}{2}}\psi_{a_{1} \ldots a_{q}}^{(2)}+\im^{\frac{q}{2}} \psi_{a_{1} \ldots a_{q}}^{(3)}-(-\im)^{\frac{q}{2}}\psi_{a_{1} \ldots a_{q}}^{(4)}\right)\right]}\, .
\eea  
We first compute the ensemble averaged version $\langle z_1z_2z_3 z_4\rangle$:
\bea 
\langle |z|^4\rangle &=&\int \cd^{4N}\psi \, e^{ \int_{0}^{T} d t\left[-\frac{1}{2} \psi_{a}^{(i)} \partial_{t} \psi_{a}^{(i)}-\frac{\cj^2}{2} \left(\im^{\frac{q}{2}} \psi_{a_{1} \ldots a_{q}}^{(1)}-(-\im)^{\frac{q}{2}}\psi_{a_{1} \ldots a_{q}}^{(2)}+\im^{\frac{q}{2}} \psi_{a_{1} \ldots a_{q}}^{(3)}-(-\im)^{\frac{q}{2}}\psi_{a_{1} \ldots a_{q}}^{(4)}\right)^2\right]}\nonumber \\
&=&\int \cd^{4N} \psi [\prod_{a<b} \cd G_{ab} \frac{\cd \Sigma_{ab}}{2\pi\im /N }]\exp\left\{ \int_{0}^T \d t\left[-\frac{1}{2}\psi_{a}^{(i)} \partial_{t} \psi_{a}^{(i)}-\frac{4 JNT}{q2^q} +\right. \right.\nonumber \\
&&\qquad \left. \left. 2\sum_{a<b}\(  \frac{JN}{q}s_{ab}G_{ab}^q - N\frac{\Sigma_{ab}}{2} G_{ab}+\sum_i\frac{\Sigma_{ab}}{2}\psi_i^a\psi_i^b  \)\right]\right\} \label{z44}
\eea 
where $s_{12}=s_{14}=s_{23}=s_{34}=\im^q,\quad s_{13}=s_{24}=-1$ and the other orders of $(a,b)$ have been absorbed into the factor of 2. We again focus on the time-independent saddle points, and the integration over fermions gives 
\bal 
&\int \d^{4N}\psi \exp\( \sum_{a<b}\Sigma_{ab}\psi_i^a\psi_i^b\)\\
&\quad =2^N\left[ \cos\(\frac{1}{2} \sqrt{(\Sigma_{14}-\Sigma_{23})^2+(\Sigma_{13}+\Sigma_{24})^2+(\Sigma_{12}-\Sigma_{34})^2}\) \right. \nonumber \\
&\qquad \left. +\cos\(\frac{1}{2} \sqrt{(\Sigma_{14}+\Sigma_{23})^2+(\Sigma_{13}-\Sigma_{24})^2+(\Sigma_{12}+\Sigma_{34})^2}\) \right]^{N}\,, \label{i4} 
\eal 
thus
\bal 
\langle |z|^4\rangle &= \int [\prod_{a<b} \cd G_{ab} \frac{\cd \S_{ab}}{2\pi\im/(NT) }]e^{-\frac{4 JNT}{q2^q}}\exp\( S_{\text{eff}}\)\,,\\\label{z4avg}
S_{\text{eff}}&=2NT\sum_{a<b}\(\frac{J}{q}s_{ab}G_{ab}^q -\frac{ \S_{ab}}{2} G_{ab}\)+N \log (e^{\im Tf_+}+e^{-\im T f_+}+e^{\im Tf_-}+e^{-\im T f_-})\,,
\eal 
with
\bea 
f_\pm=\frac{1}{2}\sqrt{(\S_{14}\pm\S_{23})^2+(\S_{13}\mp\S_{24})^2+(\S_{12}\pm\S_{34})^2}\ .\label{fpm}
\eea

\subsubsection{Exact evaluation}\label{z4e}
Similar to the exact calculation of $\langle |z|^2\rangle$ we can integrate $\Sigma_{AB}$ first to obtain
\bea \label{z4e}
\langle |z|^4\rangle &=&e^{-\frac{4J NT}{q 2^q}}\(e^{\frac{\im \hat{f}_+}{N} }+e^{-\frac{\im \hat{f}_+}{N}}+ e^{\frac{\im \hat{f}_-}{N} }+e^{-\frac{\im \hat{f}_-}{N}}\)^N e^{\frac{2NTJ}{q}\sum_{a<b}s_{ab}G_{ab}^q}\, \Big|_{G_{ab}=0}\,\\
&=&e^{-\frac{4J NT}{q 2^q}}\sum_{0\leq n_i\leq N, \sum_i n_i=N}e^{ \frac{\im(n_1-n_2)}{N} \hat{f}_+}e^{\frac{\im (n_3-n_4)}{N} \hat{f}_-}e^{\frac{2NTJ}{q}\sum_{a<b}s_{ab}G_{ab}^q}\, \Big|_{G_{ab}=0}\, ,
\eea 
where we have introduced the differential operators
\bea 
\hat{f}_\pm=\frac{1}{2}\sqrt{(\partial_{G_{14}}\pm \partial_{G_{23}})^2+(\partial_{G_{13}}\mp\partial_{G_{24}})^2+(\partial_{G_{12}}\pm\partial_{G_{34}})^2}.
\eea 
Expanding the exponentials into Taylor series and keeping only the non-vanishing terms we get
\bal
&e^{\frac{\im (n_1-n_2)}{N} \hat{f}_+}e^{\frac{\im (n_3-n_4)}{N} \hat{f}_-}e^{\frac{2NTJ}{q}\sum_{a<b}s_{ab}G_{ab}^q}\, \Big|_{G_{ab}=0},\nonumber \\
&=\sum_{n=0}^\infty\sum_{m=0}^\infty\(\frac{\im(n_1-n_2)}{N}\)^{mq}\(\frac{\im(n_3-n_4)}{N}\)^{nq}\frac{\hat{f}_+^{mq}\hat{f}_-^{nq}}{(mq)!(nq)!}e^{\frac{2NTJ}{q}\sum_{a<b}s_{ab}G_{ab}^q}\, \Big|_{G_{ab}=0}\, ,
\eal 
with
\bal 
\hat{f}_+^{mq}\hat{f}_-^{nq}&=\frac{1}{2^{(n+m)q}}\sum_{\substack{0\leq k^+_i\leq m,\sum k_i^+=m\\0\leq k^-_i\leq n,\sum k^-_i=n}}\left( (\partial_{G_{14}}+\partial_{G_{23}})^{k_1^+q}(\partial_{G_{14}}-\partial_{G_{23}})^{k_1^-q} \right.\nonumber\\
&\times\left.(\partial_{G_{13}}-\partial_{G_{24}})^{k_2^+q}(\partial_{G_{13}}+\partial_{G_{24}})^{k_2^-q}(\partial_{G_{12}}+\partial_{G_{34}})^{k_3^+q}(\partial_{G_{12}}-\partial_{G_{34}})^{k_3^-q} \right)\, \label{diffo}. 
\eal 
For each pair of differential operators in \eqref{diffo} the contribution can be obtained for example as
\bea 
&&(\partial_{G_{14}}+\partial_{G_{23}})^{k_1^+q}(\partial_{G_{14}}-\partial_{G_{23}})^{k_1^-q}e^{-\frac{2NTJ}{q}(G_{14}^q+G_{23}^q)}\Big|_{G_{ab}=0} \nonumber \\
&&=\sum_{l^+_1=0,l^-_1=0}^{k_1^+,k_1^-}{k_1^+ q \choose l_1^+ q}{k_1^- q \choose l_1^- q}\partial_{G_{14}}^{(l_1^++l_1^-)q}\partial_{G_{23}}^{(k_1^++k_1^--l_1^+-l_1^-)q}e^{\im^q\frac{2NTJ}{q}(G_{14}^q+G_{23}^q)}\Big|_{G_{ab}=0}\\
&&=\(\im^q \frac{2NJT}{q}\)^{(k_1^++k_1^-)q}\sum_{l^+_1=0,l^-_1=0}^{k_1^+,k_1^-}{k_1^+ q \choose l_1^+ q}{k_1^- q \choose l_1^- q}\frac{[(l_1^++l_1^-)q]![(k_1^++k_1^--l_1^+-l_1^-)q]!}{(l_1^++l_1^-)!(k_1^++k_1^--l_1^+-l_1^-)!} \nonumber\\
&&\equiv \(\im^q\frac{2NJT}{q}\)^{(k_1^++k_1^-)q}\Delta(k_1^+,k_1^-).
\eea 
Thus the full expression of \eqref{z4e} is
\bea 
&&\langle |z|^4\rangle=e^{-\frac{4J NT}{q 2^q}}\sum_{\substack{0\leq n_i\leq N, \\\sum_i n_i=N}}\sum_{n=0}^\infty\sum_{m=0}^\infty\sum_{\substack{0\leq k^+_i\leq m,\sum k_i^+=m\\0\leq k^-_i\leq n,\sum k^-_i=n}}\sum_{l^+_i=0,l^-_i=0}^{k_i^+,k_i^-}\frac{(\frac{\im n_{12}}{N})^{mq}}{(mq)!2^{mq}}\frac{(\frac{\im n_{34}}{N})^{nq}}{(nq)!2^{nq}} \nonumber \\
&&\quad \(\im^q\frac{2NJT}{q}\)^{(k_1^++k_1^-+k_3^++k_3^-)q}\(-\frac{2NJT}{q}\)^{(k_2^++k_2^-)q}\prod_{i=1}^3\Delta(k_i^+,k_i^-)\,,
\eea 
where $n_{12}=n_1-n_2$ and $n_{34}=n_3-n_4$. This is very complicated expression but at large T and large $N$, the leading contributions come from the cases when $n_i=N$ and $ n_j=0, j\neq i$. For each of these cases, we show in Appendix \eqref{if1} that it contributes 2 when $q=4m$ and 3 when $q=4m+2$. So in total, $\langle z_4\rangle_{T\rightarrow \infty}$ approaches to 8 when $q=4m$ and 12 when $q=4m+2$.

Next we turn to the saddle point analysis and try to match the results. 

\subsubsection{Saddle point analysis}
To make the integral~\eqref{z4avg} convergent, we do the following change of variables and deform the original integral contour so that the integral over $g_{ab}$ and $\s_{ab}$ are along the real lines
\bea 
&&G_{12}=\im e^{\frac{\im \pi}{q}}g_{12},\quad G_{23}=\im e^{\frac{\im \pi}{q}}g_{23},\quad G_{14}=\im e^{\frac{\im \pi}{q}}g_{14},\quad G_{34}=\im e^{\frac{\im \pi}{q}}g_{34}, \\
&&\S_{12}=e^{-\frac{\im \pi}{q}}\s_{12},\quad \S_{23}= e^{-\frac{\im \pi}{q}}\s_{23},\quad \S_{14}=e^{-\frac{\im \pi}{q}}\s_{14},\quad \S_{34}=e^{-\frac{\im \pi}{q}}\s_{34}, \\
&&G_{13}=g_{13},\quad G_{24}=g_{24},\quad \S_{13}=\im \s_{13},\quad \S_{24}=\im \s_{24}\ .
\eea 
The effective action then becomes
\bea 
S=2NT\sum_{ab}\(-\frac{J}{q}g_{ab}^q-\im \frac{\s_{ab}}{2}g_{ab}\)+N \log (e^{\im Tf_+}+e^{-\im T f_+}+e^{\im Tf_-}+e^{-\im T f_-})\ .
\eea 
Let us again focus on the large $T$ limit. In this limit, we expect only one of the four exponentials $e^{\pm \im T f_\pm}$ dominates the integral.
To be explicit, let us assume the dominant one to be
\bal 
e^{ \im a \frac{T}{2}  \sqrt{(\Sigma_{14}+b\Sigma_{23})^2+(\Sigma_{13}-b\Sigma_{24})^2+(\Sigma_{12}+b\Sigma_{34})^2}}=e^{ \im a \frac{T}{2}  \sqrt{e^{-\frac{2\im \pi}{q}}(\sigma_{14}+b\sigma_{23})^2-(\sigma_{13}-b\sigma_{24})^2+e^{-\frac{2\im \pi}{q}}(\sigma_{12}+b\sigma_{34})^2}}\,,
\eal 
where $a,b$ can be $\pm 1$.
The saddle point equations leads to 
\bal 
g_{34}=bg_{12},\quad g_{24}=-bg_{13},\quad g_{23}=b g_{14},\quad \sigma_{ab}=2 \im  Jg^{q-1}_{ab},\quad g_{12}^2e^{\frac{2\im \pi}{q}}+g_{14}^2e^{\frac{2\im \pi}{q}}-g_{13}^2=1\ .\label{moduli}
\eal 
There is always a trivial saddle solution
\bea \label{trivialB}
 g_{ab}=\sigma_{ab}=0,\quad \langle |z|^4\rangle_{\text{trivial}}=2^{2N}e^{-\frac{ JNT}{q 2^{q-2}}}\,,
\eea 
which corresponds to the disconnected topology.

There is in addition a large number of non-trivial saddle solutions.  ones with largest contributions to the on-shell actions, including the 1-loop corrections, are
\bal
&g_{12}=bg_{34}=a\frac{1}{2} e^{-\frac{\im \pi}{q}},\quad g_{13}=g_{14}=g_{23}=g_{24}=0,\quad \langle |z|^4\rangle_{12}=1,\label{wh1}\\
&g_{13}=-bg_{24}=a\frac{\im}{2},\quad g_{12}=g_{14}=g_{23}=g_{34}=0,\quad \langle |z|^4\rangle_{13}=e^{-\frac{2 (1+(\im)^q)N J T}{q}},\label{wh2}\\
&g_{14}=bg_{23}=a\frac{1}{2} e^{-\frac{\im \pi}{q}},\quad g_{13}=g_{12}=g_{34}=g_{24}=0,\quad \langle |z|^4\rangle_{14}=1\,,\label{wh3}
\eal
where the last equation in each line is the  on-shell action of the corresponding solution. Apparently \eqref{wh1} and \eqref{wh3} correspond to the  wormhole saddles appearing in $\langle |z|^2\rangle$ and \eqref{wh2} correspond to the possible wormhole saddle appearing in $\langle z^2\rangle$. Therefore we find that in the late time
\bea 
\langle z^4\rangle_{WH}=\begin{cases} 3\langle z^2\rangle^2_{WH} & q=4k+2,\\
	2\langle z^2\rangle^2_{WH} & q=4k.
\end{cases}
\eea 
Notice that for $a=\pm 1$ the real parts of $e^{iTf_p}$ and $e^{-iTf_p}$ are the same, so it is not possible that only the $a=1$ term dominate; what happens is that when the $e^{iaTf_p}$ dominates, the $e^{-iaTf_p}$ term also dominates and the resulting path integral result is just twice of the above results \eqref{wh1}-\eqref{wh3}. Further taking into account that $b$ can be $\pm 1$ we find that total saddle point contributions are  8  when $q=4k$ and 12  when $q=4k+2$ as we found in the exact evaluation.

The interesting $q \bmod 4$ relation is consistent with the time reversal symmetry.

\subsection{$z^2$ at fixed coupling in the Brownian SYK model}
In the following we consider the non-average expression~\eqref{z1}, which we recall here
\bal
|\operatorname{Tr} U(T)|^{2}=\int \mathcal{D} \psi_{a}^{(L)} \mathcal{D} \psi_{a}^{(R)} e^{\im \int_{0}^{T} d t\left[\frac{\im}{2} \psi_{a}^{(j)} \partial_{t} \psi_{a}^{(j)}-J_{a_{1} \ldots a_{q}}(t)\left(\im^{\frac{q}{2}} \psi_{a_{1} \ldots a_{q}}^{(L)}-(-\im)^{\frac{q}{2}}\psi_{a_{1} \ldots a_{q}}^{(R)}\right)\right]}\ .
\eal
We can again introduce
\bal
1=\int_\mathbb{R} \mathcal{D}G_{LR}\int_{\im \mathbb{R}}\frac{\mathcal{D}\S_{LR}}{ 2\times 2\p \im/N} e^{-\int \d t dt'{\frac{\S_{LR}(t,t')}{2}}\left(NG_{LR}(t,t')-\sum_{a}\psi^L_a(t)\psi^R_a(t') \right)} e^{f_{LR}\left(NG_{LR}\right)-f_{LR}\left(\sum_a \psi^L_a\psi^R_a\right)}\ .
\eal
The quantity we would like to compute is
\bal
|\operatorname{Tr} U(T)|^{2}&=\int_\mathbb{R} \mathcal{D}G_{LR}\int_{\im \mathbb{R}}\frac{\mathcal{D}\S_{LR}}{
 2\times 2\p \im/N} \int \mathcal{D} \psi_{a}^{(L)} \mathcal{D} \psi_{a}^{(R)} \\
&\times \exp \left\{\im \int_{0}^{T} d t\left[\frac{\im}{2} \psi_{a}^{(j)} \partial_{t} \psi_{a}^{(j)}-J_{a_{1} \ldots a_{q}}(t)\left(\im^{\frac{q}{2}} \psi_{a_{1} \ldots a_{q}}^{(L)}-(-\im)^{\frac{q}{2}} \psi_{a_{1} \ldots a_{q}}^{(R)}\right)\right]\right\}\\
 &\times e^{-\int \d t \d t'{\frac{\S_{LR}(t,t')}{2}}\left(NG_{LR}(t,t')-\sum_{a}\psi^L_a(t)\psi^R_a(t') \right)} e^{f_{LR}\left(NG_{LR}\right)-f_{LR}\left(\sum_a \psi^L_a\psi^R_a\right)}\ .
\eal
We further rewrite 
\bal\label{sigc}
|\operatorname{Tr} U(T)|^{2}=\int_{\mathbb{R}}\frac{\mathcal{D}\S_{LR}}{2\p /N} \F(\S_{LR})\Psi(\S_{LR})\ .
\eal
The $\psi_a$ independent part reads
\bal
\Psi(\S_{LR})=\int_\mathbb{R} \mathcal{D}G_{LR}e^{-\int \d t \d t'N{\frac{\S_{LR}(t,t')}{2}}G_{LR}(t,t')} e^{f_{LR}\left(NG_{LR}\right)}\,,
\eal
where 
\bal \label{flr}
f_{LR}(NG_{LR})=\frac{2J }{N^{q-1}q}(-1)^{\frac{q}{2}}  \left(NG_{LR,ii}\right)^q\ .
\eal 
The  $\psi_a$ dependent part is
\bal
\F(\S_{LR})=& \int \mathcal{D} \psi_{a}^{(L)} \mathcal{D} \psi_{a}^{(R)}e^{\frac{1}{2}\int \d t \d t'  \S_{LR}(t,t')\left(\psi^{L}_a(t)\psi^{R}_a(t')\right)-f_{LR}(\sum_a \psi^L_a\psi^R_a)}\\
&\times \exp \left\{\im \int_{0}^{T} d t\left[\frac{\im}{2} \psi_{a}^{(j)} \partial_{t} \psi_{a}^{(j)}-J_{a_{1} \ldots a_{q}}(t)\left(\im^{\frac{q}{2}} \psi_{a_{1} \ldots a_{q}}^{(L)}-(-\im)^{\frac{q}{2}} \psi_{a_{1} \ldots a_{q}}^{(R)}\right)\right]\right\}\ .
\eal 
For the integral over $G_{LR}$ to converge, we  rotate the contour so that
\bal
\S_{LR} =  e^{-\im \frac{\p}{q}} \s\,, \quad G_{LR} = \im  e^{\im \frac{\p}{q}} g\ .
\eal 
If we now compute the average of $\F(\s)$, we get
\bal
\<\F(\s)\>&=\int \mathcal{D} \psi_{a}^{(L)} \mathcal{D} \psi_{a}^{(R)}e^{\frac{1}{2}\int \d t \d t'  \S_{LR}(t,t')\left(\psi^{L}_a(t)\psi^{R}_a(t')\right)-f_{LR}(\sum_a \psi^L_a\psi^R_a)}\\
&\times \exp \left\{ \int_{0}^{T} \d t\left[\frac{-1}{2} \psi_{a}^{(j)} \partial_{t} \psi_{a}^{(j)}-\frac{\cj^2}{2}\left(\im^{\frac{q}{2}} \psi_{a_{1} \ldots a_{q}}^{(L)}-(-\im)^{\frac{q}{2}} \psi_{a_{1} \ldots a_{q}}^{(R)}\right)^2\right]\right\}\\
&=\int \mathcal{D} \psi_{a}^{(L)} \mathcal{D} \psi_{a}^{(R)}e^{\frac{1}{2}\int d t d t'  \S_{LR}(t,t')\left(\psi^{L}_a(t)\psi^{R}_a(t')\right)}e^{- \int_{0}^{T} d t\left[\frac{1}{2} \psi_{a}^{(j)} \partial_{t} \psi_{a}^{(j)}+\frac{\cj^2}{2^q}{N \choose q} \right]}\ .
\eal
Integrating over the fermions, we get
\bal
\<\F(\s)\>&=\exp \left\{-\frac{\cj^2}{2^q} T{N \choose q} +N\log\left[2\cos({\frac{T\s e^{-\frac{\im \pi}{q}}}{4}})\right]\right\}\\
&\sim \exp \left\{-\frac{2JT N}{q2^q}+N\log\left[2\cos({\frac{T\s e^{-\frac{\im \pi}{q}}}{4}})\right]\right\}\,,\label{Favg}
\eal
where in the last line we substitute \eqref{SYK1j} and  adopt the leading large-$N$ approximation.
For example, at $\S=0$,
\bal
\<\F(0)\>&\sim 2^N e^{-\frac{2JT N}{q2^{q}}}\,,
\eal 
and at $T=0$, 
\bal
\<\F(0)\>&\sim 2^N \,,
\eal
which is independent of $J$ and $q$. 

We still want to find the region in the $\sigma$ plane where $\F$ is self-averaging.
So next we compute the square
\bal
\F(\s)^2=& \int \mathcal{D} \psi_{a}^{(L,1)} \mathcal{D} \psi_{a}^{(R,1)}\mathcal{D} \psi_{a}^{(L,2)} \mathcal{D} \psi_{a}^{(R,2)}  \exp\left\{\frac{1}{2}\int \d t  \S_{LR}(t,t')\left(\psi^{L,1}_a\psi^{R,1}_a+\psi^{L,2}_a\psi^{R,2}_a\right)\right.\nonumber \\
&-f_{LR}(\sum_a \psi^{L,1}_a\psi^{R,1}_a)-f_{LR}(\sum_a \psi^{L,2}_a\psi^{R,2}_a)- \int_{0}^{T} \d t \left(\frac{1}{2} \psi_{a}^{(j,1)} \partial_{t} \psi_{a}^{(j,1)}+\frac{1}{2} \psi_{a}^{(j,2)} \partial_{t} \psi_{a}^{(j,2)}\right)\nonumber\\
& \left.-\im \int_{0}^{T} \d t \d t' J_{a_{1} \ldots a_{q}}(t)\left(\im^{\frac{q}{2}} \psi_{a_{1} \ldots a_{q}}^{(L,1)}-(-\im)^{\frac{q}{2}} \psi_{a_{1} \ldots a_{q}}^{(R,1)}+\im^{\frac{q}{2}} \psi_{a_{1} \ldots a_{q}}^{(L,2)}-(-\im)^{\frac{q}{2}} \psi_{a_{1} \ldots a_{q}}^{(R,2)}\right)\right\}\ .
\eal 
Its average over the random coupling is
\bal
\<\F(\s)^2\>& =\int \mathcal{D} \psi_{a}^{(L,1)} \mathcal{D} \psi_{a}^{(R,1)}\mathcal{D} \psi_{a}^{(L,2)} \mathcal{D} \psi_{a}^{(R,2)}  \exp\left\{\frac{1}{2}\int \d t   \S_{LR}(t,t')\left(\psi^{L,1}_a\psi^{R,1}_a+\psi^{L,2}_a\psi^{R,2}_a\right)\right.\nonumber \\
&-f_{LR}(\sum_a \psi^{L,1}_a\psi^{R,1}_a)-f_{LR}(\sum_a \psi^{L,2}_a\psi^{R,2}_a)- \int_{0}^{T} \d t \left(\frac{1}{2} \psi_{a}^{(j,1)} \partial_{t} \psi_{a}^{(j,1)}+\frac{1}{2} \psi_{a}^{(j,2)} \partial_{t} \psi_{a}^{(j,2)}\right)\nonumber \\
&\left.-\frac{\cj^2}{2}\left(\im^{\frac{q}{2}} \psi_{a_{1} \ldots a_{q}}^{(L,1)}-(-\im)^{\frac{q}{2}} \psi_{a_{1} \ldots a_{q}}^{(R,1)}+\im^{\frac{q}{2}} \psi_{a_{1} \ldots a_{q}}^{(L,2)}-(-\im)^{\frac{q}{2}} \psi_{a_{1} \ldots a_{q}}^{(R,2)}\right)^2\right\}\ .
\eal
Expanding out the square, and introducing the extra $G$, $\S$ variables for the quantities between the two copies, we get 
\bal
\<\F(\s)^2\>&=\int \cd G_{ab}\frac{\cd \S_{ab}}{2\pi \im/N} \mathcal{D} \psi_{a}^{(L,1)} \mathcal{D} \psi_{a}^{(R,1)}\mathcal{D} \psi_{a}^{(L,2)} \mathcal{D} \psi_{a}^{(R,2)}  \exp\left\{\frac{1}{2}\int \d t  \S_{LR}(t,t')\left(\psi^{L,1}_a\psi^{R,1}_a+\psi^{L,2}_a\psi^{R,2}_a\right)\right.\nn \\
&-f_{LR}(\sum_a \psi^{L,1}_a\psi^{R,1}_a)-f_{LR}(\sum_a \psi^{L,2}_a\psi^{R,2}_a)- \int_{0}^{T} \d t \left(\frac{1}{2} \psi_{a}^{(j,1)} \partial_{t} \psi_{a}^{(j,1)}+\frac{1}{2} \psi_{a}^{(j,2)} \partial_{t} \psi_{a}^{(j,2)}\right)\nn \\
&+\int \d t   \left(\S_{13}\psi^{(L,1)}_A\psi^{(L,2)}_A+ \S_{23}\psi^{(R,1)}_A\psi^{(L,2)}_A+\S_{14}\psi^{(L,1)}_A\psi^{(R,2)}_A+\S_{24}\psi^{(R,1)}_A\psi^{(R,2)}_A\right)\nn \\ 
&-N\int \d t \left(\S_{13}G_{13}+\S_{14}G_{14}+\S_{23}G_{23}+\S_{24}G_{24}\right)\nn  \\
&\left.+\int_{0}^{T} \d t \left[-\frac{2\cj^2}{2^q}{N\choose q} -\frac{2JN}{q} \left( G_{LL,12}^q+ G_{RR,12}^q \right)\right.\right.\nonumber\\
&+\left.\left.\frac{2JN}{q}\im ^{q} \left(G_{LR,11}^q+G_{LR,22}^q+G_{LR,12}^q+G_{LR,21}^q\right)  \right]\right\}\ .
\eal
Notice that in the first line the fermion bilinears are all in the same copy; these terms come from the $\F$ itself.  In the second line, we added in a few other terms that couple the fermions between the two copies.
Then we choose the $f_{LR}$ to cancel the $G_{LR,11}^q$ and $G_{LR,22}^q$ terms in the square, namely
\bal
f_{LR}(\sum_a\psi^L_a\psi^R_a)=\frac{2J N}{q} (-1)^{\frac{q}{2}}  G_{LR,ii}^q\,,
\eal
so that the above result simplifies to
\bal
\<\F(\s)^2\>&=\int \int \cd G_{ab}\frac{\cd \S_{ab}}{4\pi \im/N}\mathcal{D} \psi_{a}^{(L,1)} \mathcal{D} \psi_{a}^{(R,1)}\mathcal{D} \psi_{a}^{(L,2)} \mathcal{D} \psi_{a}^{(R,2)}  \exp\left\{\frac{1}{2}\int \d t  \S_{LR}(t,t')\left(\psi^{L,1}_a\psi^{R,1}_a+\psi^{L,2}_a\psi^{R,2}_a\right)\right.\nn \\
&- \int_{0}^{T} \d t \left(\frac{1}{2} \psi_{a}^{(j,1)} \partial_{t} \psi_{a}^{(j,1)}+\frac{1}{2} \psi_{a}^{(j,2)} \partial_{t} \psi_{a}^{(j,2)}\right)\nn \\
&+\frac{1}{2}\int \d t   \left(\S_{13}\psi^{(L,1)}_A\psi^{(L,2)}_A+ \S_{23}\psi^{(R,1)}_A\psi^{(L,2)}_A+\S_{14}\psi^{(L,1)}_A\psi^{(R,2)}_A+\S_{24}\psi^{(R,1)}_A\psi^{(R,2)}_A\right)\nn \\ 
&-\frac{N}{2}\int \d t \left(\S_{13}G_{13}+\S_{14}G_{14}+\S_{23}G_{23}+\S_{24}G_{24}\right)\nn  \\
&+\int_{0}^{T} \d t \left[-\frac{2\cj^2}{2^q}{N\choose q} -\frac{2JN}{q} \left( G_{LL,12}^q+ G_{RR,12}^q \right) +\frac{2JN}{q}\im^q \left( G_{LR,12}^q+ G_{RL,12}^q \right) \right]\ .
\eal
Integrating out the fermions with the help of the relation \eqref{i4}, shortening the labels according to  $(L,1)\to 1$, $(R,1)\to 2$, $(L,2)\to 3$, $(R,2)\to 4$, and using the fact that $\S_{12}=\S_{34}=\S_{LR}$ by construction,  
we get
\bal
\<\F(\s)^2\>&=\int  \frac{\d G_{AB} \d\S_{AB}}{4\pi \im /(NT)}\exp \left\{ N\left[ \log\left(2\cos\(\frac{T}{4}\sqrt{(\S_{14}-\S_{23})^2+(\S_{13}+\S_{24})^2}\)\right) \right.\right.\nn \\
&\left.+2 \cos\left(\frac{T}{4} \sqrt{(\S_{14}+\S_{23})^2+(\S_{13}-\S_{24})^2+4\S^2} \right) \right]-\frac{N T}{2} \sum_{AB} \S_{AB}G_{AB}\nn  \\
&\left.-\frac{2\cj^2}{2^q} T {N\choose q}-\frac{2JN T}{q} \left( G_{13}^q+ G_{24}^q \right) +\frac{2JN T}{q}\im^q \left( G_{14}^q+ G_{23}^q \right)\right\}\,,
\eal
where again we have focused on the time-independent saddles.
\subsubsection{Exact computation}
We can first evaluate the integral explicitly. The calculation is similar to the one of $\langle |z|^4\rangle$
\bal
\<\F(\s)^2\>&=e^{-\frac{4J NT}{q 2^q}}\(e^{\frac{\im \hat{f'}_+}{N} }+e^{-\frac{\im \hat{f'}_+}{N}}+ e^{\frac{\im \hat{f'}_-}{N} }+e^{-\frac{\im \hat{f'}_-}{N}}\)^N e^{\frac{2NTJ}{q}\sum_{}s_{AB}G_{AB}^q}\, \Big|_{G_{AB}=0}\,\label{pp2}\\
&=e^{-\frac{4J NT}{q 2^q}}\sum_{0\leq n_i\leq N, \sum_i n_i=N}e^{ \frac{\im(n_1-n_2)}{N} \hat{f'}_+}e^{\frac{\im (n_3-n_4)}{N} \hat{f'}_-}e^{\frac{2NTJ}{q}\sum_{}s_{AB}G_{AB}^q}\, \Big|_{G_{AB}=0}\, ,
\eal 
where $s_{14}=s_{23}=\im^q,s_{13}=s_{24}=-1$ and we have introduced the differential operators
\bea 
&&\hat{f'}_+=\frac{1}{2}\sqrt{(\partial_{G_{14}}+ \partial_{G_{23}})^2+(\partial_{G_{13}}-\partial_{G_{24}})^2+4T^2\Sigma^2},\\
&&\hat{f'}_-=\frac{1}{2}\sqrt{(\partial_{G_{14}}- \partial_{G_{23}})^2+(\partial_{G_{13}}+\partial_{G_{24}})^2},
\eea 
Following a similar calculation as in section~\eqref{z4e}, we get
  \bea 
&&\langle \Phi(\sigma)^2\rangle=e^{-\frac{4J NT}{q 2^q}}\sum_{\substack{0\leq n_i\leq N, \\\sum_i n_i=N}}\sum_{n=0}^\infty\sum_{m=0}^\infty\sum_{\substack{0\leq k^+_1,k^+_2,k_3\leq m,\\ 0\leq k^-_1,k^-_2\leq n \\\sum k_1^++k_2^++k_3=m\\\sum k^-_i=n}}\sum_{l^+_i=0,l^-_i=0}^{k_i^+,k_i^-}\frac{(\frac{\im n_{12}}{N})^{mq}}{(mq)!2^{mq}}\frac{(\frac{\im n_{34}}{N})^{nq}}{(nq)!2^{nq}} \nonumber \\
&&\quad \(\im^q\frac{2NJT}{q}\)^{(k_1^++k_1^-)q}\(-\frac{2NJT}{q}\)^{(k_2^++k_2^-)q}(2Te^{-\frac{\pi}{q}}\sigma)^{k_3}\prod_{i=1}^2\Delta(k_i^+,k_i^-)\,,
\eea 
where $n_{12}=n_1-n_2$ and $n_{34}=n_3-n_4$. In the large $T$ limit, as we show in the exact computation of $\langle z^4\rangle$, the leading contributions are the summation of the contribution obtained by keeping only one exponential differential operator in \eqref{pp2}. The operator $e^{\pm\frac{\im \hat{f'}_-}{N}}$ contributes 1 when $q=4m$ and 2 when $q=4m+2$. The result of operator $e^{\pm\frac{\im \hat{f'}_+}{N}}$ will be a monomial of $\sigma$ while its expression is not very illuminating so we omit here.
\subsubsection{Saddle point computation }

We deform the contour so that the integral converge. In the current case, the contours are rotated as
\bal
G_{13}=g_{13}\,, \quad G_{24}=g_{24}\,, \quad \S_{13}=i \s_{13}\,,\quad \S_{24}=i \s_{24}\label{odd}
\eal 
and 
\bal
&G_{14}=\im e^{\im \frac{\p}{q}}g_{14}\,, \quad
 G_{23}=\im e^{\im\frac{\p}{q}}g_{23}\,, \quad \S_{14}=e^{-\im\frac{\p}{q}} \s_{14}\,,\quad \S_{23}= e^{-\im\frac{\p}{q}}\s_{23}\,,\label{bp}
\eal
so that the effective action for computing $\langle \Phi(\sigma)^2\rangle$ is \footnote{To get rid of the factor $1/2$ we have scaled the variables as $\sigma_{AB}\rightarrow 2 \sigma_{AB} $ .}
\bea
S=-\frac{4 NTJ}{q2^q}-2NT\sum_{ab}\(\frac{J}{q}g_{ab}^q+ \frac{\im}{2}\s_{ab}g_{ab}\)+N \log (2\cos(Tf'_+)+2\cos(Tf'_-)),\nonumber \\
\eea 
where we have defined
\bea 
&&f'_+=\frac{1}{2}\sqrt{e^{-\frac{2\im \pi}{q}}(\sigma_{14}+\sigma_{23})^2-(\sigma_{13}-\sigma_{24})^2+e^{-\frac{2\im \pi}{q}}4\s^2},\\
&&f'_-=\frac{1}{2}\sqrt{e^{-\frac{2\im \pi}{q}}(\sigma_{14}-\sigma_{23})^2-(\sigma_{13}+\sigma_{24})^2}.
\eea 
The equation of motion of $g_{AB}$ gives universally
\bal
-\im \s_{ab}=2J g_{ab}^{q-1},\quad (ab)=(13),(24),(14),(23)\ .\label{gsaddle0}
\eal
As discussed in the previous $\<|z|^2\>$ computation, the equation of motion of $\s_{ab}$ depends on the value of time $T$. 

\subsubsection*{Very short time}
When time $T$ is very short, the cosine function can be approximated by a constant. Then the only saddle point is 
\bal
g_{ab}=0\,,\qquad \s_{ab}=0\ .\label{trivialsaddle0}
\eal 
The on-shell action, including a 1-loop determinant $\frac{1}{(TN)^4}$, around this saddle point is 
\bal
\<\F(\s)^2\>=\left(2\cos\left(T e^{- \frac{\im \pi}{q}}\s \right)\right)^{N}e^{-\frac{4JTN}{q 2^{q}}}=\left(2\cos\left(T\S\right)\right)^{N}e^{-\frac{4JTN}{q2^{q}}}\ .\label{trivialsaddle1}
\eal
The results agree with $\<\F\>^2$ as can be seen from~\eqref{Favg}. 
Notice that the trivial saddle~\eqref{trivialsaddle0} remains a saddle point for a large range of $T$, and the on-shell action around this saddle point, ie~\eqref{trivialsaddle1}, is true within this large range.

In summary, we have shown that at very short time the trivial saddles dominate the $\<\F^2\>$ which approximately equals to $\<\F\>^2$, we conclude that at short time the trivial saddle dominates and $\Phi(\s )$ is self-averaging.

\subsubsection*{Short time}
When the time is larger, the determinant term in the action cannot be approximated by a constant, and thus we expand it to the second order of $T$.  
The saddle point equation for $\s_{ab}$ is now
\bal
&\s_{13} T = 4\im g_{13} \,, \qquad \s_{24} T= 4\im g_{24} \\
&\s_{14} T=-4\im  e^{\frac{2\pi \im}{q}} g_{14}\,,\qquad \s_{23} T=-4\im  e^{\frac{2\pi \im}{q}} g_{23}\ .
\eal
Combining with~\eqref{gsaddle0}, we get the non-trivial saddle solutions
\bal
\frac{2}{JT}=g_{13}^{q-2}=g_{24}^{q-2}=-e^{-\frac{2\im \pi}{q}}g_{23}^{q-2}=-e^{-\frac{2\im \pi}{q}}g_{14}^{q-2}
\eal
Each solution gives a contribution to $\<\F(\s)^2\>$ as 
\bal
2^{2N}e^{-\frac{2JTN}{q2^{q-1}}-Ne^{-\frac{2\im \pi}{q}}\s^2 T^2/4}e^{4N\frac{q-2}{2q} (JT/2)^{\frac{2}{2-q}}(e^{\frac{4m_1 \pi \im}{2-q}}+e^{\frac{4m_2 \pi \im}{2-q}})} e^{-4N\frac{q-2}{2q} (JT/2 )^{\frac{2}{2-q}}(e^{\frac{4m_3 \pi \im}{2-q}}+e^{\frac{4m_4 \pi \im}{2-q}})}\,,
\eal
where $m_i$ are integers. 
It is cumbersome in the following to discuss the most generic case with arbitrary $q$, we therefore focus on the $q=4$ case. Then the above contribution uces to
 \bal\label{nons}
 2^{2N}e^{-\frac{2JTN}{q2^{q-1}}+\im N \s^2 T^2/4}\ .
 \eal
Recall that in the end we need to perform the integral \eqref{sigc}, as we argued in last section we have to deform the contour to a steepest contour such that the \eqref{nons} vanishes at infinity. It means that this saddle point behaves like
\bea
2^{2N}e^{-\frac{2JTN}{q2^{q-1}}- N |\text{Im}(\s^2) T^2/4}\ . 
\eea 
so it is sub-dominant comparing with the trivial saddle. Therefore in the regime of time, $\Phi$ is still self-averaging.
\subsubsection*{Long time}
At very long time we rewrite the $\log$ term as
\bal
\log\left(e^{\im T f'_+}+e^{-\im T f'_+}+e^{\im T f'_-}+e^{\im T f'_-}\right)\,.
\eal
When $T$ is sufficiently large, only one term in the above expression is dominant in the $\log$ function, there are thus two  different cases to be discussed; either the term that is independent of $\s$ or the term that depend on $\s$. When the $\s$ independent term dominates, the saddle point contribution is independent on $\s$, which is not related to either the wormhole or half-wormhole contribution that we are interested in. 
Therefore it is tentative to consider the case where the $\s$ dependent term dominates
\bea 
&&\log\left(e^{\im T f'_+}+e^{-\im T f'_+}+e^{\im T f'_-}+e^{\im T f'_-}\right) \approx { \frac{\im a T}{2}  \sqrt{e^{-\frac{2\im \pi}{q}}(\sigma_{14}+\sigma_{23})^2-(\sigma_{13}-\sigma_{24})^2+e^{-\frac{2\im \pi}{q}}4\s^2}},\nn
\eea   
where $a=\pm 1$. The saddle point equations are
\bea
&&2g_{13}+\frac{\im a g_{13}^{q-1}}{\sqrt{g_{13}^{2q-2}-e^{-\frac{2\im\pi}{q}}g_{14}^{2q-2}+e^{-\frac{2\im\pi}{q}}\tilde{\s }^2}}=0,\\
&&2g_{14}-\frac{\im a e^{-\frac{2\im\pi}{q}} g_{14}^{q-1}}{\sqrt{g_{13}^{2q-2}-e^{-\frac{2\im\pi}{q}}g_{14}^{2q-2}+e^{-\frac{2\im\pi}{q}}\tilde{\s }^2}}=0,\\
&&g_{14} = g_{23}\,,\quad g_{24} = -g_{13}\,,\quad \s_{14} = \s_{23}\,,\quad \s_{24} = -\s_{13}\ . \label{sun}
\eea 
where $\tilde{\s}^2=\s^2/J^2$ and the equation of motion~\eqref{gsaddle0} has been used. Solutions of this set of equations are complicated in general, here we only provide the solutions for $q=4$. First let us consider the non-trivial solutions $g_{ab}\neq 0$. There is a set of 8 solutions 
\bal 
&g_{13}=\pm e^{\frac{\im \pi}{8}}\sqrt{\frac{\tilde{\s }}{a}}\,,\quad g_{14}=\pm e^{\frac{7\im \pi}{8}}\sqrt{\frac{\tilde{\s }}{a}}\,,\quad g_{13}=\pm e^{\frac{\im \pi}{8}}\sqrt{\frac{\tilde{\s }}{a}}\,,\quad g_{14}=\mp e^{\frac{7\im \pi}{8}}\sqrt{\frac{\tilde{\s }}{a}},\\
&g_{13}=\pm e^{\frac{5\im \pi}{8}}\sqrt{\frac{\tilde{\s }}{a}}\,,\quad g_{14}=\pm e^{\frac{3\im \pi}{8}}\sqrt{\frac{\tilde{\s }}{a}}\,,\quad g_{13}=\pm e^{\frac{5\im \pi}{8}}\sqrt{\frac{\tilde{\s }}{a}}\,,\quad g_{14}=\mp e^{\frac{3\im \pi}{8}}\sqrt{\frac{\tilde{\s }}{a}}\,,
\eal 
where all the unlisted variables are given by \eqref{sun}.
The first four solutions lead to the same on-shell action 
\bea 
\frac{1}{2} e^{\im a T N\sqrt{ e^{-\frac{2\im \pi }{q}}\s^2}}e^{-\frac{4JTN}{q 2^{q}}}\,,
\eea 
while the last four saddles lead to another on-shell action
\bea 
\frac{1}{4} e^{\im a T N\sqrt{ e^{-\frac{2\im \pi }{q}}\s^2}}e^{-\frac{4JTN}{q2^{q}}}\ .
\eea 
Comparing with the contribution from the trivial saddles 
\bea 
\langle \F^2\rangle = \left(2\cos\left(T e^{- \frac{\im \pi}{q}}\s \right)\right)^{N}e^{-\frac{2JTN}{q }}\sim e^{\pm \im  T N\sqrt{ e^{-\frac{2\im \pi }{q}}\s^2}}e^{-\frac{4JTN}{2^{q}q}}\,,
\eea 
we find that all of them are comparable with the trivial saddle. We have seen this phenomenon in the 0-dimensional SYK model. We expect these non-trivial saddles to not contribute to the path integral, which could be checked in the Lefschetz thimble analysis.

\begin{figure}[h]
	\centering
	\includegraphics[scale=0.3]{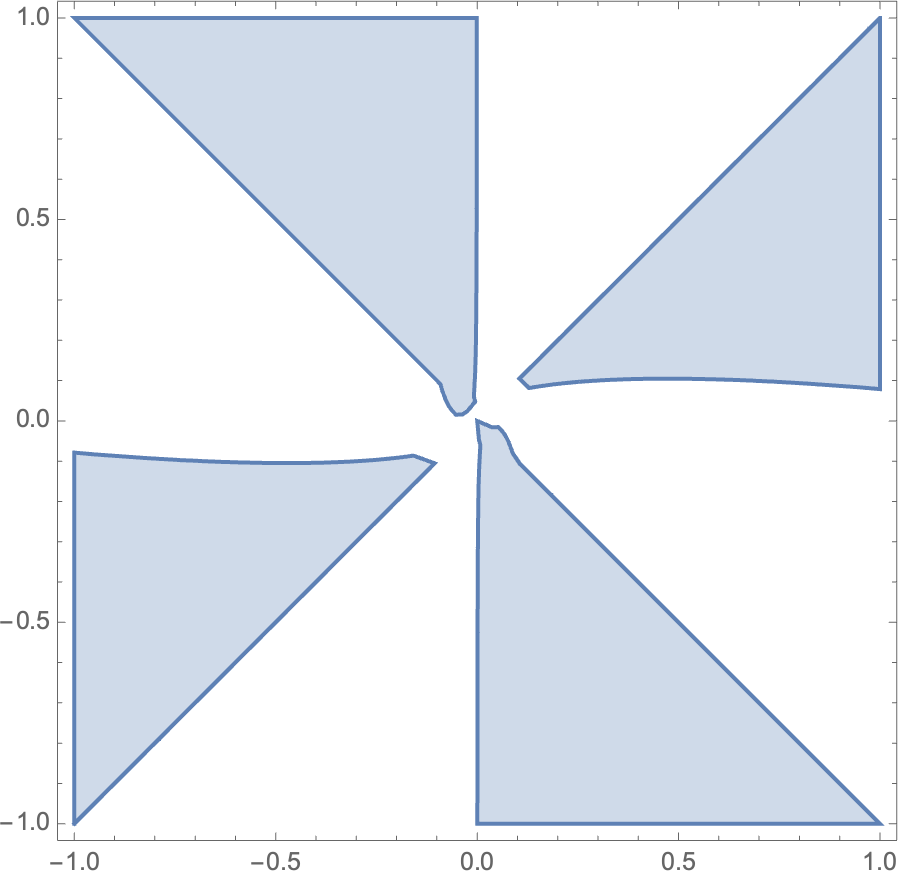}
	\caption{The shaded region is where non-trivial saddle dominates. We have set $J=1/2$.}
	\label{plot2}
\end{figure}

In the end let us consider the special non-trivial solution with $g_{23}=0=g_{14}$. Focusing on the case of $a=1$, the  saddle point which has the proper fall-off behavior at infinite of $\s $ is
\bea 
\exp\(-\frac{3}{2}NT \s^{\frac{4}{3}}\),\quad \sigma \rightarrow \infty.
\eea 
Then we  plot the region where this non-trivial saddle dominates over the trivial saddle in Fig. \ref{plot2}.  
It turns out that wormhole saddle $\sigma_{wh}=\pm \frac{1}{2}\im e^{\frac{\im \pi}{4}}$ is not in this region. For example at $J=1/2$, $\text{Re}(\langle \Phi^2(\sigma_{wh})\rangle_{\text{non-trivial}}-\langle \Phi^2(\sigma_{wh})\rangle_{\text{trivial}})=-0.23$.
It suggests that like the 0-dimensional model, the wormhole saddle of $\langle |z|^2\rangle$ are  within the self-averaging region of the $\sigma$ plane.

\subsection{Half-wormholes}
 With all the results in the previous sections, we expect that the integral \eqref{sigc} can be approximated by
\bea 
|z|^2&\approx&\int_{\mathbb{R}-0} \frac{\d \sigma}{2\pi}\Psi(\sigma) \F(\s )+\(\text{ half-wormholes at } \sigma=0 \), \\
&=& \int_{\mathbb{R}-0} \frac{\d \sigma}{2\pi}\Psi(\sigma) \langle \F(\s )\rangle+\(\text{ half-wormholes at } \sigma=0 \), \\
&=& \langle |z|^2\rangle+\Phi(0)\, , \label{hwu}
\eea 
at late time. First we notice that in the late time
\bea 
\langle \Phi(0)\rangle=2^N e^{-\frac{4JTN}{q2^{q}}}\rightarrow 0
\eea 
therefore at least \eqref{hwu} is consistent. To confirm \eqref{hwu} we need to compute $\langle \Phi(0 )^2\rangle$, $\langle \Phi(0 )|z|\rangle $ and the error:
\bea
\text{Error}=\langle (|z|^2-\langle |z|^2\rangle-\Phi(0))^2\rangle=\langle |z|^4\rangle-\langle |z|^2\rangle^2+\langle \Phi(0)^2\rangle-2\langle |z|^2\Phi(0)\rangle. \label{er}
\eea 
In the late time, each term in \eqref{er} is given by the non-trivial saddle points. It is clear that $\langle \Phi(0 )^2\rangle$, $\langle \Phi(0 )|z|\rangle $ can be obtain from $\langle |z|^4\rangle$ by setting $g_{12}=0$. Therefore we have
\bea 
\text{Error}=\begin{cases} (3-1+2-4)\langle z^2\rangle^2_{WH}=0 & q=4k+2,\\
	(2-1+1-2)\langle z^2\rangle^2_{WH}=0 & q=4k,
\end{cases}
\eea 
therefore \eqref{hwu} indeed is good in the large $T$ limit.

\section{Modified Brownian SYK model}
\label{SYK2}
In this section, we study the wormhole and half-wormholes contributions in some  modified (Brownian) SYK model.

\subsection{Brownian SYK with non-vanishing mean value}
Let us first consider to turn on the mean value of the random couplings:
\bea 
\langle J_A\rangle=J_A^{(0)}=\mu,\quad \langle J_A(t)J_B(t')\rangle=\delta(t-t')(\delta_{AB}\t^2+\mu^2)\,,
\eea 
and in this section we use
the convention $\{\psi_i,\psi_j\}=2h \delta_{i,j}$.

Taking the disorder averaging of the coupling we obtain the averaged theory
\bal 
\langle z(T)\rangle_J&=\int d^N \psi \,e^{-S_{a}}\,,\\
 S_a&=\frac{1}{2}\int_0^T dt \sum_i^N \psi_i \partial_t \psi_i-\im^{q/2} \int dt \sum_{A}J_A^{(0)}\psi_A-\frac{\t^2}{2}\int dt (\sum_{A}\psi_A^2)
\eal 
We can convert the effective Hamiltonian of  the averaged theory as a spin system
\bal 
\langle z\rangle_J=\text{Tr}(e^{-T \ch})\,, \quad \ch=-\im^{q/2} \sum_A J_A^{(0)}\psi_A-\frac{t^2}{2} \sum_A h^q=-\im^{q/2} \sum_A J_A^{(0)}\psi_A-\frac{\t^2}{2}{N \choose q }h^q \ . \label{zj}
\eal 
When $\mu=0$, the averaged partition function is given by
\bea 
\langle z\rangle_J=e^{T\frac{\t^2}{2}{N\choose q}h^q}\equiv 2^N e^{T E_0},\quad E_0=\frac{\t^2}{2}{N\choose q}h^q\sim \frac{\t^2}{2}N^q h^q\,.
\eea 
When $\mu\neq 0$, we have to evaluate the trace
\bea \label{trace}
\langle z \rangle_J=e^{TE_0}\text{Tr}(e^{T \im^{q/2}\mu \sum_A \psi_A})=e^{TE_0}\int \d^N \psi_i \exp(T \im^{q/2} \sum_i \psi_i)\ .
\eea 
However there is no simple expression for $\langle z\rangle$. We first consider the simplest case with $q=1$
\bea 
I_f=\int \d^N \psi_i \exp(a \sum_i \psi_i)\ .
\eea 
The idea is to transfer the Majorana fermions to Dirac fermions which have a well-defined rules of integrals. Assuming the total number of fermions is even $N=2K$ then we introduce $K$ Dirac fermions as
\bea 
&&c_i=\frac{1}{2\sqrt{h}}(\psi_{2i-1}-\im \psi_{2i}),\quad c_i^\dagger=\frac{1}{2\sqrt{h}}(\psi_{2i-1}+\im \psi_{2i}),\quad i=1,\dots, K\, ,\\
&&\psi_{2i-1}=\sqrt{h}(c_i+c_i^\dagger),\quad \psi_{2i}=\im \sqrt{h}(c_i-c_i^\dagger)\, ,
\eea 
which obey
\bea 
\{c_i,c_j\}&=&\{c_i^\dagger,c_j^\dagger\}=0,\quad \{c_i,c_j^\dagger\}=\delta_{ij}\, \\
\eea 
The integration measure changes as
\bea 
\cd\psi_{2i}\cd\psi_{2i-1}=2h\cd c_i \cd c_i^\dagger\, .
\eea 
Thus the integral can be evaluated as
\bea 
I_1 &=&(2h)^K \int \prod_i\cd c_i \cd c_i^\dagger \exp\(a \sum_i^K \sqrt{h}[(1+\im)c_i+(1-\im)c_i^\dagger]\) \\
&=&(2h)^K(2 \cosh (\sqrt{2}ah))^K=(2h)^N[\cosh(\sqrt{2}ah)]^{N/2}\,.
\eea 
Now we let us consider the case of $q=2$
\bea\label{i2}
I_2(a)=\int \d^N\psi_i \exp(\frac{a}{2}\sum_{i\neq j}\psi_i A_{ij}\psi_j),\quad \text{with }\quad \(A_{ij}=-A_{ij}=a,\quad i<j\),
\eea 
which looks like a Gaussian but we need to replace $\psi_i$ with $c_i$:
\bea \label{generalc}
&&I_2=(\sqrt{2h})^N \int \prod_i\cd c_i \cd c_i^\dagger \, e^{\ch} \\
 \nn \\ &&\quad   \ch=\( \im a h(\sum_i^{K}[c_i^\dagger c_i-c_ic_i^\dagger]  +2 \sum_{i<j}[c_ic_j-c_i^\dagger c_j^\dagger])+2ah\sum_{i<j}[c_i^\dagger c_j+c_i c_j^\dagger]\) 
\eea 
To get an idea how to compute this integral let us consider a simple case of $N=4$:
\bea 
&& \psi_1= \sqrt{h}(c_1+c_1^\dagger),\psi_2=\im \sqrt{h}(c_1-c_1^\dagger),\psi_3=\sqrt{h}(c_2+c_2^\dagger),\psi_4=\im \sqrt{h}(c_2-c_2^\dagger), \\
&& \sum_{i<j}\psi_i\psi_j= \im h(c_1^\dagger c_1-c_1 c_1^\dagger+c_2^\dagger c_2-c_2 c_2^\dagger+2c_1c_2-2c_1^\dagger c_2^\dagger )+2h(c_1^\dagger c_2+c_1c_2^\dagger).
\eea 
We have four different states $|\Psi_i\rangle$:
\bea 
|00\rangle,\quad c_{1}^{\dagger}|00\rangle,\quad c_{2}^{\dagger}|00\rangle,\quad c_{1}^{\dagger}c_{2}^{\dagger}|00\rangle.\label{numstate}
\eea 
So the operator $\sum_{i<j}\psi_i\psi_j$ can be written as a $4\times 4$ matrix :
\bea
\sum_{i<j}\psi_i\psi_j=\left(
\begin{array}{cccc}
 -2 \im h & 0 & 0 & -2 \im h \\
 0 & 0 & -2 h & 0 \\
 0 & 2 h & 0 & 0 \\
 -2 \im h & 0 & 0 & 2 \im h \\
\end{array}
\right)
\eea 
with 4 eigenvalues $\{\pm \im 2 h, \pm \im 2\sqrt{2}h\}$ so path integral over $c_i$ and $c_i^\dagger$ can be computed as
\bea 
&&\sum_i \langle \Psi_i|e^{a \sum_i\psi_i\psi_j}|\Psi_i\rangle =2\(\cos(2ah)+\cos(2\sqrt{2}a h)\)\, .\label{n4q2trace}
\eea 
For example of $N=6$, the corresponding matrix is 
\bea 
\sum_{i<j}\psi_i\psi_j=\left(
\begin{array}{cccccccc}
 -3 \im h & 0 & 0 & 0 & -2 \im h & -2 \im h & -2 \im h & 0 \\
 0 & -\im h & 2 h & 2 h & 0 & 0 & 0 & -2 \im h \\
 0 & -2 h & -\im h & 2 h & 0 & 0 & 0 & 2 \im h \\
 0 & -2 h & -2 h & -\im h & 0 & 0 & 0 & -2 \im h \\
 -2 \im h & 0 & 0 & 0 & \im h & 2 h & -2 h & 0 \\
 -2 \im h & 0 & 0 & 0 & -2 h & \im h & 2 h & 0 \\
 -2 \im h & 0 & 0 & 0 & 2 h & -2 h & \im h & 0 \\
 0 & -2 \im h & 2 \im h & -2 \im h & 0 & 0 & 0 & 3 \im h \\
\end{array}
\right)
\eea 
which can be divided into two blocks. We get the eigenvalues by directly diagonalizing the matrix:
\begin{align}
    \pm 5\im h,\quad \pm (2\sqrt{3}+1)\im h,\quad \pm 3\im h,\quad \pm (2\sqrt{3}-1)\im h.\label{n6eigen}
\end{align}
Similarly for general $N$, we can write effective Hamiltonian defined in \eqref{generalc} as
\bea
\ch=\sum_{i\leq j=1}^k\(\alpha_{ij}c_i^\dagger c_j+\beta_{ij}c_i c_j^\dagger+\gamma_{ij}c_i^\dagger c_j^\dagger+\theta_{ij} c_i c_j\),\label{effectiveh}
\eea
with
\bea
\alpha_{ii}=\im h,\quad \beta_{ii}=-\im h,\quad \alpha_{ij}=2h,\quad \beta_{ij}=2h,\\ \gamma_{ij}=-2\im h,\quad \theta_{ij}=2\im h,\quad \gamma_{ii}=0,\quad \theta_{ii}=0.
\eea
This Hamiltonian is quadratic and famously can be diagonalized by the Bogoliubov and Valatin's method \cite{Bogolyubov:1947zz,Valatin:1958ja}.  Explicitly we can do the transformation by taking an operator basis for the Hamiltonian 
\begin{align}
H=c^{\dagger}Mc
\end{align}
where we have 
\begin{align}
c^{\dagger}=\left(c_{1}^{\dagger},c_{2}^{\dagger},\dots,c_{1},c_{2},\dots\right).
\end{align} 
In the simple case with $N=4$  the matrix can be expressed as 
\bea
M=\left(
\begin{array}{cccc}
 \im h & h & 0 &  -\im h \\
 -h & \im h & \im h & 0 \\
 0 & \im h & -\im h & h \\
 -\im h  & 0 & -h &  -\im h \\
\end{array}
\right),
\eea
we can directly take the diagonalization and get the eigenvalues
\begin{align}
\im(1+\sqrt{2})h,\quad -\im(1+\sqrt{2})h,\quad -\im(1-\sqrt{2})h,\quad -\im(-1+\sqrt{2})h.
\end{align} 
For simplicity we take the notation as 
\begin{align}
    \lambda_{1}=\im(\sqrt{2}+1)h,\quad \lambda_{2}=\im(\sqrt{2}-1)h,
\end{align}
 then the resulting effective Hamiltonian becomes 
 \begin{align}
     H=\lambda_{1}\left( d_{1}^{\dagger}d_{1}-d_{1}d_{1}^{\dagger} \right)+\lambda_{2}\left( d_{2}^{\dagger}d_{2}-d_{2}d_{2}^{\dagger} \right).
 \end{align}
 To evaluate the trace we still take the states as \eqref{numstate} therefore we have 
 \begin{align}
     \text{Tr}(e^{H})=e^{-a(\lambda_{1}+\lambda_{2})}+e^{a(\lambda_{1}-\lambda_{2})}+e^{a(-\lambda_{1}+\lambda_{2})}+e^{a(\lambda_{1}+\lambda_{2})},
 \end{align}
so we can recover the result \eqref{n4q2trace}.  For general $N$ the operator \eqref{effectiveh} can be expressed as a block matrix
\begin{align}
    M=\left(
    \begin{array}{cc}
         A+\im hI_N&-\im A \\
         \im A&A-\im h I_N 
    \end{array}
    \right),
\end{align}
with
\bea
    A=\left(
    \begin{array}{cccc}
         0&h&h &\cdots \\
         -h&0&h &\cdots\\
         -h&-h&0&\cdots\\
        \vdots&\vdots&\vdots&\ddots 
    \end{array}
    \right)
\eea    
The characteristic equation is
\bea 
\det\(A+(\im h-\lambda))(A-(\im h+\lambda)-H^2\)&=&\det\( (h^2+\lambda^2)I_N-2\lambda A \) \\
&=& (\lambda+h)^N+(\lambda-h)^N=0\, .
\eea 
So the eigenvalues are
\bea
\lambda_m=\im h\tan(\frac{m \pi}{2 N}),\quad m=1,3,\dots,N-1\, .
\eea 
then the Hamiltonian becomes 
\begin{align}
    H=\sum_{i=1}^{N}\lambda_{i}(d_{i}^{\dagger}d_{i}-d_{i}d_{i}^{\dagger}).
\end{align}
and  the trace will have  the form 
\begin{align}
    \Tr(e^{H})=\sum_{\sigma=\pm 1} e^{a\sum_{i=1}^{k}\sigma_{i}\lambda_{i}}=\sum_{\sigma}\prod_{i=1}^k e^{a\sigma_{i}\lambda_{i}}=\prod_{i=1}^k \sum_{\sigma}e^{a\sigma_{i}\lambda_{i}}=2^k\prod_{i=1}^k\cosh\(a\l_i\)\,,
\end{align}

Now let us consider the function
\bea 
X_n=\sum_{1\leq i_1<\dots i_n\leq N}\psi_{i_1}\dots \psi_{i_N}.
\eea 
We would like to argue that in the large $N$ limit, we have the approximation
\bea 
n! X_{2n}\approx (X_2)^n, 
\eea 
as we find for the 0-dimensional theory. Note that unlike the situation of the 0-dimensional theory, $\{X_n\}$ do not form a basis for $X_2^n$. For example, let us take $N=6$, there is indeed the identity
\bea 
X_2^2=-15+2! X_4
\eea 
but we find that 
\bea 
X_2^3 &=&3!X_6+15 X_2+12(\psi_1\psi_2+\psi_1\psi_6+\psi_3\psi_4+\psi_4\psi_5+\psi_5\psi_6) \nn \\
&&\qquad -4(\psi_1\psi_4+\psi_2\psi_4+\psi_3\psi_6).
\eea 
Let us focus on the second last term in $X_2^n$
\bea 
X_2^n \approx \dots c_1 X_{2n-4}+n! X_{2n},\quad c_1= (n-2)!{n \choose 2} {N \choose 2}\,,
\eea 
where $c_1$ is computed as follows. We need to pick 2 $X_2$  out of $n$  and contract them, and the $(n-2)$  $X_2$'s remain not contracted and gives $(n-1)!$ $X_{2n-4}$. Notice that the subleading term is $X_{2n-4}$ instead of $X_{2n-2}$, since if we contract one fermion in $X_2$ to get
\bea 
\psi_1\psi_2 \psi_1\psi_3 \mapsto \psi_3\psi_2\,,
\eea 
there is going to be another contraction that gives
\bea 
\psi_1\psi_3\psi_1\psi_2 \mapsto \psi_2\psi_3\ .
\eea 
The two outcomes simply cancel with each other. The main conclusion of this computation is, given that $X_{2n} \sim N^{2n}$, the subleading terms can be safely neglected and approximate $X_{2n}$ by $X_2^n$. So in the large $N$ limit, we can use the $G,\Sigma$ trick to compute the fermionic integral
\bea 
I_q(a) &=&\int \d^N \psi_i \exp (a\sum_A \psi_A),\quad A=\{1\leq a_1<\dots<a_q\leq N\} ,\\
&\approx&\int \d^N \psi_i \, e^{a\frac{G^{\frac{q}{2}}}{\frac{q}{2}!}} e^{\im \sigma (G-\sum_{i<j}\psi_i\psi_j)}\d G\d \sigma \\
&=& \int \d G \d \sigma I_2(-\im \sigma )e^{a\frac{G^{\frac{q}{2}}}{\frac{q}{2}!}}e^{\im \sigma G} =I_2(\im \pa_{G})e^{a\frac{G^{\frac{q}{2}}}{\frac{q}{2}!}} |_{G=0}\, .
\eea 
where the function $I_2$ is defined in \eqref{i2}. We can evaluate this expression and we expect the half-wormhole contributions to be similar as the $0$-SYK model
\bal 
z\approx\langle z\rangle+\Theta\,, \qquad \Theta=\int d^N \psi\, e^{ -\int_0^T dt\, \frac{1}{2}\sum_i^N \psi_i \partial_t \psi_i+\im^{q/2} \int_0^T dt \sum_A (J_A-\mu) \psi_A  }\ .
\eal  
The detailed analysis is similar to the Brownian SYK model as we have shown above, but it is not particularly illuminating, so we omit them here. 

In the next section, we instead consider a modified SYK-like model where half-wormhole saddle can be verified explicitly.

\subsection{ Random coupling from product of Grassmann variables $J_A^{(0)}=J\prod_i\theta_{A_i}$}
A modified SYK-like model dubbed as partially disorder-averaged SYK model is proposed in~\cite{Goto:2021wfs}. In this model, the random coupling $\widetilde{J}_A$ consists of two pieces
\bea 
\widetilde{J}_A=J_A+J_A^{(0)}
\eea 
where $J_A$ is the standard random coupling of the SYK model while $J_A^{(0)}$ is specially chosen as
\bea 
J_{i_i\dots i_q}^{(0)}= \im^{q/2} q!\mu\, \theta_{i_1}\dots \theta_{i_q},\quad \text{with}\quad \{\theta_i,\theta_j\}={2}\delta_{ij}\label{theta}
\eea 
so we can think of it as coupling the fermions $\psi_i$ in the original model with some background Majorana fermions $\q_i$ (or non-dynamical fermions living in another universe~\cite{Goto:2021wfs}). Note that $J_A^{(0)}$ is not a c-number which is different from our models studied in the previous section. 

\subsubsection{0d model}
Let us first consider the $0$-dimensional model to see the difference explicitly. In this case the integral \eqref{z0}  can be written as
\bea \label{z1theta}
z&=&\int \d^N \psi \exp (\im^{q/2}\sum \widetilde{J}_{i_1\dots i_q}\psi_{i_1\dots i_q})\, \nn\\
&=&\int \d^N \psi \exp \(\im^{q/2}\sum_A J_A\psi_A+\mu(\sum_i \theta_i\psi_i)^q \).
\eea 
The averaged quantity $\langle z\rangle$ 
\bea \label{zat}
\langle z\rangle=\int \d^N \psi \exp \(\mu(\sum_i \theta_i\psi_i)^q\)\, ,
\eea 
can be computed in two ways. One can integrate out the fermions $\psi_i$ directly. The result is 
\bea 
\langle z\rangle= \frac{\mu^{N/q} N!}{(N/q)!} \int \d^N\psi (\theta_1\psi_1)\dots (\theta_N \psi_N)=\left[\prod_i \theta_i\right]\frac{\mu^{N/q} N!}{(N/q)!}\equiv\left[\prod_i \theta_i\right]\mathfrak{m}_p\, \label{MP} .
\eea 
Note that $z$ is not a c-number and depends on the background fermions living in other universe. Here we will not think of this as a problem but a feature since the model is not exactly the original SYK model. Alternatively we can compute this average quantity by the $G,\Sigma$ trick:
\bea 
&&G_\sigma=\sum_i \theta_i\psi_i\, ,\nn \\
&&\langle z\rangle=\int \d^N \psi \int \d G_\sigma \frac{\d \Sigma_\sigma}{2\pi} e^{\im [\Sigma_\sigma( G_\sigma-\sum_i \theta_i\psi_i)]}e^{ \mu G_\sigma^q} \nn \\
&&\qquad = \left[\prod_i \theta_i\right]\int \d G_\sigma \frac{\d \Sigma_\sigma}{2\pi}\Sigma_\sigma^N e^{\im \Sigma_{\sigma} G_{\sigma}+\mu G_\sigma^q} \label{eaz1} \\
&&\qquad = \left[\prod_i \theta_i\right] (\pa_{G_\sigma})^N e^{\mu G_\sigma^q}|_{G_\sigma=0}=\left[\prod_i \theta_i\right]\frac{\mu^{N/q}N!}{(N/q)!}\,.
\eea 
One can also use the effective action \eqref{eaz1} to derive the large $N$ result of \eqref{zat} as shown in \cite{Goto:2021wfs}. We will not repeat that analysis here. Instead,  we would like to consider the half-wormhole saddle of $z$ 
\bea \label{hwze}
z\approx \langle z\rangle+\Theta
\eea 
as we did in last section. The subtlety is that as we stressed $z$ is not a c-number so the approximation \eqref{hwze} is in the sense
\bea 
\langle [z-(\langle z\rangle+\Theta)]^2\rangle \approx 0\,,
\eea  
which is a c-number due to \eqref{theta} is small. Let us proceed by computing the averaged quantity $\langle z^2\rangle$
\bea\label{z2kenta}
\langle z^2\rangle &=& \int \d^{2N}\psi \, \exp\({\frac{\t^2}{q!}} (\sum_i \psi_i^L\psi_i^R)^q+\mu(\sum_i\theta_i \psi_i^L)^q+\mu(\sum_i\theta_i \psi_i^R)^q\) \nn \\
&=&\int \d^{2N}\psi \sum_k \(\frac{\t^2}{q!}\)^k \sum_{{i_1<\dots < i_{kq}}}(\psi^{LR}_{i_1}\dots\psi^{LR}_{i_{qk}})\frac{(qk)!}{k!} \(\frac{(N-kq)!}{(N/q-k)!}\)^2 \nn\\
&&\mu^{2p-2k} \sum_{j_1<\dots < j_{N-qk}\neq \{i\}}(\theta_{j_1}\psi_{j_1}^L)\dots(\theta_{j_{N-qk}}\psi^L_{j_{N-kq}}) (\theta_{j_1}\psi_{j_1}^R)\dots(\theta_{j_{N-k}}\psi^R_{j_{N-kq}})  \nn \\
&=&\int \d^{2N}\psi \sum_k \(\frac{\t^2}{q!}\)^k\mu^{2p-2k}\frac{(qk)!}{k!} \(\frac{(N-kq)!}{(N/q-k)!}\)^2{N\choose kq}\psi^{LR}_1\dots \psi_N^{LR} \\
&=&\sum_k \(\frac{\t^2}{q!}\)^k\mu^{2p-2k}\frac{(qk)!}{k!} \(\frac{(N-kq)!}{(N/q-k)!}\)^2{N\choose kq}\nn \\
&=&\sum_k^p \t^{2k}\mu^{2(p-k)}c_k\mathfrak{m}_{p-k}^2\equiv \sum_k \mathfrak{z}_2^{(k)}\, , \label{z2ktheta}
\eea 
where $c_k$ of defined in \eqref{ck} and $\mathfrak{m}_p$ is defined in \eqref{MP}. The result \eqref{z2kenta} is in the same form of \eqref{z2our}. So the analysis of the half-wormhole saddle will be similar; we insert the a suitable identity to \eqref{z1theta}
\bea 
z&=&\int \d^N \psi \exp (\im^{q/2}\sum \widetilde{J}_{i_1\dots i_q}\psi_{i_1\dots i_q}) \int \d G_{\sigma} \delta(G_\sigma- \sum_{i}\theta_i\psi_i)\exp(\frac{\mu}{q!}(G_\sigma^q-(\sum_{i}\theta_i\psi_i)^q))\nn \\
&=&\int \d^N \psi \frac{\d \Sigma_\sigma \d G_\sigma}{2\pi \im }\exp(\im^{q/2}\sum_A J_A\psi_A+\Sigma_\sigma \sum_i\theta_i\psi_i)\exp(-\Sigma_\sigma G_\sigma+\frac{\mu}{q!}G_\sigma^q)\, .
\eea 
Following the arguments below \eqref{hidentity} one can obtain the half-wormhole saddle\footnote{Here the factor $\prod_i \theta_i$ should be present.}
\bea 
\Theta=\left[\prod_i \theta_i\right]\int \d^N \psi \exp(\im^{q/2}\sum_A J_A\psi_A)\, .
\eea 
Then it is easy to find that the half-wormhole saddle satisfies
\bea 
\langle \Theta\rangle=0,\quad \langle \Theta^2\rangle=\langle \Theta z\rangle=\mathfrak{z}_2^{(p)}\, ,
\eea 
so the approximation \eqref{hwze} will be sufficient if $\mathfrak{z}_2^{(p)}$ is the dominant term in \eqref{z2ktheta} as we have shown in last section. When $\mathfrak{z}_2^{(p)}$ is not the dominant term we have to consider the contribution of fluctuation of $\Sigma_\sigma$. To finish our analysis of the half-wormhole saddle for $z$, let us redo the computation of  $\langle z^2\rangle$ with the $G,\Sigma$ trick. We need introduce three $G$ variables
\bea 
G_{LR}=\frac{1}{N}\sum_i \psi_i^L \psi_i^R,\quad G_L=\frac{1}{{N}}\sum_i\theta_i\psi_i^L,\quad G_R=\frac{1}{{N}}\sum_i\theta_i \psi_i^R
\eea 
then $\langle z^2\rangle$ can be written as
\bea 
\langle z^2\rangle &=& \int \d^{2N}\psi \prod_{a} \d G_{a}\exp \(\frac{N}{q}(t G_{LR}^q+u G_L^q+u G_R^q)\) \int \prod_a \frac{d \Sigma_a}{2\pi \im/N } \exp(-\Sigma_{LR}(NG_{LR}-\sum_i\psi_i^L \psi_i^R))\nn \\
&& \exp(-\Sigma_{L}(NG_{L}-\sum_i\theta_i\psi_i^L ))\exp(-\Sigma_{R}(NG_{R}-\sum_i\theta_i\psi_i^R ))\nn \\
&=&\int [\prod_{a} \d G_{a} \frac{d \Sigma_a}{2\pi \im /N}]\exp\(N( \frac{t}{q} G_{LR}^q+\frac{u}{q} G_L^q+\frac{u}{q} G_R^q-\sum_a\Sigma_a G_a)\)(\Sigma_{LR}+\Sigma_L\Sigma_R)^N\, ,
\eea 
where in order to have a well-defined large $N$ scaling we have introduced 
\bea 
t=\frac{\t^2}{(q-1)!}N^{q-1},\quad u=q\mu N^{q-1}.
\eea 
The saddle point equations are
\bea 
&& tG_{LR}^{q-1}=\Sigma_{LR},\quad  u G_L^{q-1}=\Sigma_L,\quad  u G_R^{q-1}=\Sigma_R \, ,\\
&& G_{LR}=\frac{1}{\Sigma_{LR}+\Sigma_L\Sigma_R},\quad G_L=-\frac{ \Sigma_R}{\Sigma_{LR}+\Sigma_L\Sigma_R},\quad G_R=-\frac{ \Sigma_L}{\Sigma_{LR}+\Sigma_L\Sigma_R}\, .
\eea 
The obvious solutions are the ``wormhole" saddles with 
\bea 
G_L=G_R=\Sigma_L=\Sigma_R=0,
\eea 
which corresponds to $\mathfrak{z}_2^p$. There are also other saddles corresponding to other $\mathfrak{z}_2^k$. For the simplest case $q=2$, these solutions can be written explicitly. The ``wormhole" saddles are
\bea 
&&G_L=G_R=\Sigma_L=\Sigma_R=0, G_{LR}=\pm \frac{1}{\sqrt{t}},\quad \Sigma_{LR}=\pm\sqrt{t},\\
&&\langle z^2\rangle_{\text{WH}}= e^{-\frac{N}{2}}t^{N/2}\,,
\eea 
which do not depend on $\mu$ and the other four solutions are
\bea 
&&\Sigma_L=\Sigma_R=u G_L =u G_R =\pm\sqrt{\frac{u^2-t}{u}},\quad \Sigma_{LR}=\frac{t}{u},\quad G_{LR}=\frac{1}{u}\\
&&\Sigma_L=u G_L=-\Sigma_R=-u G_R=\pm\sqrt{\frac{u^2-t}{u}},\quad \Sigma_{LR}=-\frac{{t}}{u},\quad  G_{LR}=-\frac{1}{u}\, ,\\
&& \langle z^2\rangle_{\text{new}}=e^{-\frac{N}{2}(2-\frac{t}{u^2})}u^N
\eea 
Apparently when $u \rightarrow \infty$, $\Sigma_{LR},G_{LR}\rightarrow 0,$ then we expect that in this limit the dominant saddle will correspond to $\mathfrak{z}_2^0$ since in this limit saddle point value does not depend on $t$.  Comparing these two saddle values we find
\bea 
\frac{\langle z^2\rangle_{\text{WH}}}{\langle z^2\rangle_{\text{new}}}=\exp\(\frac{N}{2}(1-x+\log x)\)\leq 1,\quad x=\frac{t}{u^2} \, .
\eea 
Note that when $x=1$ such that $\langle z^2\rangle_{\text{WH}}=\langle z^2\rangle_{\text{new}}$ the new saddle just reduces to the wormhole saddle. Therefore it implies that the new saddle always dominates.

This new saddle is named as ``unlinked half-wormhole" in~\cite{Goto:2021wfs} to distinguish it from the half-wormhole saddle which was found in \cite{Saad:2021rcu}. One interpretation of this new saddle is that it is the analogue of the disconnected saddle in this model; indeed, we do not find other disconnected saddle with $G_{LR}=0$, $\S_{LR}=0$ and $G_{L/R}, \S_{L/R}\neq 0$, in addition, this saddle is present only when $u\neq 0$, and this saddle is more and more important as $u$ increases.

The analysis of the half-wormhole saddle for $z^2$ will be similar to one we did in last section so we will not repeat here.

\subsubsection{1d model}
 Now we come back to the 0+1d model that is a variant of the Brownian SYK model. Let us begin by deriving the wormhole saddle of $\langle z^2 \rangle$\footnote{Here we have assumed the large $N$ limit, the exact treatment can be found in \cite{Saad:2021rcu}}
\bea 
&& z_L z_R = \int \d^{2N}\psi \exp\left\{-\int_{0}^T \d t\frac{1}{2}\sum_i(\psi_i^L\pa_t \psi_i^L+\psi_i^R\pa_t \psi_i^R)+\im^{q/2}\int_0^T\d t\sum_A \tilde{J}_A (\psi_A^L+\psi_A^R)   \right\} \nn \\
&&\langle z_L z_R\rangle=\int \d^{2N} \psi \exp\left\{ -\int_{0}^T \d t\frac{1}{2}\sum_i(\psi_i^L\pa_t \psi_i^L+\psi_i^R\pa_t \psi_i^R)+ \right. \nn \\
&&\qquad \qquad \qquad \left. \int_{0}^T \d t \({\frac{\t^2}{q!}} (\sum_i \psi_i^L\psi_i^R)^q+\mu(\sum_i\theta_i \psi_i^L)^q+\mu(\sum_i\theta_i \psi_i^R)^q\)+\t^2E_0T\right\}\, \label{brownian2}. \\
&&\qquad \qquad=\int \d^{2N} \psi [\prod_{a} \d G_{a} \frac{d \Sigma_a}{2\pi \im }]\exp\left\{ -\int_{0}^T \d t\frac{1}{2}\sum_i(\psi_i^L\pa_t \psi_i^L+\psi_i^R\pa_t \psi_i^R)+\t^2TE_0 +\right. \nn \\
&&\qquad \left. \int_{0}^T \d t \(  \frac{\t^2}{q!}G_{LR}^q+\mu G_L^q+\mu G_R^q -\sum_a\Sigma_a G_a+\sum_i[\Sigma_{LR}\psi_i^L\psi_i^R+\Sigma_L \theta_i\psi_i^L+\Sigma_R \theta_i\psi_i^R]\) \right\}\,,\nn \\
\eea 
where $E_0={ N\choose q }$ is the constant term coming from $\psi_A^{L(R)} \psi_A^{L(R)}=(-1)^{\frac{q}{2}}$. As explained in \cite{Saad:2021rcu}, we can focus on the time-independent saddles then the fermions can be simply integrated out. The result is \footnote{This is result is different from the one derived in \cite{Goto:2021wfs}. It seems that they used a wrong formula for the fermion integral.}
\bea 
\langle z_L z_R\rangle &=&e^{T \t^2 E_0}\int[\prod_{a} \d G_{a} \frac{d \Sigma_a}{2\pi \im }]e^{TN(\frac{t}{q}G_{LR}^q+\frac{u}{q} G_L^q+\frac{u}{q} G_R^q -\sum_a\Sigma_a G_a)}[\cosh(T\sqrt{\Sigma_L^2+\Sigma_R^2-\Sigma_{LR}^2})]^N. \nn \\
&=&\int[\prod_{a} \d G_{a} \frac{d \Sigma_a}{2\pi \im }]e^{\t^2 TE_0}e^{S_{eff}}
\eea 
For general $T$, the saddle equation is very hard to solve due to the complicity of $\cosh$ function. However the equations simplify in the large $T$ limit because of the following approximations
\bea 
\log(\cosh(T\sqrt{\Sigma_L^2+\Sigma_R^2-\Sigma_{LR}^2}))\approx \pm \im T \sqrt{\Sigma_{LR}^2-\Sigma_{L}^2-\Sigma_{R}^2}\, .
\eea 
Then in this limit the effective action becomes
\bea 
S_{eff}=TN(\frac{t}{q}G_{LR}^q+\frac{u}{q} G_L^q+\frac{u}{q} G_R^q -\sum_a\Sigma_a G_a)\pm \im NT\sqrt{\Sigma_{LR}^2-\Sigma_{L}^2-\Sigma_{R}^2}\, ,
\eea 
and corresponding saddle point equations are
\bea 
&& tG_{LR}^{q-1}=\Sigma_{LR},\quad u G_L^{q-1}=\Sigma_L,\quad u G_R^{q-1}=\Sigma_R  ,\\
&&G_{LR}=\pm \frac{\im  \Sigma_{LR}}{\sqrt{\Sigma_{LR}^2-\Sigma_{L}^2-\Sigma_{R}^2}}\, , \\
&&G_L=\mp \frac{\im  \Sigma_{L}}{\sqrt{\Sigma_{LR}^2-\Sigma_{L}^2-\Sigma_{R}^2}},\quad G_R=\mp \frac{\im  \Sigma_{R}}{\sqrt{\Sigma_{LR}^2-\Sigma_{L}^2-\Sigma_{R}^2}}\, .
\eea 
So the wormhole saddle still presents \cite{Saad:2018bqo}
\bea 
&&G_L=G_R=\Sigma_L=\Sigma_R=0,\quad G_{LR}=\pm \im ,\\
&&e^{S_{eff}}\Big|_{\text{WH}}=e^{\im^qTN  \frac{t}{q}}.
\eea 
The unlinked half-wormhole saddle is:
\bea 
&&G_{LR}=\Sigma_{LR}=0,\quad G_L=\sin\alpha,\quad G_R= \cos\alpha,\\
&&e^{S_{eff}}\Big|_{\text{unlink}}=e^{TN  \frac{u}{q}(\cos^q\alpha+\sin^q\alpha)}\leq e^{S_{eff}}\Big|_{\text{unlink},\alpha=0,\pi/2}=e^{TN  \frac{u}{q}}\,
\eea 
where the relation
\bea 
G_L^2+G_R^2-G_{LR}^2=1\,,
\eea 
is fulfilled and $\alpha$ satisfies
\bea 
\cos\alpha=\pm \frac{\cos^{q-1}\alpha}{\sqrt{\cos^{2q-2}\alpha+\sin^{2q-2}\alpha}}.
\eea 
In the late time $(T\rightarrow \infty)$, there is indeed a wormhole saddle so it possible to include a linked half-wormhole saddle for $z$. We also assume that the half-wormhole saddle is time independent since the wormhole saddle is time independent. Then the analysis is completely same as the one for the 0-dimensional model. So the half-wormhole saddle will be given by 
\bea 
&&\Theta=\left[\prod_i \theta_i\right]\int \d^N \psi \exp(T\im^{q/2}\sum_A J_A\psi_A)\, ,\\
&&\langle \Theta^2\rangle\approx \langle \Theta z\rangle\approx \langle z_L z_R\rangle|_{\text{Wormhole saddle}}\, .
\eea

\section{Discussion}
\label{diss}
In this paper we consider the half-wormhole proposal in some statistical models and simple SYK-like models. We showed that in all statistical models we have consider the half-wormhole conjecture \eqref{Phi} is valid almost for all the distributions except for some special cases where the mean value of the random variable vanish. In the 0-dimensional SYK model which is introduced in \cite{Saad:2021rcu} we have shown that the half-wormhole construction depends on the distribution of the couplings. When the mean value of the coupling is very large then only the disconnected saddles dominate therefore the correlation functions automatically factorize. If the mean value is very small such that only the wormhole  saddles dominate then factorization can be restored by adding half-wormhole saddles. When the disconnected saddles and wormhole saddles are comparable, we have to modified the half-wormhole saddle to restore the factorization. We also generalized $G,\Sigma$ trick to compute $\langle z\rangle$. As a by-product, we can construct a a new saddle, the single half-wormhole saddle, for $z$. Moreover we argued if the random couplings satisfy a general distribution, new half-wormhole saddles can be constructed \footnote{Interestingly, irrelevant deformation of 0-SYK model is studied in \cite{Das:2022uhj} where they show after deformation half-wormhole saddle survives. It is very possible that our new half-wormholes will also survive under the same irrelevant deformation.}. We also generalize the construction of half-wormhole saddles to (modified) Brownian SYK model.

\subsection*{Partially averaged models and spacetime branes}
The meaning of higher cumulants of the random coupling can be understood from the idea of Coleman's \cite{Coleman:1988cy} and Giddings's and Strominger's \cite{Giddings:1988wv,Giddings:1988cx}. Just as we showed they are related to the non-local interaction induced by the spacetime wormholes. However the first cumulant or the mean value seems to be puzzling. In \cite{Blommaert:2021gha,Goto:2021wfs}, ensemble theories with non-vanishing mean value random couplings are also consider where they call such models partially averaged models. In these models, the mean values of random couplings can be understood as external sources or spacetime branes which describe the non-perturbative corrections. It is shown in \cite{Blommaert:2021fob} by fine tuning these non-perturbative corrections the JT gravity can factorize for all orders. But it seems that the original half-wormhole \cite{Saad:2021rcu} constructed in 0-SYK model is not related to the branes but just a result of applying $G,\Sigma$ trick in the non-averaged theory. However we can understand this construction in an opposite way: the half-wormhole is constructed by adding eigenbranes \cite{Blommaert:2019wfy,Goel:2020yxl} in the averaged theory. This opposite point of view can also be viewed as an explicit realization of the idea \cite{Saad:2021uzi} about factorization. 
\subsection*{Standard SYK model}
There are already proposals \cite{Saad:2021uzi,Mukhametzhanov:2021hdi, Mukhametzhanov:2021nea,Goto:2021wfs} of the half-wormhole saddle for $z^2$ in the standard SYK model. But due to technical difficulty it has not been confirmed. It would be interesting to generalize our single half-wormhole saddle to the SYK model since $z$ is much simpler than $z^2$.  It would be also interesting to generalize the new half-wormholes we found in section \ref{SYK0} to the standard SYK model and understand their possible relations to saddles in JT gravity.
\subsection*{Non-trivial saddles and Null states}
In the simple 0-dimensional SYK model and Brownian SYK model, we find some non-trivial saddles whose on-shell values are comparable with one of the trivial self-averaging saddles. It would be very interesting if such non-trivial saddle also exists in the standard SYK model. It implies that there are also some non-trivial solutions in the dual (deformed) JT gravity. In the semiclassical physics, this coexistence of bulk description can be understood as the consequence of null states. In \cite{Blommaert:2022ucs}, the null states of (deformed) JT gravity are proposed. However these null states do not show up in the dual matrix model. It seems to be promising to identify these null states in the SYK model.

\begin{acknowledgments}
We thank many of the members of KITS for interesting related discussions. We also want to thank Kenta Suzuki for comments on a draft of this paper. 
CP  is supported by the Fundamental Research Funds for the Central Universities, by funds from the University of Chinese Academy of Science (UCAS), and funds from the Kavli Institute for Theoretical Science (KITS). JT is supported by  the National Youth Fund No.12105289 and funds from the UCAS program of special research associate.
\end{acknowledgments}

\appendix

\section{Explicit examples: simple observables }
Thanks to the central limit theorem (CLT), the simplest choice of $Y(X)$ is just the summation of $N$ independent and identical random samples. We will first check proposal with three explicit distributions: the Gaussian distribution, the exponential distribution and the Poisson distribution and then give a general proof for general cases. Readers who are bored with these examples can jump into the general proof directly. \par
Let $Y$ to be a summation of $N$ independent and identical random variables, i.e.
\bea
Y=\sum_{i=1}^N X_i, \quad Y^2=\sum_{i,j}^N X_i X_j.
\eea

$\bullet$ {\bf Gaussian distribution}

The PDF of Gaussian distribution $\mathcal{N}(\mu,t^2)$ is 
\bea \label{PDFG}
P(x)=\frac{1}{t\sqrt{2\pi}}e^{-\frac{1}{2}(\frac{x-\mu}{t})^2}.
\eea
Given \eqref{PDFG} one can straightforwardly compute the averaged quantities
\bea
&&\langle X_i\rangle=\mu,\quad \langle e^{\text{i}k_i X_i}\rangle=e^{-\frac{k_i^2 t^2}{2}+\text{i}k_i\mu},\quad \langle Y^2\rangle=N t^2+N^2\mu^2,\\
&&\frac{\langle Y e^{\sum_i \im k_i X_i}\rangle}{\langle e^{\im \sum_i k_i X_i}\rangle}=\sum_i\frac{\langle x_i e^{\im k_i x_i}\rangle}{\langle e^{\im k_i x_i}\rangle}\equiv \sum_i k_i[1],\quad k_i[1]=(u+\im k_i t^2).
\eea
where we have defined
\bea\label{kn}
 k_i[n]=\frac{\langle x^n_i e^{\im k_i x_i}\rangle}{\langle e^{\im k_i x_i}\rangle}.
\eea
Let us introduce another convenient quantity
\bea \label{tildek}
&&\widehat{k_i[n]^m}=\frac{1}{2\pi}\int dk_i \frac{e^{-\im k_i X_i}}{P(X_i)}\langle e^{\im k_i x_i}\rangle k_i[n]^m,\\
&&\widehat{k_i[1]}=X_i,\quad \wh{k_i[1]^2}=X_i^2-t^2,\quad \dots
\eea
then the half-wormhole can be written as
\bea 
\Phi &=&\sum_i \widehat{k_i[1]^2}+\sum_{i\neq j}\widehat{k_i[1]}\widehat{k_j[1]} =\sum_i(X_i^2-t^2)+\sum_{i\neq j}X_iX_j, \\
&=& Y^2-N t^2\label{ph1} 
\eea
Substituting into \eqref{error} one can computed the error and the ration $\rho$ directly
\bea
\text{Error}&=&Y^2-\(Nt^2+N^2 \mu^2-N^2\mu^2+Y^2-Nt^2\) \label{error1}\\
&=&0.
\eea
The proposal is exact as expected. 

$\bullet$ {\bf Exponential distribution}

The PDF of exponential distribution is given by
\bea
P_{\lambda}(x)=\begin{cases}\lambda e^{-\lambda x},\quad & x\geq0, \\
0, & x<0, \end{cases}
\eea
and the moments are given by
\bea
\langle x^n\rangle=\frac{n!}{\lambda^n}.
\eea
The relevant averaged quantities are
\bea
&&\langle X_i\rangle=\frac{1}{\lambda},\quad  \langle e^{\im k X_i}\rangle=\frac{\lambda}{\lambda-\text{i}k},\quad \langle Y^2\rangle=\frac{N(N+1)}{\lambda^2}.
\eea
From the example of Gaussian distribution we have shown that to compute $\Phi$ \eqref{ph1} and the error \eqref{error1} we only need to compute 
\bea
\wh{k_i[1]}=X_i,\quad \wh{k_i[1]^2}=\frac{X_i^2}{2},
\eea
which lead to
\bea 
&&\text{Error}=\sum_i\(\frac{X_i^2}{2}-\frac{1}{\lambda^2}\),\quad \rho\approx\frac{\langle \text{Error}^2\rangle}{\langle Y^2\rangle^2}\sim \frac{1}{N^3}.
\eea
So the proposal is correct.

$\bullet$ {\bf Poisson distribution}

Next let us examine the proposal for a discrete probability distribution: the Poisson distribution. The PDF is 
\bea
 P_\lambda(k)=\frac{e^{-\lambda}\lambda^k}{k!},\quad k=0,1,2,\dots \label{poissonpdf}
\eea
and the moments are given by
\bea
\langle x^n\rangle=B_n(\lambda), \label{poissonms}
\eea
where $B_n(\lambda)$ is the Bell polynomial.
The relevant averaged quantities can be easily computed
\bea
&&\langle X_i\rangle={\lambda},\quad  \langle e^{\im k X_i}\rangle=e^{-\lambda+e^{\im k} \lambda},\quad \langle Y^2\rangle=N\lambda+N^2\lambda^2,\\
&&k_i[1]=e^{\im k_i}\lambda.
\eea
The computation of $\wh{k_i[1]}$ and $\wh{k_i[1]^2}$ is a little subtle and needs some explanation. According to the definition \eqref{tildek}, we have
\bea
\widehat{k_i[1]}&=&\frac{\lambda}{2\pi P(X_i)}\int dk_i {e^{-\im k_i X_i}}{}\langle e^{\im k_i x_i}\rangle e^{\im k_i},\\
&=&\frac{\lambda}{P(X_i)} P(X_i-1)=X_i,
\eea
where in the second line we have used the fact that the inverse Fourier transformation of the characteristic function is the PDF
\bea
\frac{1}{2\pi} \int dk_i {e^{-\im k_i X_i}}{}\langle e^{\im k_i x_i}\rangle=P(X_i).
\eea
Similarly one can derive
\bea
 \widehat{k_i[1]^2}&=&\lambda^2\frac{P(X_i-2)}{P(X_i)}=X_i(X_i-1).
\eea
Then using \eqref{error1} we can obtain the error
\bea
\text{Error}=\sum_i(X_i-\lambda),\quad \rho\approx\frac{\langle \text{Error}^2\rangle}{\langle Y^2\rangle^2}\sim\frac{1}{N^{3}},
\eea
so the proposal is correct while the effective parameter is $\lambda N$ instead of $N$.

\section{Explicite Examples: composite observables}\label{exam}
\subsection*{Gaussian distribution}
We still start from the simple model 
\bea
&& Y=\sum_{i=1}^N X_i^2,\quad \langle X_i^2\rangle=t^2,\quad Y^2=\sum_{i,j} X_i^2 X_j^2 .
\eea
It is easy to evaluate 
\bea
&&\langle e^{\im k_i X_i}\rangle=\exp(-\sum_i \frac{k_i^2t^2}{2}),\\
&&\langle Y e^{\im \sum_{i}k_i X_i}\rangle/\langle e^{\im \sum_i k_i X_i}\rangle=\sum_i (-k_i^2 t^4+t^{2}).
\eea
Then we find 
\bea
\Phi=\sum_{i}\left(X_{i}^{4}-4t^{2}X_{i}^{2}+2t^{4} \right)+\sum_{i\neq j} X_{i}^{2}X_{j}^{2},
\eea
and the error is given by
\bea
\text{Error}=4Nt^{4}- 4t^2\sum_i X_i^2 .
\eea
From this we can easily find that the proposal is correct.

Let us consider 
\bea
&&Y=\sum_i X_i^3,\quad \langle X_i^3\rangle=0,\\
&&Y^2=\sum_{i,j}X_i^3 X_j^3,\quad \langle Y^2\rangle=15N t^6.
\eea
Then we find can evaluate
\bea
\langle Y e^{\im k_i X_i}\rangle/ \langle  e^{\im k_i X_i}\rangle=\sum_i \text{i}k_it^4(3-k_i^2 t^2),
\eea
and 
\bea
\Phi=\sum_i (X_i^6-9 t^2 X_i^4+18 t^4 X_i^2-6 t^6)+\sum_{i\neq j} X_i^3 X_j^3.
\eea
So that the error is given by
\bea
\text{Error}=\sum_i ( 9t^2 X_i^4-18 t^4 X_i^2-9t^6),
\eea
the leading order of $\langle \text{Error}^2\rangle$ is
\bea
0 N^2.
\eea

Let us consider 
\bea
&&Y=\sum_i X_i^4,\quad \langle X_i^4\rangle=3t^4,\quad Y^2=\sum_{i,j}X_i^4 X_j^4
\eea
then we find can evaluate
\bea
\langle Y e^{\im k_i X_i}\rangle/ \langle  e^{\im k_i X_i}\rangle=\sum_i t^{4}(k_{i}^{4}t^{4}-6k_{i}^{2}t^{2}+3)
\eea
and 
\bea
\Phi=8\sum_i (3t^8-12t^6 X_i^2+9 t^3 X_i^4-2t^2 X_i^6)+Y^2
\eea
such that the error is 
\bea
\text{Error}=-8\sum_i (15t^8-12t^6 X_i^2+9 t^3 X_i^4-2t^2 X_i^6).
\eea
We can also find the leading order of $\langle \text{Error}^{2}\rangle$ is zero.

In general let us consider 
\bea
&& Y=\sum_i e^{m X_i},\quad\langle e^{m X_i}\rangle=e^{\frac{m^2 t^2}{2}},\quad Y^2=\sum_{i,j}e^{m X_i}e^{m X_j},
\eea
and
\bea 
\Phi=\sum_i(e^{m(-m t^2+2 X_i)}-e^{2m X_i})+Y^2,
\eea
such that
\bea 
\text{Error}=\sum_i (e^{-m^2 t^2}-1)(e^{2m^2t^2}-e^{2m X_i}),
\eea
\bea
\langle \text{Error}^2\rangle=Ne^{4m^2 t^2}(e^{4m^2t^2}-1)(e^{-m^2 t^2}-1)^2,
\eea
which is sub-dominant comparing to $\langle Y^{2}\rangle^{2}$.

\subsection*{Exponential distribution}
First let us choose
\bea 
&& Y=\sum_i X_i^2,\quad \langle X_i^2\rangle=\frac{2}{\lambda^2}, \quad Y^2=\sum_{i,j}X_i^2X_j^2.
\eea
Correspondingly we find
\bea
 \frac{\langle Y e^{\im k X_i} \rangle }{\langle e^{\im k X_i} \rangle}=-\frac{2}{(k+\text{i}\lambda)^2}
\eea
and 
\bea
\Phi&=&-\frac{5}{6}\sum_i X_i^4+\sum_{i, j}X_i^2X_j^2.
\eea
So the error is given by
\bea
&& \text{Error}=\sum_i\(\frac{5 X_i^4}{6}-\frac{20}{\lambda^4}\),\\
&& \langle \text{Error}^2\rangle= \frac{25 N}{36 \lambda^8}(8!-4!^2).
\eea
In general, let us consider 
\bea
&&Y=\sum_i e^{\beta X_i},\quad \langle e^{\beta X_i}\rangle=\frac{\lambda}{\lambda-\beta},\quad Y^2=\sum_{i,j}e^{\beta X_i}e^{\beta X_j}.
\eea
Then we can obtain
\bea
 \Phi&=& \sum_i \(e^{\beta X_i}(1-e^{\beta X_i}+\beta X_i)    \)+\sum_{i,j}e^{\beta X_i}e^{\beta X_j}.
\eea
So the error is 
\bea 
&& \text{Error}=\sum_i\( e^{2\beta X_i}-e^{\beta X_i}(1+\beta X_i)  \)-\frac{N \beta^2 \lambda}{(\beta-\lambda)^2(\lambda-2\beta)}, \\
&&\langle  \text{Error}^2\rangle=N\(   \frac{2\beta^4(3\lambda-8\beta)}{(2\beta-\lambda)^3(4\beta-\lambda)(\lambda-3\beta)}   -\frac{\lambda^2\beta^4}{(\lambda-\beta)^4(\lambda-2\beta)^2}        \),
\eea
given that the condition
\bea
\beta<\frac{\lambda}{4} ,
\eea
to ensure that all the integrals to be convergent.

\subsection*{Poisson distribution}
Let us choose
\bea
&& Y=\sum_i X_i^2,\quad \langle X_i^2\rangle=\lambda^2+\lambda,\quad Y^2=\sum_{i,j} X_i^2X_j^2
\eea
Then we find
\bea
 \frac{\langle Y e^{\im k_i X_i} \rangle }{\langle e^{\im k_i X_i} \rangle}=\sum_{i}\lambda e^{\im k_i} (1+\lambda e^{\im k_i})
\eea
and
\bea
&& \Phi=\sum_i \(X_i(-3+6X_i-4X_i^2)\)+\sum_{i,j}X_i^2 X_j^2.
\eea
Therefore the error is
\bea 
&& \text{Error}=\sum_i \(X_i (3-6X_i+4 X_i^2)-\lambda(1+6\lambda+4\lambda^2)\) \\
&&\langle \text{Error}^2\rangle=N \lambda  (24 \lambda  (\lambda  (6 \lambda  (\lambda +4)+23)+4)+1) \\
&&\frac{\langle \text{Error}^2\rangle}{\langle Y^2\rangle^2}\sim \frac{1}{N^{3}}
 \eea

In general let us consider
\bea
&&Y=\sum_i e^{\beta X_i},\quad \langle e^{\beta X_i}\rangle=\exp\lambda(e^\beta-1),\quad Y^2=\sum_{i,j}e^{\beta X_i}e^{\beta X_j}.
\eea
Then we find
\bea 
&& \frac{\langle Y e^{\im k_i X_i} \rangle }{\langle e^{\im k_i X_i} \rangle}=e^{\left(e^{\beta }-1\right) e^{\im k_i} \lambda }
\eea
and
\bea
 \Phi&=&\sum_i (\left(2 e^{\beta }-1\right)^{X_i}-e^{2 \beta  X_i})+Y^2.
\eea
Therefore the error is
\bea
&&\text{Error}=\sum_i\( -\left(2 e^{\beta }-1\right)^{X_i}+e^{2 \beta  X_i}-e^{-2 \lambda } \left(e^{e^{2 \beta } \lambda +\lambda }-e^{2 e^{\beta } \lambda }\right) \) \\
&&\langle \text{Error}^2\rangle=-e^{-4 \lambda } N \left(e^{4 e^{\beta } \lambda }-2 e^{\left(e^{\beta }+1\right)^2 \lambda }+e^{2 \left(e^{2 \beta }+1\right) \lambda }-e^{\left(e^{4 \beta }+3\right) \lambda }-e^{4 \left(e^{\beta } \left(e^{\beta }-1\right)+1\right) \lambda }+2 e^{\left(e^{2 \beta } \left(2 e^{\beta }-1\right)+3\right) \lambda }\right) \nn \\
\eea
and 
\bea
 \frac{\langle \text{Error}^2\rangle}{\langle Y^2\rangle^2}\sim \frac{1}{N^{3}}.
\eea

\section{Explicte Examples: generalized statistical models}\label{exam2}
\subsubsection*{The exponential distribution}
First let us consider the exponential distribution. It is straightforward to derive
\bea
&&\langle Y\rangle =\langle \sum_{i\neq j}X_i X_j\rangle=N(N-1)\frac{1}{\lambda^2},\\
&&\langle Y^2\rangle=\frac{2N(N-1)(2N-1)+N^{2}(N-1)^{2}}{\lambda^4}
\eea
and
\bea
\Phi&=&\sum_{i\neq j\neq p\neq q }X_i X_j X_p X_q+4\sum_{i\neq j\neq p}\frac{X_i^2}{2}X_j X_p+2\sum_{i\neq j}\frac{X_i^2}{2} \frac{X_j^2}{2}
\eea
in particular as a consistency check
\bea
\langle \Phi\rangle=N^{2}(N-1)^{2}\frac{1}{\lambda^{4}}. 
\eea
Therefore the error is
\bea
&&\text{Error}=2\sum_{i\neq j\neq p}X_i^2 X_j X_p+\frac{3}{2}\sum_{i\neq j}X_i^2 X_j^2-\langle Y^2\rangle ,\\
&&\langle \text{Error}^2\rangle\sim (4*2*2+16-4*2*4) N^6+\# N^5=\# N^5/\lambda^8. 
\eea

\subsubsection*{Poisson distribution}
First we compute
\bea
&&\langle \sum_{i\neq j}X_i X_j\rangle=N(N-1)\lambda,\\
&&\langle Y^2\rangle=2N(N-1)(1+2(N-1)\lambda)\lambda^2+N^{2}(N-1)^{2}\lambda^{2},
\eea
and 
\bea
 \Phi&=&\sum_{i\neq j\neq p\neq q }X_i X_j X_p X_q+4\sum_{i\neq j\neq p}(X_i^2-X_i) X_j X_p+2\sum_{i\neq j}(X_i^2-X_i) (X_j^2-X_j).
\eea
One can check that
\bea
\langle \Phi\rangle=N^{2}(N-1)^{2}\lambda^{2}.
\eea
Therefore the error is 

\bea\label{PoiE}
\text{Error}=-\langle Y^2\rangle+4\sum_{i\neq j\neq p}X_i X_j X_p+4\sum_{i\neq j}X_i^2 X_j-2 \sum_{i\neq j}X_i X_j
\eea
and it is easy to check (note that only the first two term in \eqref{PoiE} will contribute)
\bea 
\langle \text{Error}^2 \rangle\sim 0 N^6+\# N^5.
\eea

\section{Exact evaluation of $\langle z^4\rangle$ for Brownian SYK at large T}
\label{if1}
We can only consider one $f$ term in the logarithm, we expect it gives the non-trivial contribution in the large $T$ limit. For instance we have 
\begin{align}
   & e^{\frac{\im}{2} \sqrt{(\partial_{G_{14}}+\partial_{G_{23}})^{2}+(\partial_{G_{13}}-\partial_{G_{24}})^{2}+(\partial_{G_{12}}+\partial_{G_{34}})^{2}}}e^{2NTJ/q \sum_{a,b}s_{ab}G_{ab}^{q}} \label{z4exa1}\\
    &=\sum_{m}^{\infty}\frac{\left(\frac{\im}{2}\right)^{m}}{m!}\left[ (\partial_{G_{14}}+\partial_{G_{23}})^{2}+(\partial_{G_{13}}-\partial_{G_{24}})^{2}+(\partial_{G_{12}}+\partial_{G_{34}})^{2}\right]^{m/2}e^{2NTJ/q \sum_{a,b}s_{ab}G_{ab}^{q}}\\
    &=\sum_{m=0}^{\infty}\frac{\left(\frac{\im}{2}\right)^{m}}{m!}\sum_{k_{1}k_{2}k_{3}}\frac{\frac{m}{2}!}{\frac{k_{1}}{2}!\frac{k_{2}}{2}!\frac{k_{3}}{2}!}(\partial_{G_{14}}+\partial_{G_{23}})^{k_{1}}(\partial_{G_{13}}-\partial_{G_{24}})^{k_{2}}(\partial_{G_{12}}+\partial_{G_{34}})^{k_{3}}e^{\frac{2NTJ}{q} \sum_{a,b}s_{ab}G_{ab}^{q}},
\end{align}
and 
\begin{align}
    &(\partial_{G_{14}}+\partial_{G_{23}})^{k_{1}}e^{-2NTJ/q\left(G_{14}^{q}+G_{23}^{q}\right)}=\sum_{l_{1}+l_{2}=k_{1}}\frac{k_{1}!}{l_{1}!l_{2}!}\partial_{G_{14}}^{l_{1}}\partial_{G_{23}}^{l_{2}}e^{-(2NTJ/q) G_{14}^{q}}e^{-(2NTJ/q) G_{23}^{q}}\\
    &=\sum_{l_{1}+l_{2}=k_{1}}\left(-\frac{2NTJ}{q}\right)^{k_{1}/q}\frac{k_{1}!}{(l_{1}/q)!(l_{2}/q)!},
\end{align}
where now $k_{1}$ is a multiple of $q$.  Then \eqref{z4exa1}  becomes  
\begin{align}
  &  \sum_{m=0}^{\infty}\frac{(\im/2)^{m}}{m!}\sum_{k_{1}k_{2}k_{3}}\frac{(m/2)!}{(k_{1}/2)!(k_{2}/2)!(k_{3}/2)!}\sum_{l_{1}+l_{2}=k_{1}}\left(-\frac{2NTJ}{q}\right)^{k_{1}/q}\frac{k_{1}!}{(l_{1}/q)!(l_{2}/q)!}\\
   &\times \sum_{r_{1}+r_{2}=k_{2}}\left(\im^{q}\frac{2NTJ}{q}\right)^{k_{2}/q}\frac{k_{2}!}{(r_{1}/q)!(r_{2}/q)!}\sum_{s_{1}+s_{2}=k_{3}}\left(\im^{q}\frac{2NTJ}{q}\right)^{k_{3}/q}\frac{k_{3}!}{(s_{1}/q)!(s_{2}/q)!} ,\\
   &=\sum_{m=0}^{\infty}\frac{(\frac{\im}{2})^{m}}{m!}\sum_{k_{1}+k_{2}+k_{3}=m}\frac{\frac{m}{2}!}{\frac{k_{1}}{2}!\frac{k_{2}}{2}!\frac{k_{3}}{2}!}\left(-\frac{4NTJ}{q}\right)^{\frac{k_{1}+k_{2}+k_{3}}{q}}\frac{k_{1}!k_{2}!k_{3}!}{\frac{k_{1}}{q}!\frac{k_{2}}{q}!\frac{k_{3}}{q}!}(-\im^{q})^{\frac{k_{2}+k_{3}}{q}}
\end{align}
where $m$ is also a multiple of $q$. To evaluate it we can take an approximation when $m>q$
\begin{align}
  \sum_{k_{1}+k_{2}+k_{3}=m} (-\im^{q})^{\frac{k_{2}+k_{3}}{q}} \frac{k_{1}!k_{2}!k_{3}!}{m!}\frac{\frac{m}{2}!}{\frac{k_{1}}{2}!\frac{k_{2}}{2}!\frac{k_{3}}{2}!}\frac{\frac{m}{q}!}{\frac{k_{1}}{q}!\frac{k_{2}}{q}!\frac{k_{3}}{q}!}\cong \begin{cases}
	3\,, \qquad q=4k+2\\
	3\,, \qquad q=4k,\quad m/q \text{ is even}\\
	-1\,,~\quad q=4k,\quad m/q \text{ is odd}
\end{cases}\ 
\end{align}
then up to a constant \eqref{z4exa1} becomes 
\begin{align}
\begin{cases}
    3e^{\frac{4NTJ}{q2^{q}}},\hspace{4.5cm} q=4k+2\\
    3\cosh\left(\frac{4NTJ}{q2^{q}}\right)+\sinh\left(\frac{4NTJ}{q2^{q}}\right),\quad q=4k.
    \end{cases}
\end{align}

\section{Lefschetz Thimbles}
\label{Lefschetz}
In this appendix, we review the method of Lefschetz thimble~\cite{Witten:2010cx}.
Suppose we would like to evaluate the integral
\bal
Z=\int_{\mathcal{M}_{\mathbb{R}}} dx^i e^S\,,
\eal
where the integration contour is $\mathcal{M}_{\mathbb{R}}$. Then we complexify the  manifold on which the integration is done to $\mathcal{M}_\mathbb{C}$. If we choose $\Re(S)$ to be the Morse function. The saddle points of the integral are the critical points of the Morse function. Around each critical point on $\mathcal{M}_{\mathbb{C}}$ we  introduce a set of local coordinates $\{z_i\}$. The Morse flow is determined by the flow equations
\bea \label{cflow}
\frac{d z^i}{dt}=-g^{i\bar{j}}\frac{\partial \bar{S}}{\partial \bar{z}^j},\quad \frac{d\bar{z}^i}{dt}=-g^{i\bar{j}}\frac{\partial S}{\partial z^j}\ .
\eea 
We find 
\bea 
\frac{d(S-\bar{S})}{dt }=\frac{\partial S}{\partial z^i}\frac{dz^i}{dt}-\frac{\partial \bar{S}}{\partial \bar{z}^i}\frac{d\bar{z}^i}{dt}=0\,,
\eea 
which implies that the imaginary part of $S$ is a constant along the flow. Each of the critical points is associated with a pair of flows, the thimble and the anti-thimble. The thimble is the ``stable'' direction such that the Morse function $\mathcal{M}_{\mathbb{R}}$ decays along the thimble and the integral of $\exp(S)$ along the thimble converges. On the contrary, the anti-thimble is the ``unstable'' direction. Explicitly the boundary conditions for a particular critical point $p_{\sigma}$ are 
\begin{align}
    \lim_{t\to-\infty}z(t)=p_{\sigma},\quad\text{for thimbles},\\
    \lim_{t\to+\infty}z(t)=p_{\sigma},\quad\text{for anti-thimbles}.
\end{align}

The main statement in~\cite{Witten:2010cx} that we will use repeatedly is that the integral can be approximated by a weighted sum over integrals along the thimbles of each critical point 
\bea 
Z=\sum_i n_i \int _{\mathcal{J}_i} dt\, e^{S[t]}\,,
\eea 
where $i$ runs over all the critical points,   $\mathcal{J}_i$ is the Lefschetz thimble attaching to the $i^{\text{th}}$ critical point, and the weight $n_i$ is given by the intersection number between the anti-thimble and the original integration contour $\mathcal{M}_{\br}$. 

\subsection{some examples}
To illustrate how this works, we first go through some simple examples.

\subsubsection{The Gaussian function}
\begin{figure}
\begin{center}
  \includegraphics[width=5 cm]{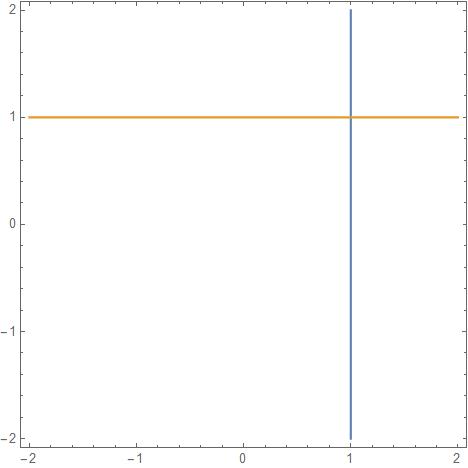}
  \caption{The red line denotes the thimble and the blue line denotes the anti-thimble. The anti-thimble intersects with the real line, so this saddle point contributes} \label{thimble}
  \end{center}
\end{figure}

Let us consider a simple example with  
\bea
S=-x^2/2+\sigma x\ .
\eea 
The integral can be regarded as a zero-dimension theory with quadratic interaction and a complex source $\s$. The only critical point is at $x=\sigma=a+\im b$.  The flow equation is 
\bea 
\frac{\d x}{\d t}=\bar{x}-(a-\im b)\ .
\eea  
Expressing $x=x_1+\im x_2$, we get the following equations
\bea 
\frac{\d x_1}{\d t}=x_1-a,\quad \frac{\d x_2}{\d t}=b-x_2\ .
\eea 
The general solution can be easily solved 
\begin{align}
    x_{1}=a+c_{1}e^{t},\quad x_{2}=b+c_{2}e^{-t},
\end{align}
where $c_{1}$,$c_{2}$ are two undetermined constants.  The boundary conditions for the thimble is 
\begin{align}
    (x_{1},x_{2})\to (a,b),\quad t\to -\infty,
\end{align}
while for the anti-thimble we have 
\begin{align}
    (x_{1},x_{2})\to (a,b),\quad t\to +\infty,
\end{align}
where $(a,b)$ is the critical point. Then with these boundary conditions we can get the equations for the thimble and the anti-thimble respectively 
\begin{align}
x_{2}=b, \\
x_{1}=a.
\end{align}
We plot the thimble and the anti-thimble in this case in Figure \ref{thimble}, where for simplicity we let $\sigma=1+\im$. 

We can also compare the saddle point solution with the exact result. The integral can evaluated as 
\begin{align}
    \int \d x e^{-x^{2}/2+\sigma x}=\sqrt{2\pi}e^{\sigma^{2}/2}.
\end{align}
While the saddle point solution gives 
\begin{align}
    e^{\sigma^{2}/2},
\end{align}
with the one-loop correction $\sqrt{2\pi}$ the saddle point analysis recovers the exact result. 

\subsubsection{The Airy function}
\begin{figure}
\begin{center}
  \includegraphics[scale=0.3]{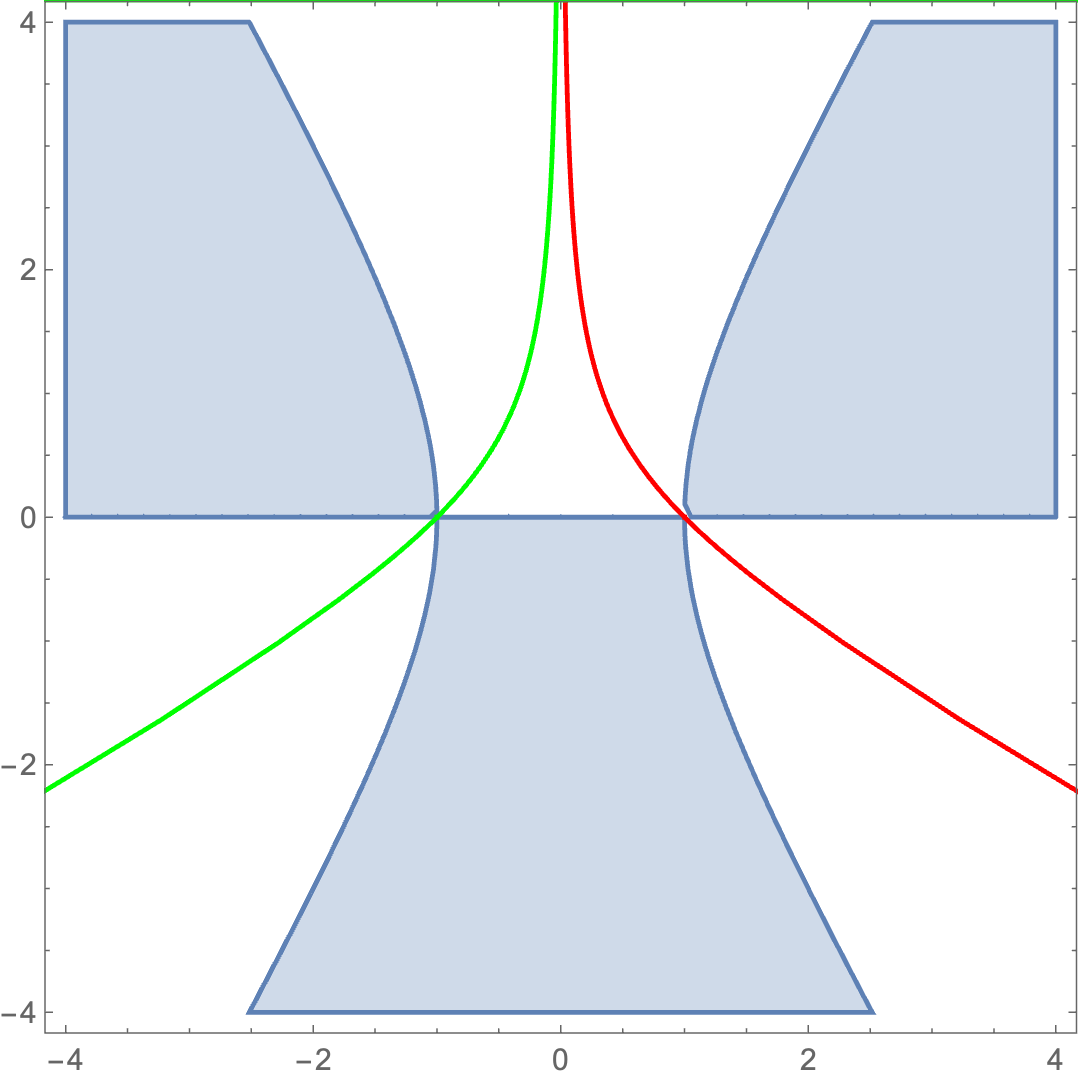}
  \caption{The red line denotes the anti-thimble of $x=-1$ and the Green line denotes the anti-thimble of $x=1$.The grey regions are the convergent regions.} \label{thimble2}
  \end{center}
\end{figure}

A slightly less trivial example is the  Airy action
\bea 
Z_\lambda=\int_{-\infty}^\infty \d x e^S\,,\quad S= \im \lambda \(\frac{x^3}{3}-x\)\ .
\eea 
It is not hard to find that for real $\lambda$ there are three ``convergent'' regions, namely $\Re(S)<\infty$, on the complex $x$-plane:
\bea 
&& x=r e^{\im \theta},\quad \frac{2n \pi}{3}\leq \theta\leq \frac{\pi}{3}+\frac{2n\pi}{3},\quad n=0,1,2.
\eea 
In each convergent region, the Airy integrand is exponentially small. As we vary $\lambda$ to complex values, we should deform the integration contour of $x$ accordingly so that the integral remains converge. This gives an analytic continuation of $Z_\lambda $. The two critical points are located at $x=\pm 1$. The values of saddle points are 
\bea 
S_\pm=\mp \frac{2\im \lambda}{3}.
\eea 
Since along the (anti-)thimbles, the imaginary part of $S$ is a constant and 
\bea 
\text{Im}(S_\pm)=\mp  \frac{2 \text{Re}(\lambda)}{3}.
\eea
Therefore the two (anti-)thimbles associated with the two critical points will not intersect except for the case of  $\text{Re}(\lambda)=0$. The thimble which connects critical points is called the Stoke ray. Using the Lefschetz thimbles $\mathcal{J}_\pm$, we can rewrite the integral as
\bea 
Z_\lambda =n_+ \int_{\mathcal{J}_+} \exp S+n_- \int_{\mathcal{J}_-} \exp S.
\eea 
To solve the thimbles, let us take $\lambda=1$, then the flow equations are
\bea 
\frac{\d x}{\d t}=\im (\bar{x}^2-1).
\eea 
Expressing $x=x_1+\im x_2$, we obtain
\bea 
\frac{\d x_1}{\d t}=2 x_1 x_2,\quad \frac{\d x_2}{\d t}=x_1^2-x_2^2-1.
\eea 
We plot the anti-thimbles in Fig. \ref{thimble2}
Therefore for $\lambda=1$ both of the saddle points contribute. This result is expected since that the two critical points are already located on the real line.

\begin{figure}
\begin{center}
  \includegraphics[scale=0.3]{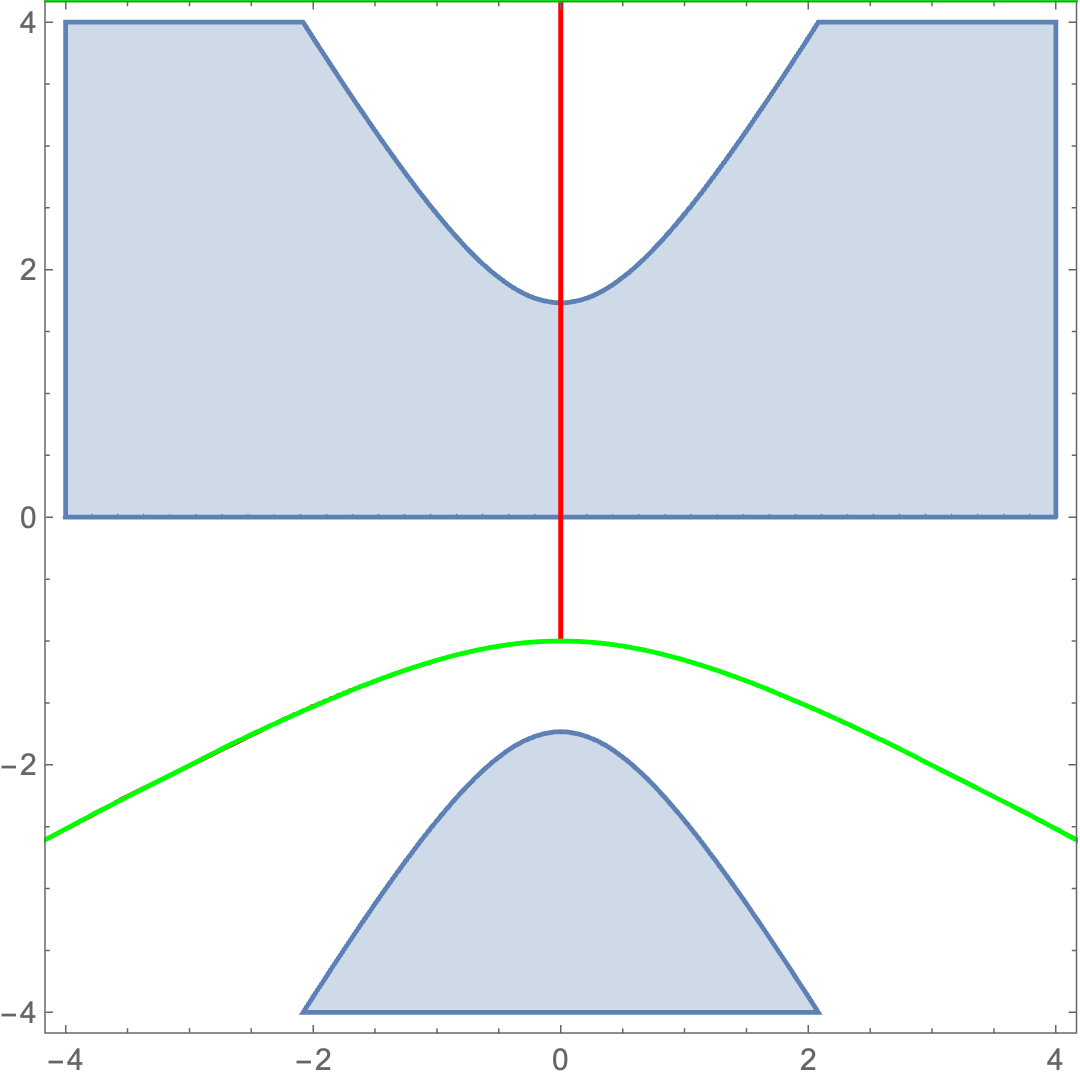}
  \caption{The red line is the anti-thimble of $x=\im $ which intersects with the real line while the Green line is the anti-thimble of $x=-\im$ which does not intersect with the real line. The grey regions are the convergent regions.} \label{thimble3}
  \end{center}
\end{figure}

The problem we met in the main text is better illustrated by the following toy model
\bea 
\tilde{Z}=\int_{-\infty}^\infty \d x \exp S, \quad S=\im \(\frac{x^3}{3}+x\)\ .
\eea  
The integral is convergent and can be expressed by the Airy function
\bea 
\tilde{Z}=\frac{2 \pi  \text{Ai}\left(\frac{1}{\sqrt[3]{3}}\right)}{\sqrt[3]{3}}=0.83\ .
\eea 
We now try to compute the integral with saddle point approximation, where the saddle points are located at $x=\pm \im$. 
The saddle point value, plus the 1-loop correction, of the integral at these two saddle points, $\tilde{Z}_\pm$ are the same, and the sum of them is larger than the exact evaluation of the integral
\bea 
\tilde{Z}_+=\tilde{Z}_-=0.733,\quad \tilde{Z}_++\tilde{Z}_->\tilde {Z}.
\eea 
This is exactly the situation we are encountering. 
In this toy model, it is easy to show that the anti-thimble associated with the saddle point $x=-\im$ does not intersect with the real axis,  Figure.~\ref{thimble3}, so the saddle point $x=-\im$ does not contribute to the integral. 

\subsection{Multi-variable cases}
\begin{figure}[h]
\begin{center}
  \includegraphics[scale=0.4]{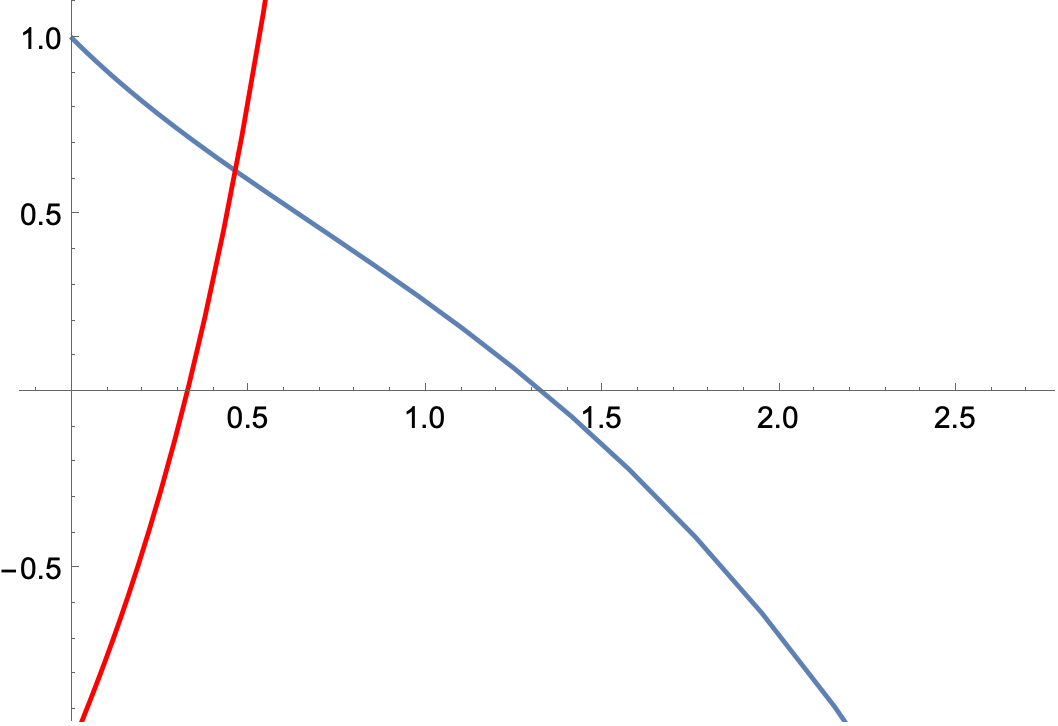}
  \caption{Since $\sigma_\pm$ are already on the real line here we only plot the $g$-plane of the anti-thimbles. Clearly both of these two anti-thimbles intersect with the real axis so these two saddle points both contribute to the integral.} \label{thimble4}
  \end{center}
\end{figure}

Let us consider another example with two variables
\bea 
Z=\int \d\sigma \frac{\d g}{2\pi } e^{S}\,,\quad S=\log \sigma -\im \sigma g-\frac{1}{2}g^2\ .
\eea 
The integral can be done directly to get
\bal
Z=0\ .
\eal
There are two saddle points 
\bea 
g_\pm=\pm \im,\quad \s_\pm=\mp 1\ .
\eea 
with saddle point contributions to the integral (on-shell action)
\bea 
Z_\pm=\mp \frac{1}{\sqrt{e}},\quad Z_-+Z_+=0\ ,
\eea 
Matching this with the exact solution suggests that $n_\pm=1$. Note that $\s_\pm=\mp 1$ are already on the real line so corresponding anti-thimbles always intersect with the original contour. The flow equations are
\bea 
\frac{d\sigma}{dt}=-\frac{1}{\bar{\sigma}}-\im \bar{g},\quad \frac{\d g}{dt}=-\im \bar{\sigma}+\bar{g}.
\eea 
Expressing $\sigma=\sigma_1+\im \sigma_2$ and $g=g_1+\im g_2$ we obtain the following differential equations
\bea 
&&\frac{\d \sigma_1}{\d t}+g_2+\frac{\sigma_1}{\sigma_1^2+\sigma_2^2}=0,\quad \frac{\d \sigma_2}{\d t}+g_1+\frac{\sigma_2}{\sigma_1^2+\sigma_2^2}=0,\\
&&\frac{\d g_1}{\d t}+\sigma_2-g_1=0,\quad \frac{\d g_2}{\d t}+\sigma_1+g_2=0.
\eea 
We find that indeed these two saddles should both be included. We plot the $g$-plane of the anti-thimbles in Fig. \ref{thimble4}.
Note that this example is special case of \eqref{z2t1} with $q=2$.

\subsection*{Flow equations in real coordinates}
Sometimes it more convenient to use real form of the flow equations \eqref{cflow}.  We start with the relations 
\begin{align}
    \frac{\partial S}{\partial z}=\frac{1}{2} \frac{\partial S}{\partial x}+\frac{1}{2\im} \frac{\partial S}{\partial y},
\end{align}
\begin{align}
    \frac{\partial \bar{S}}{\partial \bar{z}}=\frac{1}{2} \frac{\partial \bar{S}}{\partial x}-\frac{1}{2\im} \frac{\partial \bar{S}}{\partial y},
\end{align}
where 
\begin{align}
    z=x+\im y.
\end{align}
 Then we evaluate the equation as  
\begin{align}
    \frac{\d z}{\d t}+\frac{\d\bar{z}}{\d t}&=- \frac{\partial S}{\partial z}-\frac{\partial \bar{S}}{\partial \bar{z}}=-\frac{\partial \text{Re}(S)}{\partial x}-\frac{\partial \text{Im}(S)}{\partial y},
\end{align}
\begin{align}
    \frac{\d z}{\d t}-\frac{\d\bar{z}}{\d t}&=\frac{\partial S}{\partial z}-\frac{\partial \bar{S}}{\partial \bar{z}}=-\im\frac{\partial \text{Re}(S)}{\partial y}+\im\frac{\partial \text{Im}(S)}{\partial x},
\end{align}
where we work in the flat space.
Recall the Cauchy-Riemann equation we get 
\begin{align}
    \frac{\d x}{\d t}=-\frac{\partial \text{Re}(S)}{\partial x},\quad 
    \frac{\d y}{\d t}=-\frac{\partial \text{Re}(S)}{\partial y}.
    \label{fleqr}
\end{align}
To illustrate it we consider a special case in the Airy function 
\begin{align}
    S=\im \left( \frac{x^{3}}{3} +x \right).
\end{align}
In the complex plane its conjugate is 
\begin{align}
    \bar{S}=-\im \left( \frac{\bar{x}^{3}}{3} +\bar{x} \right),
\end{align}
and we can define the components 
\begin{align}
    x=x_1+\im x_2,\quad \bar{x}=x_1-\im x_2.
\end{align}
The flow equation in complex coordinates becomes 
\begin{align}
    \frac{\d x}{\d t}=-\frac{\partial \bar{S}}{\partial \bar{x}}=\im(\bar{x}^{2}+1),
\end{align}
which leads to the equations in real coordinates 
\begin{align}
    \frac{\d x_1}{\d t}=2x_1x_2,\quad     \frac{\d x_2}{\d t}=x_1^{2}-x_2^{2}+1.
    \label{fleqr2}
\end{align}
On the other hand we can get the equations with the real part of $S$:
\begin{align}
    \text{Re}(S)=-x_2-x_1^{2}x_2+\frac{x_2^{3}}{3}.
\end{align}
From the equations \eqref{fleqr} we can recover the two flow equations \eqref{fleqr2}.

\section{Averaged models} 
\label{am}
In this section, before talking about higher dimensional model we detour the main topic a little bit and consider the averaged theory in general. Let us consider a real scalar field with a chemical potential. The partition function of the theory is 
\bea 
Z=\int \cd \Phi \exp\(-\int dx\,\partial_\mu \phi(x) \partial^\mu \phi(x)+J(x)\phi(x)\)\,
\eea 
where $J(x)$ is a random source drawn from some probability distribution. If the random coupling $J$ does not depends on the spacetime i.e.
\bea \label{cj}
\langle J^n\rangle_{\text{connected}}=\kappa_n\,\quad 
\eea 
where $\kappa_n$ is the $n$-th cumulants of the probability distribution. This situation is like the regular SYK theory. To take the ensemble average we can expand the exponential as
\bea 
\langle Z\rangle_J &=&\int \cd\Phi \exp\(-\int dx\, \cl_0\)\sum_n \frac{\langle J^n \rangle }{n!} \(\int dx \,\phi(x)\)^n \\
&=& \int \cd\Phi \exp\(-\int dx\, \cl_0\)\text{MGF}\(\int dx \phi(x)\)
\eea 
where $\text{MGF}$ is the moment generating function of $J$. If $J$ is Gaussian $\cn(0,t^2)$ then 
\bea 
&&\text{MGF}\(\int dx \phi(x)\) =\exp\(\frac{t^2}{2}\int dx\, dy \, \phi(x)\phi(y)\),\\
&&\langle Z\rangle_J=\int \cd\Phi \exp\(-\int dx\, \cl_0+\frac{t^2}{2}\int dx\, dy \, \phi(x)\phi(y)\)\, 
\eea 
which is similar to SYK model that after the Gaussian average, bi-local interaction is generated. For the general distribution \eqref{cj}, the resulting averaged action is 
\bea 
\langle Z\rangle_J=\int \cd\Phi \exp\(-\int dx\, \cl_0+\sum_n \frac{\kappa_n}{n!}[\int dx \phi(x)]^n\)\, ,
\eea 
so multi-local interactions will be generated. In particular, if the distribution is Poisson we have
\bea
\langle Z\rangle_J=\int \cd\Phi \exp\(-\int dx\, \cl_0+e^{\lambda\int dx \phi(x)}-1\) \, ,
\eea 
which is a highly non-local theory. \par 
Alternatively we can require the random coupling $J$ to depend on the spacetime i.e.
\bea 
\langle \prod^n_i J(x_i)\rangle =\prod_{i=1,j=i}^n\delta(x_i-x_j) \kappa_n\, .
\eea 
This situation is like the Brownian SYK theory \cite{Saad:2018bqo}. Now we can take the ensemble average to get
\bea 
\langle Z\rangle_J &=& \int \cd\Phi \exp\(-\int dx\, \cl_0\)\sum_n \frac{\langle J^n \rangle }{n!} \(\int dx \,\phi(x)\)^n \\
&=&\int \cd\Phi \exp\(-\int dx\, \cl_0\)\sum_n \frac{\kappa_n}{n!} \(\int dx \,\phi^n(x)\) \nn \\
&=&\int \cd\Phi \exp\(-\int dx\, \cl_0\)\int dx \, \text{MGF}(\phi(x)).
\eea 
If $J$ is Gaussian $\cn(0,t^2)$ then we have
\bea 
\langle Z\rangle_J=\int \cd\Phi \exp\(-\int dx\, (\cl_0+\frac{t^2}{2}\phi^2(x))\).
\eea 
and if $J$ is Poisson then we simply get
\bea 
\langle Z\rangle_J=\int \cd\Phi \exp\(-\int dx\, [\cl_0+\lambda (e^{\phi(x)}-1)]\),
\eea 
which also coincides with results in \cite{Peng:2020rno}.

\section{Computations in large N} \label{appendixlargeN}
\subsection*{CLT with $\mu=0$}
In the appendix, we will consider the half-wormhole conjecture for statistical model in a more systematic way.
The error functions we care about are 
\begin{align}
        \text{Error}&=Y^{2}-\langle Y^{2}\rangle+\langle Y\rangle^{2}-\Phi,\\
        \langle \text{Error}^{2}\rangle&=\langle \left( Y^{2}-\langle Y^{2}\rangle+\langle Y\rangle^{2}-\Phi \right)^{2}\rangle ,
\end{align}
where $\Phi$ is defined in \eqref{Phi}. And if the approximation is proper we must have 
\begin{align}
    \langle \text{Error}\rangle=0,\quad \langle \text{Error}^{2}\rangle/\langle Y^{4}\rangle \ll 1 . \label{cltreq2}
\end{align}
The first requirement leads to 
\begin{align}
	\langle \Phi\rangle =\langle Y\rangle ^{2} 
\end{align}
which is proved around \eqref{Phi}, then we have 
\begin{align}
	\langle \text{Error}^{2}\rangle =\langle  \left( Y^{2}+\langle Y\rangle^{2}-\Phi \right)^{2} \rangle -\langle Y^{2}\rangle ^{2}.
\end{align}

In the following we'll give one type of function and try to explore the feasibility of the proposal, i.e. we compute the two error functions and check the relations \eqref{cltreq2} respectively.

First we consider the function $Y$ consisting of $N$ identical and independent variables $X_{i}$ as 
\begin{align}
    Y=\sum_{i\neq j}X_{i}X_{j},
\end{align}
and for simplicity each combination only appears once since $X_{i}$'s commute with each other. The properties for $X_{i}$ in this section are set as
\begin{align}
    \langle X\rangle=0,\quad \langle X^{2}\rangle=t^{2},\label{cltxproperty}
\end{align}
in the next section we'll consider the distribution with a non-zero mean.
But actually for any distribution the property \eqref{cltxproperty} can always be satisfied since we can always take the subtraction
\begin{align}
    \tilde{X}=X-\langle X\rangle.\label{cltmsub}
\end{align}
 therefore our discussion may be applied into the case with any distribution.

Obviously the first requirement in \eqref{cltreq2} is satisfied then we consider the computation for the second requirement 
\begin{align}
    Y^{2}=\sum_{i\neq j}X_{i}^{2}X_{j}^{2}+2\sum_{i\neq j\neq k}X^{2}_{i}X_{j}X_{k}+6\sum_{i\neq j\neq k\neq l}X_{i}X_{j}X_{k}X_{l},
    \label{gey21}
\end{align}
where the first sum contains $p$ different terms 
\begin{align}
    p={N\choose 2},\label{cltpdef}
\end{align}
while we have $2p(N-2)/2$ for the second sum and $p(p-2(N-2)-1)/6$ for the third one. For the $\Phi$ in the approximation we have 
\begin{align}
     \Phi&= \frac{1}{(2\pi)^{N}}\int d\vec{k} \frac{e^{-\im \vec{k} \vec{X}}}{P(\vec{X})} \langle e^{\im\vec{k}\vec{x}}\rangle \(\frac{\langle Y e^{\im\vec{k}\vec{x}}\rangle }{\langle e^{\im\vec{k}\vec{x}}\rangle}\)^2,\label{clt2phi}\\
     &=\frac{1}{(2\pi)^{N}}\int d\vec{k} \frac{e^{-\im \vec{k} \vec{X}}}{P(\vec{X})} \langle e^{\im\vec{k}\vec{x}}\rangle\(\sum_{i\neq j}\frac{\langle x_{i}x_{j} e^{\im\vec{k}\vec{x}}\rangle }{\langle e^{\im\vec{k}\vec{x}}\rangle}\)^2,
\end{align}
the computation is similar to that in $Y^{2}$ so that we can split the $\Phi$ function into three parts according to the three sums in  \eqref{gey21}
\begin{align}
    \Phi_{0}&=\frac{2}{(2\pi)^{N}}\int d\vec{k} \frac{e^{-\im \vec{k} \vec{X}}}{P(\vec{X})} \langle e^{\im\vec{k}\vec{x}}\rangle\sum_{i\neq j}\frac{\langle x_{i}x_{j} e^{\im\vec{k}\vec{x}}\rangle }{\langle e^{\im\vec{k}\vec{x}}\rangle}\frac{\langle x_{i}x_{j} e^{\im\vec{k}\vec{x}}\rangle }{\langle e^{\im\vec{k}\vec{x}}\rangle},\\
    &=\sum_{i\neq j}\frac{1}{\left(2\pi\right)^{2}}\int dk_{i}dk_{j}  \frac{e^{-\im k_{i} X_{i}}e^{-\im k_{j} X_{j}}}{P(X_{i})P(X_{j})} \langle e^{\im k_{i}x_{i}}\rangle  \langle e^{\im k_{j}x_{j}}\rangle\(\frac{\langle x_{i} e^{\im k_{i}x_{i}}\rangle }{\langle e^{\im k_{i}x_{i}}\rangle}\)^2\(\frac{\langle x_{j} e^{\im k_{j}x_{j}}\rangle }{\langle e^{\im k_{j}x_{j}}\rangle}\)^2,
\end{align}
\begin{align}
    \Phi_{1}&=\frac{2}{(2\pi)^{N}}\int d\vec{k} \frac{e^{-\im \vec{k} \vec{X}}}{P(\vec{X})} \langle e^{\im\vec{k}\vec{x}}\rangle\sum_{i\neq j\neq k}\frac{\langle x_{i}x_{j} e^{\im\vec{k}\vec{x}}\rangle }{\langle e^{\im\vec{k}\vec{x}}\rangle}\frac{\langle x_{i}x_{k} e^{\im\vec{k}\vec{x}}\rangle }{\langle e^{\im\vec{k}\vec{x}}\rangle},\\
    &=2\sum_{i\neq j\neq k}X_{j}X_{k}\frac{1}{2\pi}\int dk_{i} \frac{e^{-\im k_{i} X_{i}}}{P(X_{i})} \langle e^{\im k_{i}x_{i}}\rangle \(\frac{\langle x_{i} e^{\im k_{i}x_{i}}\rangle }{\langle e^{\im k_{i}x_{i}}\rangle}\)^2,
\end{align}
\begin{align}
    \Phi_{2}&=\frac{6}{(2\pi)^{N}}\int d\vec{k} \frac{e^{-\im  \vec{k} \vec{X}}}{P(\vec{X})} \langle e^{\im\vec{k}\vec{x}}\rangle\sum_{i\neq j\neq k\neq l}\frac{\langle x_{i}x_{j} e^{\im\vec{k}\vec{x}}\rangle }{\langle e^{\im\vec{k}\vec{x}}\rangle}\frac{\langle x_{k}x_{l} e^{\im\vec{k}\vec{x}}\rangle }{\langle e^{\im\vec{k}\vec{x}}\rangle},\\
    &=6\sum_{i\neq j\neq k\neq l}X_{i}X_{j}X_{k}X_{l}.
\end{align}
For simplicity we can define the function below 
\begin{align}
    \phi_{i}= \frac{1}{2\pi}\int dk \frac{e^{-\im k X_{i}}}{P(X_{i})} \langle e^{\im kx}\rangle \(\frac{\langle x e^{\im kx}\rangle }{\langle e^{\im kx}\rangle}\)^2,\label{cltphii}
\end{align}
so that the $\Phi$ functions can be expressed as 
\begin{align}
    \Phi_{0}=\sum_{i\neq j}\phi_{i}\phi_{j}, \label{cltphi0}
\end{align}
\begin{align}
    \Phi_{1}=2\sum_{i\neq j\neq k}\phi_{i}X_{j}X_{k}.\label{cltphi1}
\end{align}
And note that the $\phi$ function has the property 
\begin{align}
    \langle \phi_{i} \rangle=\langle X_{i}\rangle^{2},\label{cltphiav}
\end{align}
which is useful in the later computation, where $X_{i}$'s are identical.

Then following the procedure we have 
\begin{align}
     \langle \text{Error}^{2}\rangle&=\langle \left(Y^{2}+\langle Y\rangle^{2}-\Phi\right)^{2}\rangle-\langle Y^{2}\rangle^{2},\\
     &=\left\langle \left(\sum_{i\neq j}\left(X_{i}^{2}X_{j}^{2}-\phi_{i}\phi_{j}\right)+2\sum_{i\neq j\neq k}\left(X^{2}_{i}-\phi_{i}\right)X_{j}X_{k}\right)^{2} \right\rangle-p^{2}\langle X^{2}\rangle^{4},
\end{align}
where the last sum cancels out and $p$ is defined in \eqref{cltpdef}. The cross terms between different sums are zero due to the zero mean of $X_{i}$, therefore we can only consider the squares of each sum
\begin{align}
     \langle \text{Error}^{2}\rangle=\left\langle \left(\sum_{i\neq j}\left(X_{i}^{2}X_{j}^{2}-\phi_{i}\phi_{j}\right)\right)^{2}+\left(2\sum_{i\neq j\neq k}\left(X^{2}_{i}-\phi_{i}\right)X_{j}X_{k}\right)^{2} \right\rangle-p^{2}\langle X^{2}\rangle^{4}.
     \label{clterror22}
\end{align}
The numbers of the diagonal terms in the two sums in the first average are 
\begin{align}
    p,\quad 4p(N-2),
\end{align}
while the non-diagonal terms in the first sum have 
\begin{align}
    p(p-2(N-2)-1)\quad  \text{ for } \quad \langle X_{i}^{2}X_{j}^{2}X_{k}^{2}X_{l}^{2}\rangle \label{cltx22} \\
    2p(N-2) \quad  \text{ for } \quad \langle X_{i}^{4}X_{j}^{2}X_{k}^{2}\rangle,
\end{align}
and the non-diagonal terms in the second sum vanish due to the zero mean of $X_{i}$. 

We want to talk about \eqref{cltreq2} in this case, and the $\phi$ function has its role in the discussion. If the function $\phi$ satisfies the relation
\begin{align}
    X_{i}^{2}-\phi_{i}=f(X_{i}),\quad \langle f(X_{i})\rangle=0 \label{cltcond1}
\end{align}
where $f(X_{i})$ is an unknown function of $X_{i}$ . Actually this relation leads to 
\begin{align}
    t^{2}=\langle X_{i}^{2}-\phi_{i}\rangle=0,
\end{align}
therefore in this case the distribution should be non-trivial with $\mu=0$ and $t=0$. Usually we do not consider such cases, but we can still show how it works following this computation.
Then in this case the dominant terms in $\langle \text{Error}^{2}\rangle$ will be \eqref{cltx22} as it contains the largest number of terms. The $\phi$ terms are not included since the averages of the non-diagonal $\phi$ terms are zero  because of \eqref{cltxproperty} and \eqref{cltphiav}. Then we can find that the dominant terms of $\langle \text{Error}^{2}\rangle$  which locate at the order $p^{2}$ or $N^{4}$ cancel out in \eqref{clterror22}. Therefore the relation \eqref{cltreq2} is satisfied since the dominant terms in $Y^{2}$ are also at the order $N^{4}$, so that the proposal holds in this case when $N$ is sufficiently large.

While if $\phi$ has the relation below like what we have in the Gaussian case 
\begin{align}
    X_{i}^{2}-\phi_{i}= g(X_{i}), \quad  \langle g(X_{i})\rangle\neq 0 \label{cltcond2}
\end{align}
where $g(X_{i})$ is also unknown and we denote the average with a nonzero constant $C$. Then the diagonal terms in the second sum can contribute to the dominant terms. We can have the explicit computation as the following, 
\begin{align}
    2\sum_{i\neq j\neq k}\left(X^{2}_{i}-\phi_{i}\right)X_{j}X_{k}\sim 2(N-2)C\sum_{j\neq k}X_{j}X_{k},
\end{align}
where the approximation is valid as we only care about the maximal numbers. In the square this term is similar to \eqref{gey21} except we have an extra parameter $4(N-2)^{2}$. The non-vanishing contribution after the average will be the diagonal terms in the square as $X_{i}$ has a zero mean, which means the number of the terms is 
\begin{align}
    4(N-2)^{2}p=2N(N-1)(N-2)^{2}.
\end{align}
This contribution also locate at the order $N^{4}$ which is additional comparing to the previous case, so that the dominant terms can not cancel out in \eqref{clterror22}. Therefore the proposal fails in this case.

After having discussed a simple example  We can try to consider the general one 
 \begin{align}
     Y=\sum_{i_{1}\dots i_{q}}X_{i_{1}}\dots X_{i_{q}},
 \end{align}
 where the sum contains $p$ terms 
 \begin{align}
     p={N\choose q}
 \end{align}
and $X_{i}$'s are still identical and independent variables.  Following the procedure above we have 
\begin{align}
    Y^{2}=\sum_{i_{1}\dots i_{q}}X_{i_{1}}^{2}\dots X_{i_{q}}^{2}+\sum_{i_{1}\dots i_{q+1}}X_{i_{1}}^{2}\dots X_{i_{q-1}}^{2}X_{i_{q}}X_{i_{q+1}}+\dots+ \sum_{i_{1}\dots i_{2q}}X_{i_{1}}\dots X_{i_{2q}},
    \label{y2compo}
\end{align}
the numbers of the total terms for each sum are respectively
\begin{align}
    {N\choose q},\quad {N\choose q}{q\choose 1}{N-q \choose 1},\dots, {N\choose q}{N-q\choose q},
    \label{parameter1}
\end{align}
and the numbers for each single term in each sum or the repetitions are 
\begin{align}
    1,\quad {2\choose 1},\quad {4\choose 2},\dots,{2q\choose q}.\label{parameter2}
\end{align}
When $q=2$ we can find that the computations in the two cases match.
For the function $\Phi$ similarly we can define the functions 
\begin{align}
    \Phi_{0},\quad \Phi_{1},\dots,\Phi_{q},
\end{align}
which correspond to the terms in \eqref{y2compo} respectively.  And we can also introduce the $\phi$ functions to express each $\Phi_{s}$, substituting the $X^{2}_{i}$ by $\phi_{i}$ in \eqref{y2compo} we can get all the $\Phi_{s}$'s just like \eqref{cltphi0},\eqref{cltphi1}.

Following the previous procedure the square of the error function can be evaluated as
\begin{align}
     \langle \text{Error}^{2}\rangle&=\langle \left(Y^{2}+\langle Y\rangle ^{2} -\Phi\right)^{2}\rangle-\langle Y^{2}\rangle^{2},\\
 &=\left\langle \left(\sum_{s=0}^{q}\sum_{i_{1}\dots i_{q+s}}\left(Y^{2}_{s}-\Phi_{s}\right) \right)^{2}\right\rangle-p^{2}\langle X^{2}\rangle^{2q},\\
  &=\left\langle \sum_{s=0}^{q}\left(\sum_{i_{1}\dots i_{q+s}}\left(Y^{2}_{s}-\Phi_{s}\right) \right)^{2}\right\rangle-p^{2}\langle X^{2}\rangle^{2q},
  \label{clterror2}
\end{align}
where the index $s$ labels the components in both $Y^{2}$,$\Phi$ and the cross terms among the different sums vanish due to zero mean of $X_{i}$.

Then we'll encounter the previous problem again that there are two different conditions \eqref{cltcond1},\eqref{cltcond2} for $\phi$, and we also first consider the former one. The total numbers of the diagonal terms in the above each sum $s$ are respectively
\begin{align}
    {N\choose q},\quad {2\choose 1}{N\choose q}{q\choose 1}{N-q \choose 1},\dots, {2(q-1) \choose q-1}{N\choose q}{q\choose q-1}{N-q\choose q-1},\label{cltnumbers}
\end{align}
which are the combinations of \eqref{parameter1} and \eqref{parameter2} and the last one in \eqref{y2compo} cancels out. 
While about the non-diagonal terms as the averages of $X_{i}$ is zero, the nonzero non-diagonal terms only appears in the sum $s=0$. And as the average of $\Phi_{i}$ is also zero, the nonzero non-diagonal terms in the sum $s=0$ will only come from  $Y^{2}_{0}$. The computation of the square of this term is similar to \eqref{y2compo} except the replacement of $X_{i}$ by $X_{i}^{2}$,
\begin{align}
    Y^{4}_{nn}=\sum_{i_{1}\dots i_{q+1}}X_{i_{1}}^{4}\dots X_{i_{q-1}}^{4}X_{i_{q}}^{2}X_{i_{q+1}}^{2}+\dots +\sum_{i_{1}\dots i_{2q}}X_{i_{1}}^{2}\dots X_{i_{2q}}^{2},
    \label{y4nncompo}
\end{align}
where $nn$ means  nonzero non-diagonal.  The non-zero contributions in \eqref{clterror2} come from the square of the first sum $s=0$ and the diagonal terms in the square of the other sums $s>0$. And note that the last one labelled by $s=q$ cancels out, and  in \eqref{cltnumbers} the last one is subordinate to the square of the first term when $N$,$q$,$N/q$ are sufficiently large. Then under the condition \eqref{cltcond1} the dominant terms will be the first sum with square according to the equation \eqref{cltnumbers}. The square of the first sum in \eqref{clterror2} gives the dominant terms which are the last one in \eqref{y4nncompo}, it contains the number 
\begin{align}
    {N\choose q}{N-q\choose q}.
\end{align}
Then after the average  the dominant terms in the square of the error becomes
\begin{align}
   \langle \text{Error}^{2}\rangle=\left({N\choose q}{N-q\choose q}-{N\choose q}{N\choose q}\right)\langle X^{2}\rangle^{2q},
\end{align}
which is subordinate comparing to $\langle Y^{4}\rangle$.

While under the condition \eqref{cltcond2} the diagonal terms in the squares of the sums with $s>0$ will contribute.  The computation is similar to the previous case with $q=2$, the nonzero average increases the number of the diagonal or nonzero terms.  As an example we calculate the last non-zero one in \eqref{clterror2}
\begin{align}
    \sum_{j i_{1}\dots i_{2q-2}}(X_{j}^{2}-\phi_{j})X_{i_{1}}\dots X_{i_{2q-2}}=\alpha\sum_{i_{1}\dots i_{2q-2}}X_{i_{1}}\dots X_{i_{2q-2}}
    \label{cltnew}
\end{align}
where 
\begin{align}
\alpha=C{N\choose q}{q\choose q-1}{N-q\choose q-1}/{N\choose 2q-2},
\end{align}
and for simplicity we denote the $\phi$ term with $C$. Since we only care about the number of the terms, the denotation does not matter for the result. 
The constant $C$ is defined in \eqref{cltcond2} and the numerator is the number of the terms in the sum labelled by $s=q-1$, while the denominator is the number of the sum of the right hand side of \eqref{cltnew}. In the square of this term only the diagonal ones survive in the average, we have the number 
\begin{align}
	{N\choose q}^{2}{q\choose q-1}^{2}{N-q\choose q-1}^{2}/{N\choose 2q-2}\sim N^{2q}\sim p^2.
\end{align}
Actually we can find that except the last sum all the sums in \eqref{clterror2} will contribute to the dominant terms.  So that the proposal will fail as the first sum cancels out with the last one in \eqref{clterror2}, while we have other additional contributions to the dominant terms.

\subsection*{CLT with $\mu\neq 0$}

In the above section we have discussed the case with $\mu=0$, after that we can move to consider the case with $\mu\neq 0$. Remember that for the distribution with a non-zero mean we deal it with the subtraction \eqref{cltmsub}, now we keep this non-trivial mean. The difference between them may be illustrated by a simple example
\begin{align}
    Y=\sum_{i\neq j}\left(X_{i}-\langle X_{i}\rangle\right)\left(X_{j}-\langle X_{j}\rangle\right),\\
    Y=\sum_{i\neq j}X_{i}X_{j}-\left\langle\sum_{i\neq j}X_{i}X_{j}\right\rangle.
\end{align}
The two definitions are not equivalent which is manifest in the higher powers of $Y$. Actually the former one is equivalent to the case with $\mu=0$ in the previous section, the latter without the subtraction is what we'll consider in this section.

We still start with the simple case with $q=2$
\begin{align}
	Y=\sum_{i\neq j}X_{i}X_{j},\\
	Y^{2}=\sum_{i\neq j}X_{i}^{2}X_{j}^{2}+2\sum_{i\neq j\neq k}X^{2}_{i}X_{j}X_{k}+6\sum_{i\neq j\neq k\neq l}X_{i}X_{j}X_{k}X_{l}.\label{56cltmy2}
\end{align}
About $\Phi$ we can have similar definition to \eqref{clt2phi}, and  the definition and the requirement for the error are the same as before. We also separate each sum in both $Y^{2}$ and $\Phi$ so that the square of the error can be expressed as 
\begin{align}
      \langle \text{Error}^{2}\rangle&=\langle \left(Y^{2}+\langle Y\rangle^{2}-\Phi\right)^{2}\rangle-\langle Y^{2}\rangle^{2},\\
      &=\left\langle \left( Y_{0}^{2}+\langle Y\rangle^{2}_{0}-\Phi_{0} +Y_{1}^{2}+\langle Y\rangle^{2}_{1}-\Phi_{1} \right)^{2} \right\rangle-\langle Y^{2}\rangle^{2}.\label{clterrorm}
\end{align}
And note that here we can not separate the square of the sum into the sum of the square of each sum, since now the cross terms are not zero due to the nonzero mean of $X_{i}$.  When $N\to\infty$ we can only consider the dominant terms, but as the case is different to before the terms we want will also be different. Now the dominant terms in the two parts in \eqref{clterrorm} are both in the second sum in \eqref{56cltmy2} or $Y_{1}^{2}$ \eqref{clterrorm}, since this sum contains the largest number of terms and is non-zero under the averagre because of the non-zero mean.
Which means we can only consider the cross terms in the square of this sum which contains the largest number of terms. By the direct calculation we have 
\begin{align}
    \Phi_{1}=2\sum_{i\neq j\neq k} \phi_{i}X_{j}X_{k},
\end{align}
where $\phi$ is defined in \eqref{cltphii}. Then we have 
\begin{align}
    Y_{1}^{2}+\langle Y\rangle^{2}_{1}-\Phi_{1}=2\sum_{i\neq j\neq k}\left(\left(X_{i}^{2}- \phi_{i}\right)X_{j}X_{k}+\langle X_{i}\rangle^{2}\langle X_{j}\rangle \langle X_{k}\rangle\right),
\end{align}
while the dominant terms can be written as 
\begin{align}
    \langle \text{Error}^{2}\rangle_{d}=\left\langle\left(2\sum_{i\neq j\neq k}\left(\left(X_{i}^{2}- \phi_{i}\right)X_{j}X_{k}+\langle X_{i}\rangle^{2}\langle X_{j}\rangle \langle X_{k}\rangle\right)\right)^{2}\right\rangle-\left\langle 2\sum_{i\neq j\neq k}X_{i}^{2}X_{j}X_{k}\right\rangle^{2}.
    \label{errordomi}
\end{align}
The dominant terms of the first part are the cross terms, whose number is at the same order of $N$ as the second part. So to evaluate the approximation we can only consider the two terms below 
\begin{align}
\langle \left(X_{i}^{2}- \phi_{i}\right)X_{j}X_{k}+\langle X_{i}\rangle^{2}\langle X_{j}\rangle \langle X_{k}\rangle\rangle,\quad \langle X_{i}^{2}X_{j}X_{k}\rangle,
\end{align}
where the variables are identical and independent.  Recall that we have the equation \eqref{cltphiav}
therefore the two expressions are equal under the average.  Explicitly the cross terms also contain many parts, we can also write down the one with the largest number of terms, 
\begin{align}
	\left(X_{i}^{2}- \phi_{i}\right)X_{j}X_{k}\left(X_{l}^{2}- \phi_{l}\right)X_{m}X_{n},\quad 4{N\choose 2}(N-2){N-3\choose 2}(N-5),
\end{align}
while the second part in \eqref{errordomi} contains 
\begin{align}
	4{N\choose 2}^2(N-2)^2.
\end{align}
Therefore the dominant contribution in \eqref{errordomi} will cancel out, which makes the proposal hold.

 We can also consider the problem that occurs in the case with $\mu=0$ when the average of $X^{2}-\phi$ is not zero, and we can find that there's no such problem here. As the dominant term \eqref{errordomi} contains the largest number of $X$ terms, therefore even when $X^{2}-\phi$ is a constant the total number of $X$ terms in \eqref{errordomi} will not change.

Next we consider the case with general $q$. 
Following the previous procedure for the error we can get an expression similar to  \eqref{clterrorm} 
\begin{align}
      \langle \text{Error}^{2}\rangle=\left\langle \left(\sum_{s=0}^{q-1} (Y_{s}^{2}+\langle Y\rangle^{2}_{s}-\Phi_{s}) \right)^{2} \right\rangle-\langle Y^{2}\rangle^{2},
\end{align}
and the dominant sum will also be the last non-zero one, i.e. the sum labelled with $s=q-1$.
To find out the behavior of the dominant terms the relation of the two functions below is important,
\begin{align}
    \langle\left(X_{i_{1}}^{2}- \phi_{i_{1}}\right)X_{i_{2}}\dots X_{i_{2q-1}}+\langle X_{i_{1}}\rangle^{2}\langle X_{i_{2}}\rangle \dots\langle X_{i_{2q-1}}\rangle\rangle,\quad \langle X_{i_{1}}^{2}X_{i_{2}}\dots X_{i_{2q-1}}\rangle,
\end{align}
we want to determine whether they are equal. 
And we can find that the relation \eqref{cltphiav} still holds thus the approximation is also proper in general $q$.

\subsection*{SYK with one time point with $\mu=0$}

Like in the previous section given any distribution we can always construct a variable with zero mean, such as 
\begin{align}
	\tilde{X}=X-\langle X\rangle,
\end{align}
so that we can consider any distribution we want.

We first consider a simple case with $q=2$, the Hamiltonian has the form
\begin{align}
    Y=\sum_{i\neq j}\text{sgn}(ij)X_{i}X_{j},\label{sykz2}
\end{align}
where $X_{i}$'s are identical and independent variables and the sign term mimics the sign function in the SYK model. Note that here we have no need to compute the explicit form of the sign function, since it has no effect yet. It is very similar to the previous case, but the difference is that here all $X_{i}$'s only appear once in the sum. Therefore for the square we have 
\begin{align}
    Y^{2}=\sum_{i\neq j}X_{i}^{2}X_{j}^{2}+2\sum_{i\neq j\neq k\neq l}\text{sgn}(ijkl)X_{i}X_{j}X_{k}X_{l},
\end{align}
comparing to \eqref{gey21} it lacks the middle term. Our previous computation shows that the approximation fails in the Gaussian case with $\mu= 0$ due to the middle term, thus the approximation is proper in this case no matter what the distribution is.  

We can have an explicit computation for the proposal, but first the convention needs to be consistent with the SYK model. Now the total number $N$ counts  the whole Majorana fermions and the number $q$ is for the number of the fermions in the interaction. Then for $N/q=2$ the expression \eqref{sykz2} is still valid for the partition function, except that the $N$ and $q$  have different meanings. Then the number of $X_{i}$ in \eqref{sykz2} becomes 
\begin{align}
	\frac{N!}{\left(\left(N/2\right)!\right)^{2}},
\end{align}
and the sum contains the number of the terms 
\begin{align}
		\frac{N!}{2!\left(\left(N/2\right)!\right)^{2}}.
\end{align}
Following the previous process we have 
\begin{align}
	Y^{2}-\Phi=\sum_{i\neq j}\left(X_{i}^{2}X_{j}^{2}-\phi_{i}\phi_{j} \right)
\end{align}
where $\phi$ is defined in \eqref{cltphii} and for the error 
\begin{align}
	     \langle \text{Error}^{2}\rangle=\left\langle \left(\sum_{i\neq j}\left(X_{i}^{2}X_{j}^{2}-\phi_{i}\phi_{j}\right)\right)^{2}\right\rangle-\left\langle \sum_{i\neq j}X_{i}^{2}X_{j}^{2}\right\rangle^{2}.\label{sykerr2}
\end{align}
Since there are only two parts in the above function and the sums are the same, we can only compare the two expressions in the brackets. The average of $\phi$ is zero due to the zero mean of $X_{i}$, so the proposal holds in the case. Explicitly the dominant terms in \eqref{sykerr2} come from the cross term in the first part and the whole second part, the number becomes
\begin{align}
		\frac{N!}{2!\left(\left(N/2\right)!\right)^{2}}\left(	\frac{N!}{2!\left(\left(N/2\right)!\right)^{2}}-1 \right)-\left(	\frac{N!}{2!\left(\left(N/2\right)!\right)^{2}}\right)^{2}.
\end{align}
We can find the dominant terms cancel out so that the proposal is valid in this simple case.

Then we can consider a more general Hamiltonian with arbitrary $N$ and $q$. Here we take the more familiar convention in SYK and  still define $p=N/q$, then the partition function becomes
\begin{align}
    z=\sum_{A}\text{sgn}(A)J_{A_{1}}\dots J_{A_{p}}.
\end{align}
 Before we check the approximation we can first have some computation about the numbers of the different terms in this partition function. The total number of the $J_{A}$'s is 
\begin{align}
    {N\choose q},
\end{align}
and the number of the combinations of $J_{A}$'s is 
\begin{align}
    \frac{N!}{p!(q!)^{p}},
\end{align}
while the times for a single $J_{A}$ appearing in the Hamiltonian is 
\begin{align}
    \frac{(N-q)!}{(p-1)!(q!)^{p-1}}. 
\end{align}
The square of the partition function can be given as 
\begin{align}
    z^{2}&=\sum_{A}\text{sgn}(A)^{2}J_{A_{1}}^{2}\dots J_{A_{p}}^{2}+\sum_{A,B}\text{sgn}(A)\text{sgn}(B)J_{A_{1}}^{2}\dots J_{A_{p-2}}^{2}J_{A_{p-1}}J_{A_{p}}J_{B_{p-1}}J_{B_{p}}\nonumber\\
    &\quad +\dots +\sum_{A,B}\text{sgn}(A)\text{sgn}(B)J_{A_{1}}\dots J_{A_{p}}J_{B_{1}}\dots J_{B_{p}}
    \label{cltz2}
\end{align}
the first non-diagonal term appears with four different $J_{A}$'s as the combination $A$ should be the permutation of $1,\dots,N$. And in the sums there may be many identical terms, here we may not give the explicit forms. To figure out the explicit expressions for the numbers in each sum we can count the numbers of different combinations of $r$ different $J_{A}$'s in the product of the combinations of $A$ and $B$. Obviously there's only one combination in the case $r=1$, actually $r=1$ is identical to $r=0$ since the last $J_{A_{i}}$ in a combination $A$ can be determined by the other $p-1$ $J_{A_{i}}$'s. For the larger $r$ we have to subtract the combinations in $r-i$ $(i<r)$ to get the correct ones, as it can give $r-i$ cases in the product of two combinations. Therefore we have 
\begin{align}
    r=2,\quad \frac{(2q)!}{2!(q!)^{2}}-1,\\
    r=3,\quad \frac{(3q)!}{3!(q!)^{3}}-{3\choose 2}\left(\frac{(2q)!}{2!(q!)^{2}}-1\right)-1,
\end{align}
from the above we can introduce the function $n_{r}$ for this counting 
\begin{align}
    n_{r}=\frac{(rq)!}{r!(q!)^{r}}-\sum_{s=2}^{r-1}{r\choose s}n_{s}-1.
\end{align}
From the calculation we can take mathematical induction to derive a simpler form 
\begin{align}
    n_{r}=\sum_{s=1}^{r}\frac{(sq)!}{s!(q!)^{s}}{r\choose s}(-1)^{r-s}+(-1)^{r},\quad r>1,
    \label{syknr}
\end{align}
and 
\begin{align}
	n_{0}=1,
\end{align}
where $n_{1}$ is absent according to the equation \eqref{cltz2}.

Then we calculate the numbers of the terms in the each sum in \eqref{cltz2},
\begin{align}
    \frac{N!}{p!(q!)^{p}}n_{0},  \frac{N!}{p!(q!)^{p}}{p\choose 2}n_{2},\frac{N!}{p!(q!)^{p}}{p\choose 3}n_{3},\dots, \frac{N!}{p!(q!)^{p}}n_{p},\label{sykseris}
\end{align}
and note that the first one is different to the others. And about the last one the dominant term of the subtrahend should be at the order $(N-q)!$, therefore the subtrahend is almost at the order $N(N-q)!$. Actually this number is far larger than the accurate one, since in \eqref{syknr} there's a factor $(-1)^{r-s}$ then the accurate number is approximately at $(N-q)!$.

Here we also have two conditions similar to \eqref{cltcond1},\eqref{cltcond2}, and first we consider the former one. As in this section we assume the mean of the $J_{A_{i}}$ is zero, therefore  we have 
\begin{align}
	\langle z^{2}\rangle = \left\langle \sum_{A}\text{sgn}(A)^{2}J_{A_{1}}^{2}\dots J_{A_{p}}^{2}\right\rangle=\frac{N!}{p!(q!)^{p}}\bar{J}^{2p}
\end{align}
where 
\begin{align}
	\langle J_{A_{i}}^{2}\rangle =\bar{J}^{2}.
\end{align}
For the error we evaluate it as 
\begin{align}
	\langle \text{Error}^{2}\rangle&=\langle \left(z^{2}-\Phi\right)^{2}\rangle-\langle z^{2}\rangle^{2},\\
	&=\left\langle \sum_{s=0,2}^{p}\left(\sum_{i_{1}\dots i_{q+s}}\left(z^{2}_{s}-\Phi_{s}\right) \right)^{2}\right\rangle-\langle z^{2}\rangle^{2},\label{sykerror2}
\end{align}
where $s$ does not take value $1$ for clarity and the cross terms of different sums are zero due to the zero mean of $J_{A_{i}}$. Then we find that the computation is similar to before that for $s>0$ the nonzero contribution under the average is the diagonal terms in the square, while for $s=0$ the whole terms contribute to the computation of the error.  Which means to find out the dominant sum $s$ we should compare the square of the first term with the rest  terms in \eqref{sykseris}, and note that the last one cancels out in \eqref{sykerror2} so that the non-trivial largest number should be the one labeled by $p-1$.  And we give an upper bound for it
\begin{align}
	\frac{N!}{p!(q!)^{p}}{p\choose p-1}n_{p-1}<p^{2}\frac{N!}{p!(q!)^{p}}\frac{(N-q)!}{(p-1)!(q!)^{p-1}}.
\end{align}
Note that the number of the identical terms in this sum may have an effect but we expect the upper bound is still valid. Then we can find that when $N$,$q$ are sufficiently large it is subordinate to the square of the first term in \eqref{sykseris}. Therefore we write the dominant terms in \eqref{sykerror2} 
 \begin{align}
 	\frac{N!}{p!(q!)^{p}}n_{p}-\left(\frac{N!}{p!(q!)^{p}}\right)^{2} 
 	\label{sykzm1}
 \end{align}
where the cross terms of the first part and the whole second part contribute.  Recall the expression \eqref{syknr} the dominant terms in the above equation cancel out, which implies the proposal holds.

Then we consider the second condition \eqref{cltcond2} that for a single variable we can view $J_{A_{i}}^{2}-\Phi_{i}$ as a constant. The expression \eqref{sykerror2} is still valid so that the thing we need to consider is to find the correction to \eqref{sykzm1}.
As an example we consider the behavior of the second sum labeled by $s=2$, the diagonal contribution after the subtraction should be 
\begin{align}
  \alpha\sum J_{A_{p-1}}J_{A_{p}}J_{B_{p-1}}J_{B_{p}},
\end{align}
where $\alpha$ is determined by the quotient of the two numbers of the two sum
\begin{align}
	\alpha= \frac{N!}{p!(q!)^{p}}{p\choose 2}n_{2}/\left({N\choose 2q}\frac{(2q)!}{2!(q!)^{2}}n_{2}\right).
\end{align}
Therefore after the square the total number becomes 
\begin{align}
	\left(\frac{N!}{p!(q!)^{p}}{p\choose 2}n_{2}\right)^{2}/\left({N\choose 2q}\frac{(2q)!}{2!(q!)^{2}}n_{2}\right),
\end{align}
comparing to the two terms in \eqref{sykzm1} the above term is subordinate.  Similarly we can find that after the subtraction the largest number in the sums $s>0$ occurs in the case $s=p-1$, which gives 
\begin{align}
	\left(\frac{N!}{p!(q!)^{p}}{p\choose p-1}n_{p-1}\right)^{2}/\left({N\choose N-q}\frac{(N-q)!}{(p-1)!(q!)^{p-1}}n_{p-1}\right).
\end{align}
The above contribution is still subordinate to the two terms in \eqref{sykzm1}, which means the proposal is proper in this case.

Thus for the SYK model we can conclude that when the mean is zero the proposal works well no matter what the distribution is. 

\subsection*{SYK with one time point with $\mu\neq 0$}
\label{proof}
Like what we have discussed before the case with non-zero mean is different to last section. The whole computation is similar to the case with $\mu=0$ except the mean terms, we can take relevant expressions from the previous sections. We consider the case with general $N/q$, 
\begin{align}
    z=\sum_{A}\text{sgn}(A)J_{A_{1}}\dots J_{A_{p}},
\end{align}
the main task here is find the dominant terms.
The sign functions may have some effect which makes $s=q-1$ not dominant, such as when $N/q$=2 we have 
\begin{align}
	\left\langle\sum_{A}\text{sgn}(A)J_{A_{1}}J_{A_{2}}\right\rangle=\sum_{A}\text{sgn}(A)\langle J_{A_{i}}\rangle^{2}={N/2-1\choose [N/4]}\langle J_{A_{i}}\rangle^{2},\label{syksign1}
\end{align}
where the square bracket means the integer part. Actually here $N/4$ is $q/2$, in the later computation we always have even $q$ therefore we'll omit the square bracket. 

To derive this parameter explicitly we have two different ways, one is to find the sequence functions over different $N$ which can be solved by Mathematica.  Another is to consider different combinations, we can define the positive and the negative combinations then their difference gives the parameter. Given $N$ different numbers we can have a permutation group which can be divided into the even and odd parts, and the two parts contain the same number of elements. If the indices of $J$ have no order then the numbers of the even and odd parts are equal, so that the parameter is zero. But in the SYK model the indices are listed in a particular order, it makes the numbers of the two parts different. 

To proceed the computation we define the sign of a $q$-sequence,
\begin{align}
	\text{sgn}(J_{A_{i}})=(-1)^{a_{i_{1}}+\dots+a_{i_{q}}},\quad q=0\text{mod}(4)  \label{syksigna}
\end{align}
where when $q=2\text{mod}(4)$ the sign will be opposite. Note that here we take the $a_{i}$ from the indices of $J_{A_{i}}$, but what in the superscript should be the initial sites of the indices. When the indices take the values $1,\dots,N$, the values themselves naturally label their initial sites. But when the indices take arbitrary numbers or symbols, we can define an initial order and the sites with any  permutation. After the assignment complete we can take the definition \eqref{syksigna} to define the sign of any given $q$-sequence.

Then for a combination $A$ containing $r$ negative q-sequences we have the argument 
\begin{align}
	\text{sgn}(A)=1,\quad \text{if } \sum_{i=1}^{r}s_{i}= 0 \text{mod} (2),
\end{align}
where $s_{i}$ is the site of the $i$-th negative $q$-sequence.
We give the condition above as the negative $q$-sequences seems appearing in pairs. For $N/q=2$ the positive combination $A$ contains two positive q-sequences, so our mission is to find the first sequence with positive sign. As the combination has an order the first index in $J_{A_{1}}$ is 1, we only need to consider the others. The positive $q$-sequence has the number
\begin{align}
	{N/2-1\choose 1}{N/2\choose q-2}+ {N/2-1\choose 3}{N/2\choose q-4}+\dots+ {N/2-1\choose q-1}{N/2\choose 0},
\end{align}
where $q=N/2$ the first chooses odd number of odd integers while the second chooses even number of even integers. Then the negative $q$-sequence has 
\begin{align}
	{N/2-1\choose 0}{N/2\choose q-1}+ {N/2-1\choose 2}{N/2\choose q-3}+\dots+ {N/2-1\choose q-2}{N/2\choose 1},
\end{align}
the difference of the two numbers will give the parameter
\begin{align}
	d(p,q)=\sum_{i=0}^{q-1}(-1)^{i+1}{pq/2-1\choose i}{pq/2 \choose q-1-i},
\end{align}
 which coincides with \eqref{syksign1} when $N/q=2$. For general $p=N/q$ we can also find an expression for finding the first positive $q$-sequence
 \begin{align}
 	d(p,q)=\frac{1}{p-1}{pq/2-1\choose q/2},\label{sykrela2}
 \end{align}
where $q$ is even and it can be verified by numerics.
 
 And note that for different $q$'s there may be a difference with the minus sign, which is explained in \eqref{syksigna}.

For general $N/q$ we can try to compute $\gamma$ in the expression below by the recursion
\begin{align}
    \left\langle\sum_{A}\text{sgn}(A)J_{A_{1}}\dots J_{A_{p}}\right\rangle=\sum_{A}\text{sgn}(A)\langle J_{A_{i}}\rangle^{p}=\gamma\langle J_{A_{i}}\rangle^{p}.
\end{align}
In the previous computation we have calculated the case with $p=2$, we only need to derive the $p+1$ case from the $p$ one. Given $N$ numbers we can give the difference between the positive combinations  and the negative ones which is denoted as $D(p,q)$ and we have
\begin{align}
D(2,q)=d(2,q),\quad D(p+1)=f(D(p,q)).
\end{align}
where $f$ is an unknown function.  
When we add extra $q$ numbers to the $N$ case which is already known, in the new permutation we can fix one number in the extra $q$ ones at the $p+1$ site in case of repetition.   For simplicity we let the additional $q$ numbers be $1,\dots,q$ and put it at the first site, and we always fix the number $1$ in this site.  

We determine the first site through the procedure in the case of $p=2$, which can be divided into the positive and negative parts. After choosing $q$ numbers from the $N+q$ ones, about the left $N$ numbers we assume we already know the difference between the positive combinations and the negative ones.  To illustrate the product of the sequences, we define the positive (negative) sequences as $P_{+}$ ($P_{-}$) so that we have
\begin{align}
	P_{+}P_{+}=P_{+},\quad P_{-}P_{-}=P_{+},\quad P_{+}P_{-}=P_{-}.
\end{align}
The product of two positive or two negative sequences gives a positive one while the product of one  positive and one negative sequences gives a negative one. Then we can derive the sequences in the case  $N+q$ as 
\begin{align}
	P_{+}^{N+q}=P^{q}_{+}P^{N}_{+}+P^{q}_{-}P^{N}_{-},\quad P_{-}^{N+q}=P^{q}_{+}P^{N}_{-}+P^{q}_{+}P^{N}_{-},
\end{align}
therefore the difference of the positive and negative sequences can be illustrated as 
\begin{align}
	P_{+}^{N+q}-P_{-}^{N+q}=(P_{+}^{q}-P_{-}^{q})(P_{+}^{N}-P_{-}^{N}).
\end{align}
Which means the recursion can be expressed as  
\begin{align}
	D(p,q)=d(p,q)D(p-1,q)=\prod_{s=2}^{p}d(s,q),
\end{align}
inserting the equation \eqref{sykrela2} it becomes 
\begin{align}
	D(p,q)=\frac{(pq/2-1)!}{(p-1)!(q/2-1)!((q/2)!)^{p-1}}=\frac{(pq/2)!}{p!((q/2)!)^{p}}.
\end{align}

Then we can try to evaluate the cross terms in the square of the partition function \eqref{cltz2}. Consider the terms with $p-s$ squares in the sequence,
\begin{align}
	\sum_{A,B}\text{sgn}(A)\text{sgn}(B)J_{A_{1}}^{2}\dots J_{A_{p-s}}^{2} J_{A_{p-s+1}}\dots J_{A_{p}}J_{B_{p-s+1}}\dots J_{B_{p}},
\end{align}
after the average over $J$ the number in the sum becomes 
\begin{align}
	    d_{s}=\sum_{r=2}^{s}{p-r\choose s-r}{N\choose pq-rq}\frac{(pq-rq)!}{(p-r)!(q!)^{p-r}}D(r,q)^{2}(-1)^{s-r}+{p\choose s}(s-1)(-1)^{s-1}\frac{(pq)!}{p!(q!)^{p}},
	\label{sykds}
\end{align}
where the derivation is similar to the equation \eqref{syknr}.

We can find an upper bound for the number \eqref{sykds}
\begin{align}
	{N\choose pq-sq}\frac{(pq-sq)!}{(p-s)!(q!)^{p-s}} D(s,q)^2,
\end{align}
 Then for $s=2,\dots,p-1$ we can compare the above number to the first one in \eqref{sykseris}, from this we may find out the dominant one.
 
By a few numerical comparisons it seems that the dominant terms do not locate at a particular $s$ as the two parameters $p$,$q$ vary. In this computation we can directly take the equation \eqref{sykds} rather than the one with approximation.

\section{Some details about the fermion integral} 
The relevant integral is
\bea 
I_f^N&=&\int \d^{2N}\psi \exp\(T[\sum_i[\Sigma_{LR}\psi_i^L\psi_i^R+\Sigma_L \theta_i\psi_i^L+\Sigma_R \theta_i\psi_i^R]\)\, \\
&=& \(\int \d \psi^L \d \psi^R \exp(T \Sigma_{LR}\psi^L\psi^R)\exp(T\Sigma_L \theta  \psi^L)\exp(T\Sigma_R \theta \psi^R) \)^N
\eea 
Before evaluating this integral we need some specifications. Here $\psi^{L(R)}$ are not Grassmann numbers but can be transferred into Dirac fermions as
\bea 
&&c=\frac{\psi^L+\im \psi^R}{2},\quad c^\dagger=\frac{\psi^L-\im \psi^R}{2},\quad \psi_L=c+c^\dagger,\quad \psi_R=\im(c^\dagger-c)\, \\
&&\{c,c^\dagger\}=1,\quad \{c,c\}=\{c^\dagger,c^\dagger\}=0\, .
\eea 
Then the integral $I_f$ becomes 
\bea 
I_f&=&2 \int \d c\d c^\dagger e^{a_1 (c c^\dagger-c^\dagger c)+a_- c+a_+ c^\dagger},\quad a_1=\im T\Sigma_{LR},\quad a_\pm=T(\Sigma_L\pm \im \Sigma_R)\theta,\\
&=&4\cosh(\sqrt{a_1^2+a_+ a_-})=4\cosh(T\sqrt{\Sigma_L^2+\Sigma_R^2-\Sigma_{LR}^2})
\eea 
Next let us use the same method to compute \eqref{i4}:
\bea 
I_4^N &=&\int \d^{4N}\psi \exp\(\Sigma_{ab}\psi_i^a\psi_i^b\) \\
&=&\{\int \d^4 \psi \exp \left(\frac{1}{2}\Sigma_{12}\psi^1\psi^2+\Sigma_{13}\psi^1\psi^3+\Sigma_{14}\psi^1\psi^4
\right. \nn \\
&& \quad \left.  \Sigma_{23}\psi^2\psi^3+\Sigma_{24}\psi^2\psi^4+\Sigma_{34}\psi^3\psi^4 \)\}^N.
\eea 
Generally to compute $I_4$ let us introduce two Dirac fermions 
\bea 
\psi_1=\sqrt{h}(c_1+c_1^\dagger),\quad \psi_2=\im\sqrt{h}(c_1-c_1^\dagger),\quad \psi_3=\sqrt{h}(c_2+c_2^\dagger),\quad \psi_4=\im\sqrt{h}(c_2-c_2^\dagger),
\eea 
such that $\Sigma_{ab}\psi_i^a\psi_i^b$ can be written as a $4\times 4$ matrix $H$ with non-vanishing elements
\bea
&&H_{11}=-\im h (\Sigma_{12}+\Sigma_{34}),\quad H_{14}=-h (\Sigma_{13}-\Sigma_{24})-\im h (\Sigma_{14}+\Sigma_{23}) \\
&& H_{22}=\im h (\Sigma_{34}-\Sigma_{12}),\quad H_{23}=-h (\Sigma_{13}+\Sigma_{24})+\im h (\Sigma_{14}-\Sigma_{23})\\
&& H_{32}=h (\Sigma_{13}+\Sigma_{24})+\im h (\Sigma_{14}-\Sigma_{23}),\quad H_{33}=-\im h (\Sigma_{34}-\Sigma_{12}) \\
&& H_{41}= h (\Sigma_{13}-\Sigma_{24})-\im h (\Sigma_{14}+\Sigma_{23}),\quad H_{44}=\im h (\Sigma_{12}+\Sigma_{34}).
\eea 
By diagonalizing this matrix we can compute $I_4$ as the trace of the exponential of the matrix
\bea 
&&I_4=2\(\cos\(h \sqrt{(\Sigma_{14}-\Sigma_{23})^2+(\Sigma_{13}+\Sigma_{24})^2+(\Sigma_{12}-\Sigma_{34})^2}\)\right.  \nn \\
&&\qquad  \left.+\cos\(h \sqrt{(\Sigma_{14}+\Sigma_{23})^2+(\Sigma_{13}-\Sigma_{24})^2+(\Sigma_{12}+\Sigma_{34})^2}\)\).
\eea 
And note that the change of the variable from $\psi$ to $c$,$c^{\dagger}$ will introduce an additional coefficient in the integral 
\begin{align}
    (2h)^{2},
\end{align}
the final result is the product of the two terms.

\end{document}